\documentclass[useAMS,usenatbib]{mn2e}
\def\Msol{\hbox{M$_\odot$}}

\def\kms{\hbox{km$\,$s$^{-1}$}}
\def\cmt{\hbox{cm$^{-3}$}}
\def\Apix{\hbox{\AA$\,$pix$^{-1}$}}
\def\one{\,{\sc i}}             % for producing Na I as Na\one\ etc.
\def\two{\,{\sc ii}}

\newcommand\fsec{\hbox{$.\!\!^{\rm s}$}}

\usepackage{amsmath}
\usepackage{amssymb}
\usepackage{mathrsfs}
\usepackage{color}
\usepackage{upgreek}
\usepackage{graphicx}
\usepackage{overpic}

\title[VIMOS-IFU observations of NGC 253]{Spatially resolved optical IFU spectroscopy of the inner superwind of NGC 253\thanks{Based on observations collected at the European Organisation for Astronomical Research in the Southern Hemisphere, Chile under programmes 060.A-9123 and 081.C-0808.}}
\author[M.S.\ Westmoquette et al.] {M.\ S.\ Westmoquette$^1$\thanks{E-mail: msw@star.ucl.ac.uk}, L.\ J.\ Smith$^2$, J.\ S.\ Gallagher III$^{3}$ \\
$^1$Department of Physics and Astronomy, University College London, Gower Street, London, WC1E 6BT\\
$^2$Space Telescope Science Institute and European Space Agency, 3700 San Martin Drive, Baltimore, MD 21218, USA\\
$^3$Department of Astronomy, University of Wisconsin-Madison, 5534 Sterling, 475 North Charter St., Madison WI 53706, USA\\
}
\date{Accepted 2011 March 8.  Received 2011 February 28; in original form 2010 November 29}
\pagerange{\pageref{firstpage}--\pageref{lastpage}}
\pubyear{2010}
\begin{document}
\maketitle
\label{firstpage}
%%%%%%%%%%%%%%%%%%%%%%%%%%%%
\begin{abstract}
We present optical integral field unit (IFU) observations (VLT/VIMOS-IFU and WIYN/SparsePak), and associated archival deep H$\alpha$ imaging (MPG/ESO 2.2m WFI), of the nearby starburst galaxy NGC~253. With VIMOS we observed the nuclear region and southern superwind outflow in detail with five pointings, and with the WIYN/SparsePak IFU we observed two partially overlapping regions covering the central disk and northern halo. The high signal-to-noise of the data and spectral resolution (80--90~\kms) enable us to accurately decompose the emission line profiles into multiple components.

The combination of these datasets, together with the wealth of information on NGC~253 available in the literature, allow us to study the starburst-driven superwind in great detail. We investigate the known minor axis outflow cone, which is well-defined in the H$\alpha$ imaging and kinematics between radii of 280--660~pc from the nucleus. Kinematic modelling indicates a wide opening angle ($\sim$60$^{\circ}$), an inclination consistent with that of the disk, and deprojected outflow speeds of a few 100~\kms\ that increase with distance above the plane. The [N\two]/H$\alpha$ and [S\two]/H$\alpha$ line ratio maps imply that a significant fraction of the wind optical emission lines arise from shocked gas, with localised pockets/filaments of strongly shocked gas. From the kinematics, the cone appears partially closed in at least one place, and very broad H$\alpha$ line widths ($>$400~\kms\ FWHM) suggest there is material filling the cone in some regions. Extrapolation of the cone to its apex shows it is not centred on the starburst nucleus, suggesting the wind is deflected and collimated by the dense circumnuclear material. We discuss the implications of these findings on our understanding of the origins and evolution of the superwind. No evidence for an outflow is found on the north-western side of the disk out to $>$2~kpc in our optical data, due to obscuration by the foreground disk. The lack of an obvious connection between the inner ($r$$<$1~kpc) H$\alpha$ and X-ray bright outflow cone and the large-scale ($r$$\lesssim$10~kpc) X-ray ``horns'' is also discussed.

\end{abstract}


%%%%%%%%%%%%%%%%%%%%%%%%%%%%
\section{Introduction}\label{intro}

\begin{figure*}
\centering
\includegraphics[width=\textwidth]{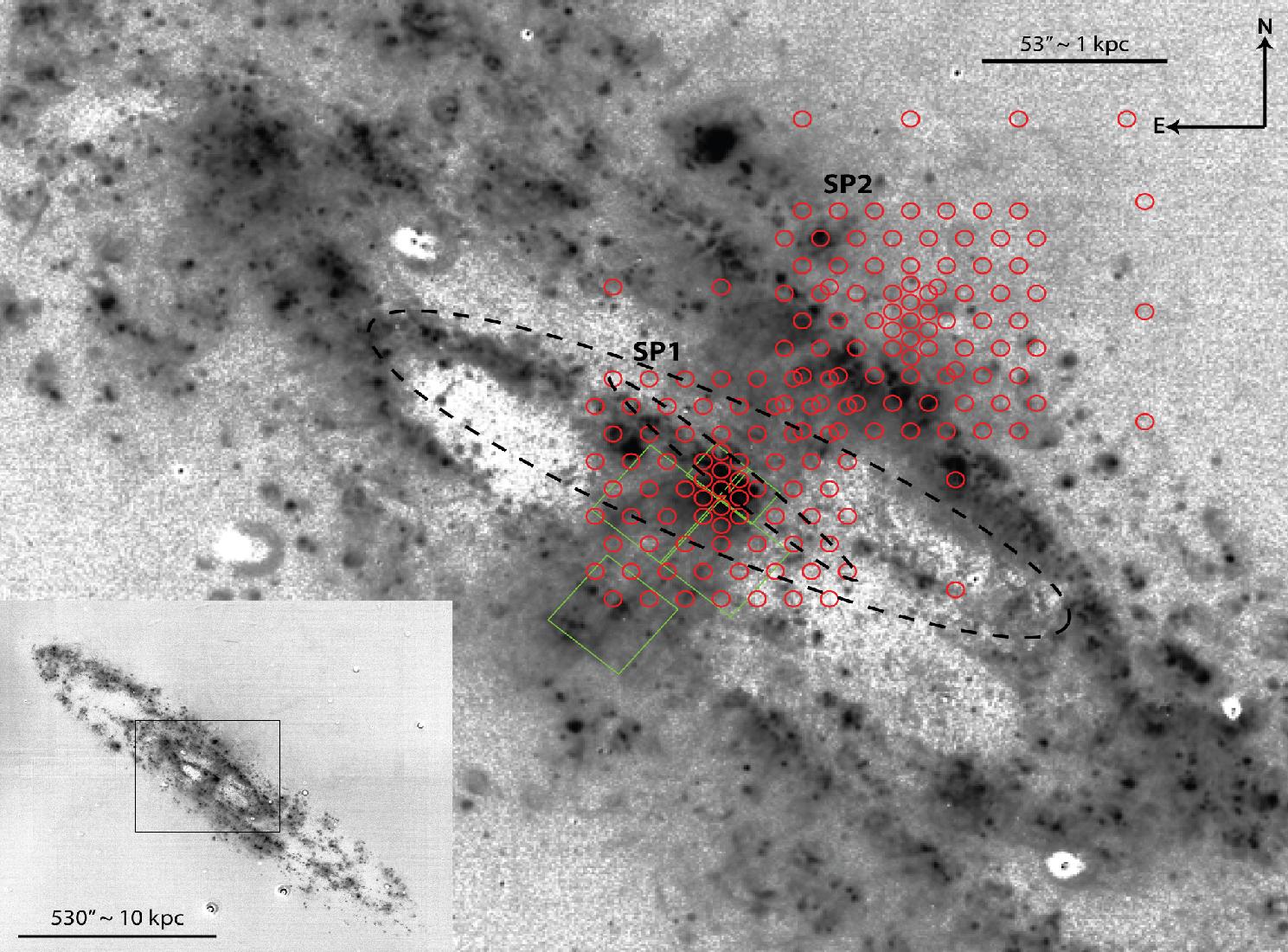}
\caption{MPG/ESO 2.2m WFI continuum subtracted H$\alpha$ image of NGC 253 (square-root scaled and inverted for clarity) with the VIMOS-IFU (green boxes) and SparsePak footprints (arrays of red circles) shown. Dashed lines indicate the extent of the projected $x_1$ and $x_2$ bar orbits \citep{sorai00, das01}. Inset shows a zoomed out version of the same H$\alpha$ image with a box outlining the field-of-view of the main figure.}
\label{fig:finder_large}
\end{figure*}

\begin{figure*}
\begin{minipage}{0.49\textwidth}
\begin{overpic}[width=\textwidth]{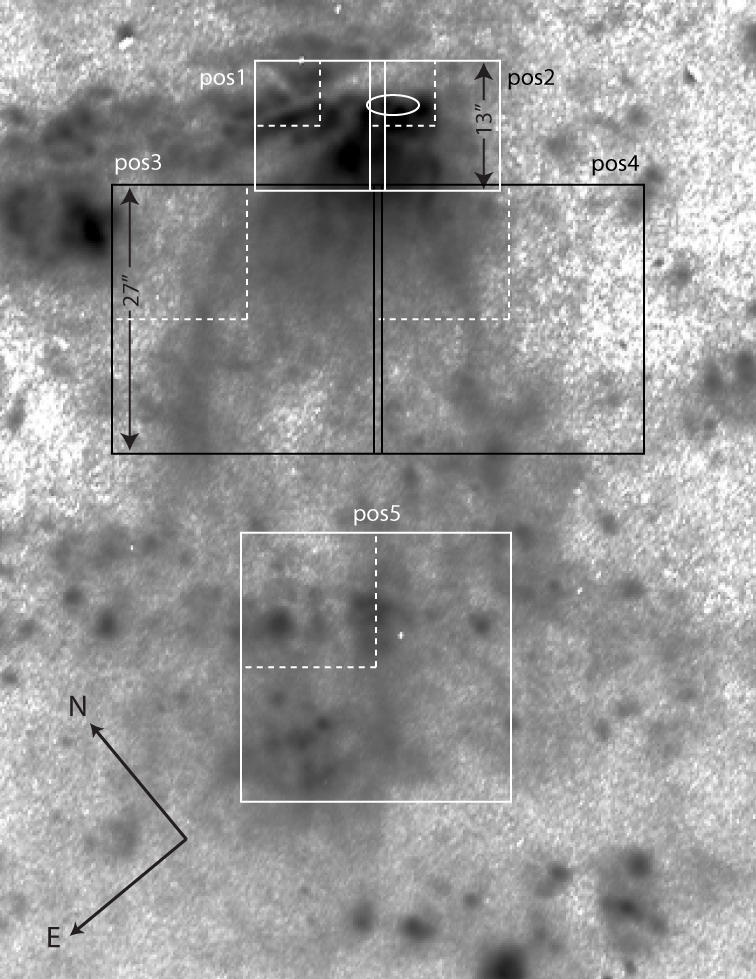}
\put(2,96){\textcolor{black}{\large (a)}}
\end{overpic}
\end{minipage}
\hspace{0.2cm}
\begin{minipage}{0.47\textwidth}
\begin{minipage}{\textwidth}
\begin{overpic}[width=\textwidth]{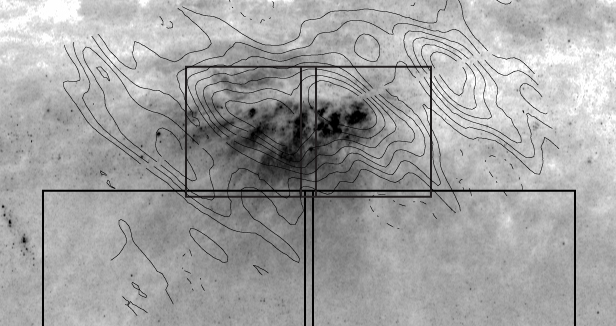}
\put(3,47){\large (b)}
\end{overpic}
\end{minipage}
\begin{minipage}{\textwidth}
\begin{overpic}[width=\textwidth]{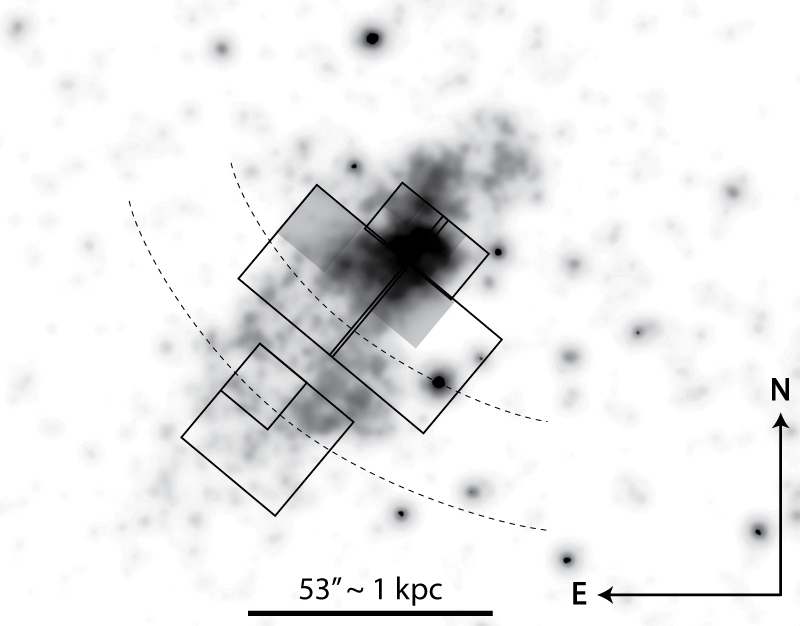}
\put(3,70){\large (c)}
\end{overpic}
\end{minipage}
\end{minipage}
\caption{(a) Zoom-in of the WFI H$\alpha$ image (again greyscale inverted and stretched to enhance low surface-brightness features) with the VIMOS-IFU footprints shown. The dashed lines enclose quadrant 3 of the detector for which instrument problems resulted in the blurring of the spectral line profiles, while the white ellipse marks the nuclear region including the radio, X-ray and IR peaks; (b) \textit{HST} F814W ($I$-band) image of the central region of NGC~253 (again in inverted greyscale) showing the VIMOS-IFU footprints and the SiO (($v$=0, $J$=2$\rightarrow$1) integrated intensity contours from \citet{garcia-burillo00} that trace the inner and outer molecular tori in the circumnuclear disk; (c) \textit{Chandra} soft X-ray image (inverted greyscale) from \citet{strickland00} again with the VIMOS-IFU footprints overlaid (the dashed curved lines delineate the region in which \citet{strickland00} identify and model the limb-brightened cone outflow).
}
\label{fig:finder}
\end{figure*}

\begin{figure}
\centering
\includegraphics[width=0.49\textwidth]{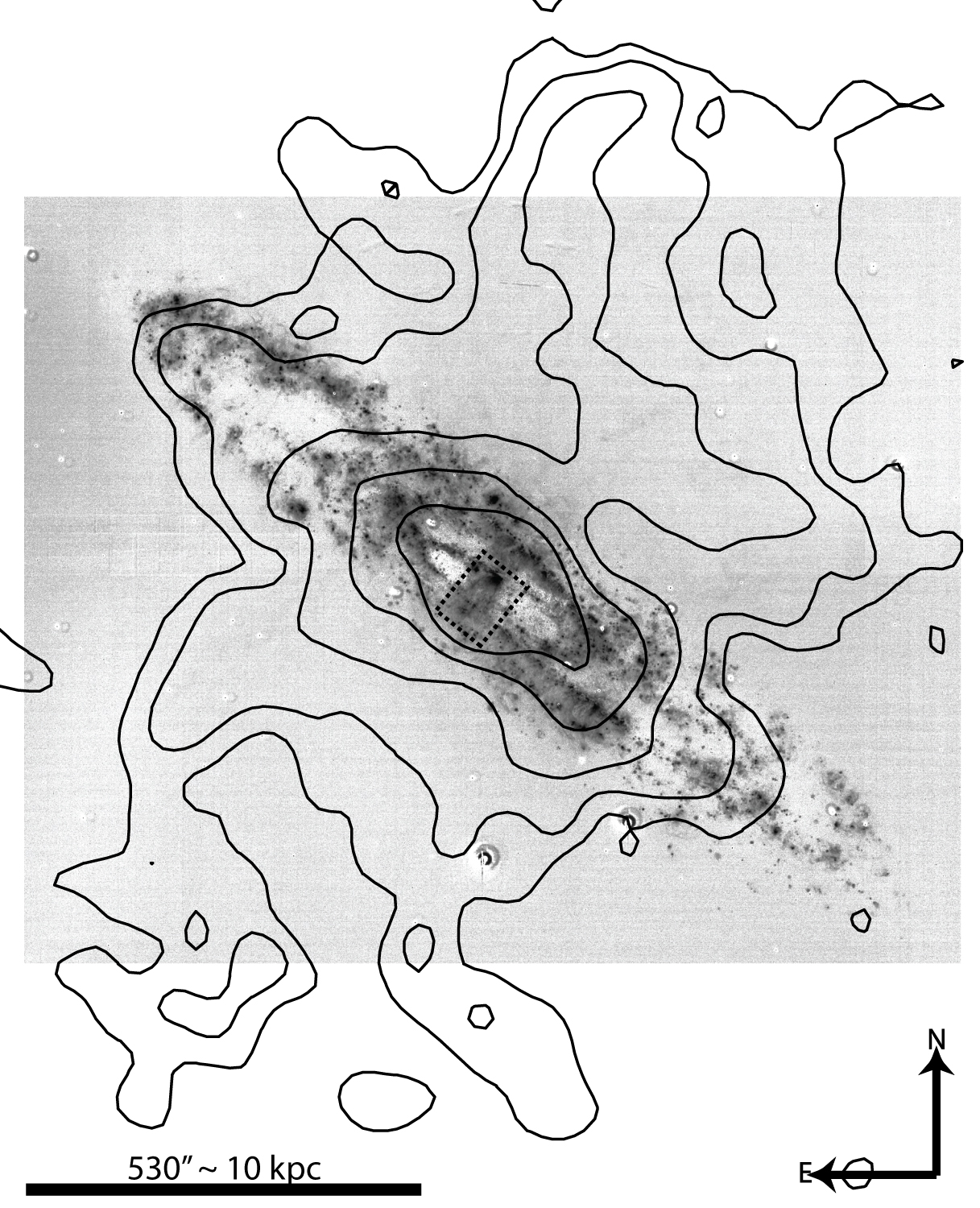}
\caption{The extent of the large-scale X-ray superwind seen with \textit{ROSAT} is shown in contours \citep{pietsch00, strickland02} overlaid on the WFI H$\alpha$ image. The dashed box around the nuclear regions indicates the coverage of Fig.~\ref{fig:finder}a and also the approximate extent of the \textit{inner} superwind.}
\label{fig:finder_ROSAT_WFI}
\end{figure}

%intro para with link to high-z universe and big questions

The consequences of a starburst can dramatically affect the evolution of the host galaxy through photoionization and mechanical/chemical feedback, and (if powerful enough) the intergalactic medium through the escape of ionizing radiation and the products of nucleosynthesis. By studying local examples of starburst-driven outflows, it is possible to investigate the underlying physics of the feedback processes, and then apply this knowledge to gain a better understanding of the phenomenon of starburst outflows. This is important since it is now recognised that starburst-driven outflows are ubiquitous at high redshift \citep{pettini01, conselice03, shapley03, martin09}.
%Since interactions/mergers and hence starbursts and starburst-driven outflows are thought to have been much more prevalent and intense in the early Universe, it is essential to study the present-day manifestations of these processes in detail where the achievable spatial resolution and sensitivity allow much sharper measurements of the key parameters.
In recent years, observations of local systems have been providing more and more evidence that superwinds are not the simple, smooth, thermally-driven outflows that were once envisaged \citep[e.g.][]{cc85}. A starburst is typically very clumped, with each clump containing multiple super-star clusters (SSCs). This seems naturally to lead to highly structured, multi-phase ($10^2$--$10^7$~K), mass loaded winds. It is not known, however, how the flow from individual star-cluster sources is collimated, where it is accelerated and importantly if/where mass is entrained/loaded into the flow. To address these issues and others, we have studied the inner wind regions of the nearby starbursting galaxies NGC~1569 and M82 using optical integral field unit (IFU) observations \citep{westm07a, westm07b, westm09a, westm09b}.

In this paper we extend this project to examine NGC~253, a much larger system with a nuclear starburst and large-scale superwind. We present optical IFU observations and associated deep H$\alpha$ imaging in an attempt to understand the inner region of the outflow in more detail. 
%Its proximity (3.9~Mpc; see below), starburst strength, and almost edge on inclination \citep[$i\sim 78^{\circ}$;][]{pence80} makes it of fundamental importance as an analogue to intensely star-forming galaxies at high-$z$.

%Deprojected view of galaxy in \citet{engelbracht98}, deprojected view of bar in \citet{das01}.

%Escape velocity $v_{\rm esc}$$\sim$280~\kms\ (derived using rotation velocity of 200~\kms) \citep{heesen09a}.

%Superwind 
Early, deep optical and X-ray imaging of NGC 253 showed the presence of an extended ($r < 1$~kpc) nuclear outflow \citep{demoulin70, ulrich78, fabbiano84, mccarthy87}. This superwind has been observed in optical emission lines \citep{schulz92, m-t93} and soft X-rays with \textit{Chandra} \citep{strickland00}. \citet{strickland02} found a strong spatial correspondence between the (\textit{Chandra}) soft X-ray emission and H$\alpha$ emission on scales down to $\lesssim$20~pc, implying that the standard adiabatic model of superwind flows \citep{cc85} cannot hold. Instead of tracing the hot component of the wind outflow as the model predicts, the X-ray emission, like H$\alpha$, primarily traces the interface zones between the hot and warm phases of the outflow.

In addition to this nuclear outflow, huge lobes of diffuse X-ray and far-IR dust emission extend up to 8--9~kpc away from the disk, and are thought to trace the walls of two large-scale bubbles containing low-density hot gas \citep{pietsch00, strickland00, strickland02, bauer08, kaneda09}. The link, however, between the inner (conical, limb-brightened) wind studied in this paper and seen with \textit{Chandra}, and the large-scale ($\sim$10~kpc) diffuse bubbles/lobes seen with the \textit{ROSAT}, \textit{XMM-Newton} and \textit{AKARI} telescopes, is still unclear. The inner superwind evidently is driven by a burst of star formation that has been taking place in the central $\sim$500~pc for the past 20--30~Myr \citep{rieke80, engelbracht98} at a rate of $\sim$3~\Msol~yr$^{-1}$ \citep{ott05}.

%Starburst + nucleus properties
The nuclear starburst region itself is highly obscured \citep[$A_{V} \sim 5$--18~mag;][]{kornei09, fernandez-ontiveros09} and morphologically complex. The resulting confusion has caused a proliferation of conflicting views of the nature of the nuclear region. On the smallest scales, optical, near-IR and mm sources appear to trace out a ring of size $\sim$50~pc \citep{forbes00}. This region is contained within two nested circumnuclear tori observed in CO, SiO and H$^{13}$CO$^{+}$ \citep{garcia-burillo00, paglione04}, and this, in turn, within a much larger ($\sim$500~pc) cold gas torus associated with the 7~kpc long stellar bar \citep{scoville85, mauersberger96, das01}. 
%The molecular circumnuclear tori are thought to trace gas associated with the bar $x_{2}$ orbits \citep{peng96, garcia-burillo00, paglione04}.
The galactic nucleus of NGC~253 has historically been associated with a powerful, compact, non-thermal radio source (TH2) often cited to be a low-luminosity AGN \citep{ulvestad97, weaver02}. However, high-resolution radio \citep{brunthaler09}, near-IR \citep{fernandez-ontiveros09}, and IR \citep{engelbracht98} imaging do not show any evidence of a suitable point source that would indicate nuclear activity. Recent measurements of the 2D stellar kinematics with VLT/SINFONI by \citet{muller-sanchez10} identify the kinematic centre to lie within a $\sim$$1\farcs2$ region containing both TH2 and the central point-like X-ray source (X-1). Water maser emission has also in the past been associated with the presumed nuclear radio source TH2 \citep{nakai95}. However, \citet{brunthaler09} showed that the 22~GHz water masers are not related to AGN activity, but almost certainly associated with star formation. Thus, although some evidence for central black hole activity remains \citep{turner85, mohan02, weaver02}, the overwhelming consensus points towards a purely starburst origin for the radio--X-ray emission seen in NGC~253.

%SSCs
A number of young (20--30~Myr) massive star clusters have been discovered within the inner star-forming ring \citep{watson96}. More recently \citet{kornei09} presented IR observations of the near-IR peak, 4$''$ from TH2 on the south-west side of the ring, and find a complex system, presumably consisting of several epochs of star formation as evidenced by the simultaneous presence of RSG, OB and WR star signatures. By comparison to population synthesis models they find a very young age of $\sim$6~Myr, a mass of $\sim$$10^7$~\Msol, and $A_{V}$ $\sim$ 18~mag for this region. 
%[S\three]/H$\alpha$ line ratios of this region, derived from \textit{HST} imaging, are consistent with photoionization rather than SNR shocks \citep{forbes00}.
\citet{fernandez-ontiveros09} detect multiple pc-sized knots in their VLT/NACO adaptive optics imaging, and conclude that they are probably young ($<$10~Myr), dust-embedded ($A_{V}$ $\sim$ 5--11) star clusters similar to those found in NGC~5253 and NGC~4038/9.

%extraplanar HI
NGC~253 also shows some large-scale peculiarities. \citet{malin97} found an extended but relatively smooth, low surface brightness, stellar halo with an asymmetric extension to the south which they interpreted as evidence for an ancient merger. The H\one\ 21~cm line study by \citet{boomsma05} revealed two further anomalies: an H\one\ disk that is considerably smaller then the flattened stellar envelope found by Malin \& Hadley, and extraplanar H\one\ around the eastern half of the galaxy extending to vertical distances of approximately 12~kpc above the thin star-forming disk and accounting for 3\% of the total H\one\ mass. They interpret the extraplanar H\one\ as a product of the starburst which has lifted interstellar material out of the disk. Evidence that interstellar gas is being entrained by the wind and lifted into the NGC~253 halo also comes from the \textit{AKARI} detection of thermal dust emission from the northern wind boundaries \citep{kaneda09}. 

%broad component
%Broad emission line wings have been detected in a number of other nearby starburst galaxies (e.g.~\citealt{izotov96, homeier99, marlowe95, mendez97}; \citealt*{sidoli06}; \citealt{vanzi06}). However, due to mismatches in spectral and spatial resolution and in the specific environments observed, the nature of the energy source for these broad lines has been contested, resulting in a large number of possible explanations being proposed. Recent detailed IFU (integral field unit) studies of the ionized ISM in NGC 1569, however, have shed a considerable amount of light on this problem \citep{westm07a, westm07b, westm08}.

%By mapping out the properties of the individual line components (including the broad underlying component), \citet{westm07a} identified a number of correlations that allowed them to determine the likely origin of the broad component. They concluded that the evaporation and/or ablation of material from cool interstellar gas clouds caused by the impact of the high-energy photons and fast-flowing cluster winds \citep{pittard05} produces a highly turbulent layer on the surface of the clouds \citep{slavin93, binette99} from which the emission arises. \citetalias{westm07c} showed that this explanation is also applicable to the broad lines in M82. They argued that since the high pressure ISM is highly fragmented, with many small clouds well mixed in with the star clusters, there are many cloud surfaces with which the copious ionizing photons and fast winds can interact.

%aims and objectives of this study
The structure and dynamics of the NGC~253 disk and halo are clearly complex. They vary both on the large and small scales, and are complicated by the inflows and outflows caused by the bar and starburst, and by the magnetic field \citep{heesen09a, heesen09b}. Unlike in M82's case, the optical outflow in NGC~253 has received comparatively little attention despite its prominence and obvious importance. We have, therefore, obtained a set of high resolution, integral field, optical spectra of the inner regions of NGC~253 at a range of spatial scales, but comparable spectral resolution. With the VLT/VIMOS-IFU we observe the nuclear region and southern outflow in detail with five pointings, and with the WIYN/SparsePak IFU we observed two partially overlapping regions covering the central disk and northern halo. These two datasets together provide a wealth of information on the state of the ionized gas, and present a unique opportunity to study the $T\sim 10^4$~K phase of the inner wind and disk-halo interface.

In Section~\ref{sect:obs} we present the three sets of observations we have used to study the NGC~253 outflow: deep H$\alpha$ imaging with MPG/ESO 2.2m WFI, and optical IFU spectroscopy with VLT/VIMOS and WIYN/SparsePak. In this section we describe the observational details and data reduction steps undertaken. Section~\ref{sect:results} contains a description of the results obtained from these observations, including the emission line maps and derived measurements. In section~\ref{sect:disc} we discuss what these results show, and present a simple kinematic model of the inner wind cone. A summary of our findings is given in Section~\ref{sect:summary}.

We adopt a distance to NGC~235 of 3.9~Mpc\footnote{We note that discrepancies exist in the literature with regards to the distance to NGC~253; some authors \citep[e.g.][]{alonso03, engelbracht98} adopt the value of $2.5\pm0.4$~Mpc determined by \citet{devaucouleurs78}, others \citep[e.g.][]{kornei09, fernandez-ontiveros09} adopt $3.9\pm0.4$~Mpc measured by \citet{karachentsev03}, while others \citep[e.g.][]{sakamoto06, brunthaler09, matsubayashi09} adopt $3.5\pm 0.2$~Mpc from \citet{rekola05}.}, meaning $1''\sim 20$~pc, and a heliocentric systemic velocity of 250~\kms \citep{devaucouleurs91}. %or 1kpc~53"

%%%%%%%%%%%%%%%%%%%%%%%%
\section{Observations} \label{sect:obs}

%%%%%%%%%%%%%%%%%
\subsection{MPG/ESO 2.2m WFI imaging}\label{sect:WFI}
Deep, narrow-band H$\alpha$(+[N\two]) images, together with corresponding narrow-band continuum images, of NGC~253 obtained with the MPG/ESO 2.2m WFI instrument \citep{baade99} were retrieved from the ESO archive to complement our IFU spectroscopy (described below). These were observed on the night of 21st September 2001 as part of programme 60.A-9123, and consist of five exposures in each filter, dithered to cover the gaps between the eight CCDs of the WFI instrument. The exposure times were 5$\times$900~secs in H$\alpha$ (NBHALPHA/7\_ESO856) and 5$\times$450~secs in the corresponding `off-band' filter (NB665/12\_ESO858). The $\sim$75~\AA\ FWHM of the H$\alpha$ filter means that emission from [N\two]$\lambda\lambda$6549,6583 is also included.

Images were processed with the imaging pipeline package \textsc{theli}\footnote{http://www.astro.uni-bonn.de/\textasciitilde mischa/theli.html}; the steps included bias subtraction, flat fielding, weighting (to identify chip overlaps and cosmetic defects), determining the astrometric solution via catalogue matching using \textsc{scamp}, background matching, and finally co-addition of the individual exposures. The on- and off-band images were then aligned using their World Coordinate System (WCS) solutions and the off-band was subtracted to leave a pure emission-line H$\alpha$(+[N\two]) image. Fig.~\ref{fig:finder_large} shows the full image in the inset, and the central disk in the main figure. Fig.~\ref{fig:finder}a shows a portion of the image covering the south-western outflow with a greyscale stretch designed to enhance the low surface-brightness wind features.

%%%%%%%%%%%%%%%%%%%%%
\subsection{VLT/VIMOS-IFU spectroscopy} \label{sect:VIMOS_obs}
Queue-mode observations of NGC 253 were made using the VIsible MultiObject Spectrograph \citep[VIMOS;][]{le-fevre03} Integral Field Unit (IFU) between August and December 2008, covering the nucleus and minor axis outflow with five positions at a position angle of $140^{\circ}$, with 0.6--0.9~arcsec seeing (11--17~pc at the distance of NGC 253). 

The IFU positions were chosen to cover the nuclear regions of NGC 253, and the known \textit{inner} minor axis wind outflow. Fig.~\ref{fig:finder_large} shows the position of the IFU fields on the MPG/ESO 2.2m WFI continuum subtracted H$\alpha$ image of the whole disk. The location of the IFU positions are shown in more detail in Fig.~\ref{fig:finder} on (a) a zoom-in of the WFI H$\alpha$ image; (b) an archive \textit{HST}/ACS F814W image; and (c) a \textit{Chandra} soft X-ray image from \citet{strickland00}. Fig.~\ref{fig:finder}b also shows the SiO ($v$=0, $J$=2$\rightarrow$1) integrated intensity contours from \citet[][see Section~\ref{sect:inner_wind}]{garcia-burillo00}. Fig.~\ref{fig:finder_ROSAT_WFI} shows the extent of the large-scale X-ray outflow as seen with \textit{ROSAT} \citep{pietsch00, strickland02} overlaid on the WFI H$\alpha$ image (for a discussion of the connection between the inner wind and and X-ray lobes see Section~\ref{sect:connection}).

Positions 1 and 2 were observed in the higher resolution $0\farcs33$ spaxel$^{-1}$ setting giving a field-of-view (FoV) of $13\times 13$~arcsecs ($\sim$$250\times 250$~pc). The remaining positions were observed in the $0\farcs67$ spaxel$^{-1}$ setting, giving a FoV of $27\times 27$~arcsecs ($\sim$$500\times 500$~pc).

The use of the high-resolution (HR) red grism (for all positions) gave a spectral coverage of 645--860 nm at a resolution of R=3100. This allowed us to cover the nebular diagnostic lines of H$\alpha$, [N\two]$\lambda\lambda 6548,6583$ and [S\two]$\lambda\lambda 6716,6731.$\footnote{The Ca\two\ triplet at $\lambda$8498.0, 8542.1, 8662.1 were also present in our wavelength range, but this part of the spectrum was affected by fringing which we could not correct for.} With the medium- and high-resolution grisms on VIMOS, the spectra are long enough that the detector no longer has enough space to accommodate the full 6400 spectra resulting from the $80\times 80$ fibres of the IFU. The IFU head is therefore partially masked by a shutter, leaving a square of $40\times 40$ fibres giving 1600 individual spectra. These spectra are reformatted by the optical fibres for imaging onto four 2k$\times$4k EEV CCDs. These CCDs are arranged to form four quadrants of a rectangle and are hereafter referred to as Q1--Q4.

We took a set of dithered exposures at each position (offset by integer spaxel shifts). The exposures for positions 3, 4 and 5 were split into 3 subsets of exposures. Total integration times varied between 2340 and 6993~secs. Table~\ref{tbl:vimos_obs} lists the dates, coordinates and exposure times for each subset of exposures for each position.

In order to remove the pixel-to-pixel sensitivity differences, and enable the wavelength and flux calibration of the data, a number of bias frames, screen-flats, twilight flats, arc calibration frames, offset sky exposures (exposure times 2$\times$200~secs for each position), and observations of photometric standard stars, were taken contemporaneously with the science fields.

\begin{table}
\centering
\caption {VLT/VIMOS-IFU observation log}
\label{tbl:vimos_obs}
\begin{tabular}{c c r @{\hspace{0.2cm}} l r @{ $\times$ } l }
\hline
Pos. & Date(s) & \multicolumn{2}{c}{Coordinates} & \multicolumn{2}{c}{Exp.\ Time} \\
No. & & \multicolumn{2}{c}{(J2000)} & \multicolumn{2}{c}{(s)} \\
\hline 
1 & 31/8/08 & $00^{\rm h}\,46^{\rm m}\,55\fsec8$, & $-25^{\circ}\,24'\,22\farcs8$ & 3 & 780 \\
2 & 31/8/08 & $00^{\rm h}\,46^{\rm m}\,55\fsec1$, & $-25^{\circ}\,24'\,30\farcs2$ & 3 & 780 \\
3 & 31/8/08, & $00^{\rm h}\,46^{\rm m}\,57\fsec1$, & $-25^{\circ}\,24'\,33\farcs0$ & 9 & 570 \\
& 01/9/08\\
4 & 01/10/08, & $00^{\rm h}\,46^{\rm m}\,55\fsec6$, & $-25^{\circ}\,24'\,50\farcs0$ & 9 & 570 \\
& 22/10/08 \\
5 & 30/11/08, & $00^{\rm h}\,46^{\rm m}\,58\fsec0$, & $-25^{\circ}\,25'\,07\farcs9$ & 9 & 777 \\
& 02/12/08, \\
& 24/12/08 \\
\hline
\end{tabular}
\end{table}

\subsubsection{Reduction} \label{sect:VIMOSreduction}
Basic data reduction was performed following the standard ESO reduction pipeline tasks implemented in \textsc{gasgano}. Briefly the steps go as follows: a master bias frame is created using the task \textsc{vmbias}, then \textsc{vmifucalib} is run to (1) create a master flat field frame; (2) trace the position of each spectrum on the CCD from the screen-flat; (3) determine the fibre-to-fibre transmission functions; and (4) determine wavelength calibration solutions. The standard star and science observations are then processed using \textsc{vmifustandard} and \textsc{vmifuscience}, respectively, to apply the aforementioned correction functions and extract the individual spectra to produce four data files per exposure (science and sky), each containing 400 reduced, flux-calibrated spectra corresponding to one of the four quadrants of the detector. 
%After extraction of the spectra, the $x$ and $y$ spatial units are termed `spaxels' to differentiate from `pixel', which refers to the spatial units of the CCD. 

The final steps of the reduction procedure were to clean the cosmic-rays using \textsc{lacosmic} \citep{vandokkum01}, mosaic the data files for each quadrant into one, scale and subtract the sky frames from the corresponding science frames, correct each cube for the effects of differential atmospheric correction \citep[using an \textsc{iraf}-based code written by J.\ R.\ Walsh based on an algorithm described in][]{walsh90}, and combine the three dithered (offset) exposures for each position. These steps were achieved using a combination of \textsc{iraf}\footnote{The Image Reduction and Analysis Facility ({\sc iraf}) is distributed by the National Optical Astronomy Observatories which is operated by the Association of Universities for Research in Astronomy, Inc. under cooperative agreement with the National Science Foundation.} and \textsc{idl}.

Due to an intermittent problem (apparently now resolved) with VIMOS that caused the focussing of light onto Q3 of the detector to fail at particular rotator angles (unfortunately including the one used in our observations), the spectra in Q3 of every observation were blurred and therefore unusable for the analysis of line profile shapes. However, we retained Q3 throughout the reduction procedure so that we could measure the integrated line fluxes, which remain unaffected by the focussing problems.

In order to determine an accurate measurement of the instrumental contribution to the line broadening, we fitted single Gaussians to isolated arc lines on a wavelength calibrated arc exposure in all 1200 properly focussed apertures. The average instrumental broadening (velocity resolution; FWHM) of the final dataset (not including Q3) is 2.0$\pm$0.1~\AA, or 90$\pm$0.5~\kms, consistent with the documented grism resolution.

The emission line profiles of H$\alpha$, [N\two] and [S\two] were fitted with multiple Gaussians as described in Appendix~\ref{sect:line_profiles}. Overall, we find that each line could be decomposed into $\le$3 components. The resulting emission line maps showing the flux, FWHM and radial velocity distributions are described in Section~\ref{sect:VIMOS_maps}.

\begin{figure*}
\centering
\begin{minipage}{7cm}
\begin{overpic}[width=7cm]
{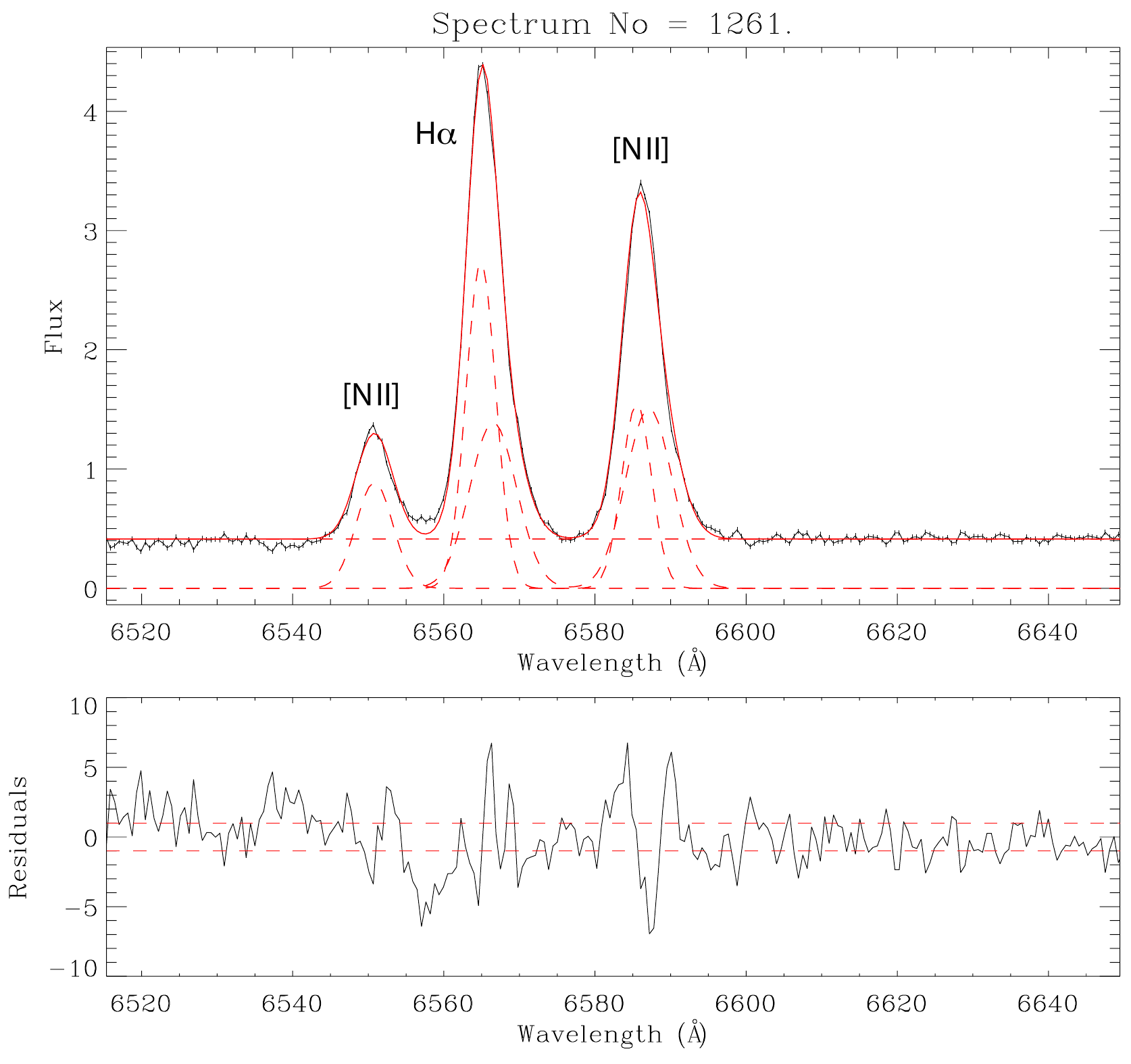}
\put(13,82){\large (a)}
\end{overpic}
\end{minipage}
\hspace{0.8cm}
\begin{minipage}{7cm}
\begin{overpic}[width=7cm]
{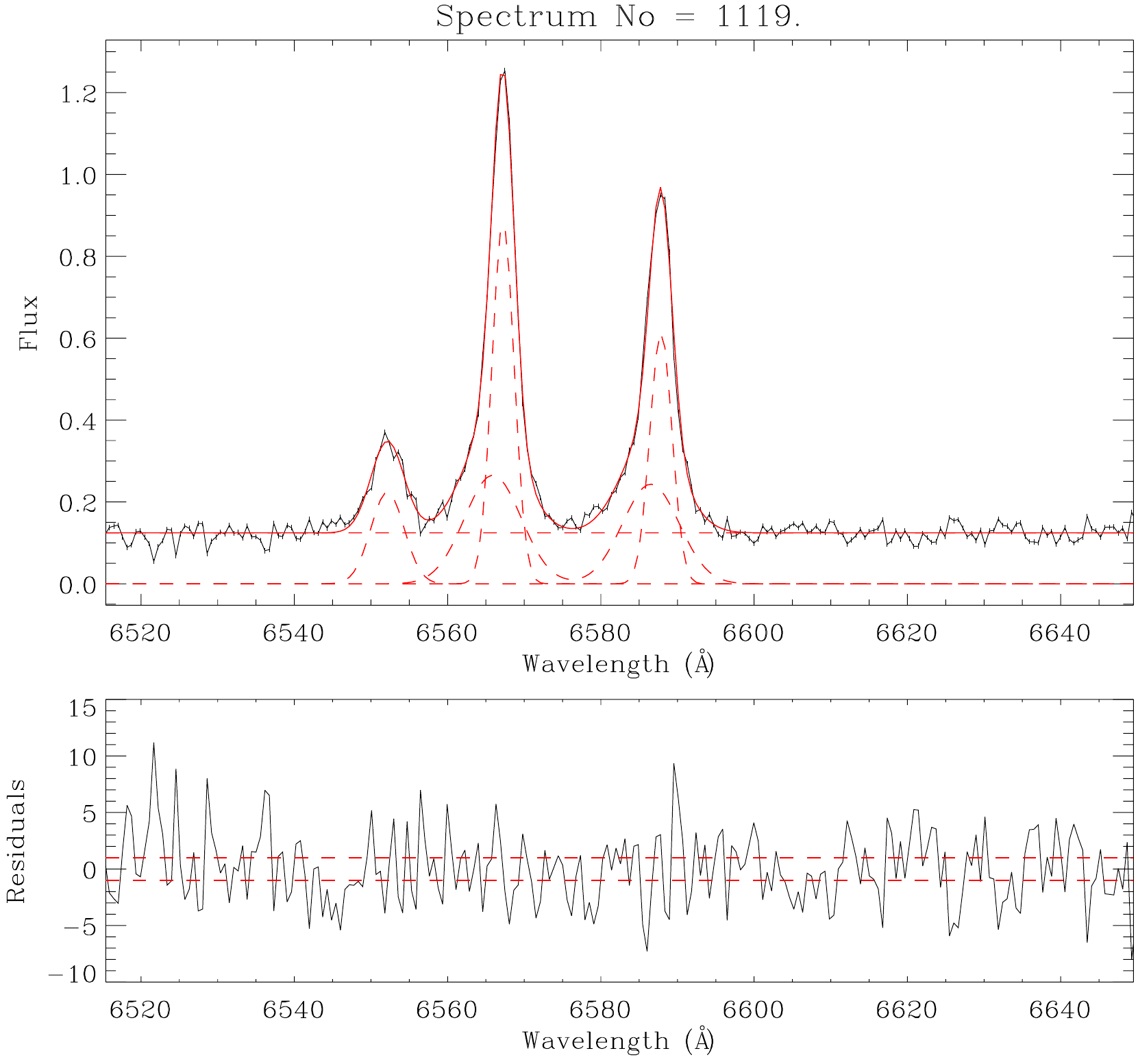}
\put(13,82){\large (b)}
\end{overpic}
\end{minipage}
\begin{minipage}{7cm}
\vspace{0.5cm}
\begin{overpic}[width=7cm]
{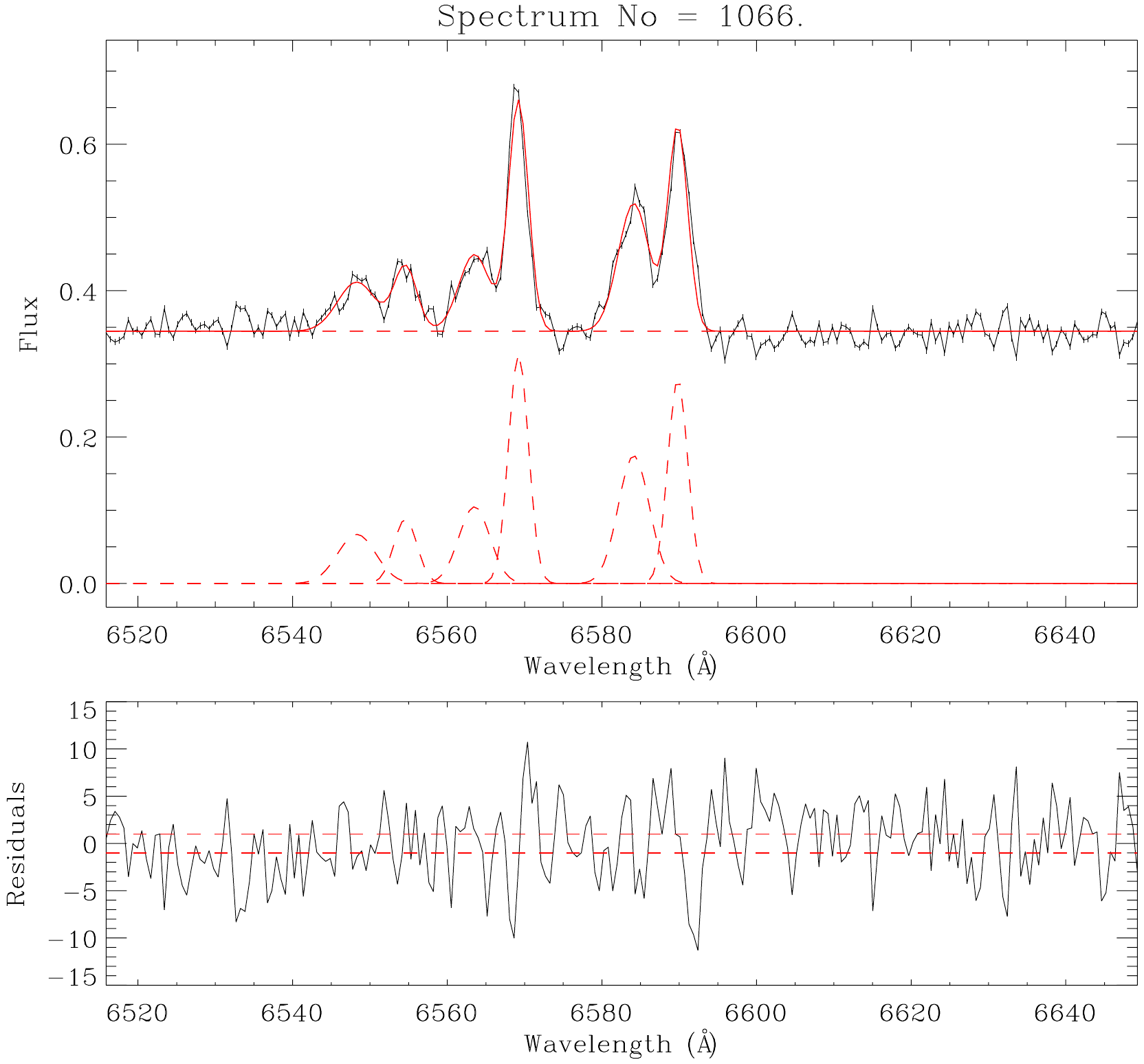}
\put(13,82){\large (c)}
\end{overpic}
\end{minipage}
\hspace{0.8cm}
\begin{minipage}{7cm}
\vspace{0.5cm}
\begin{overpic}[width=7cm]
{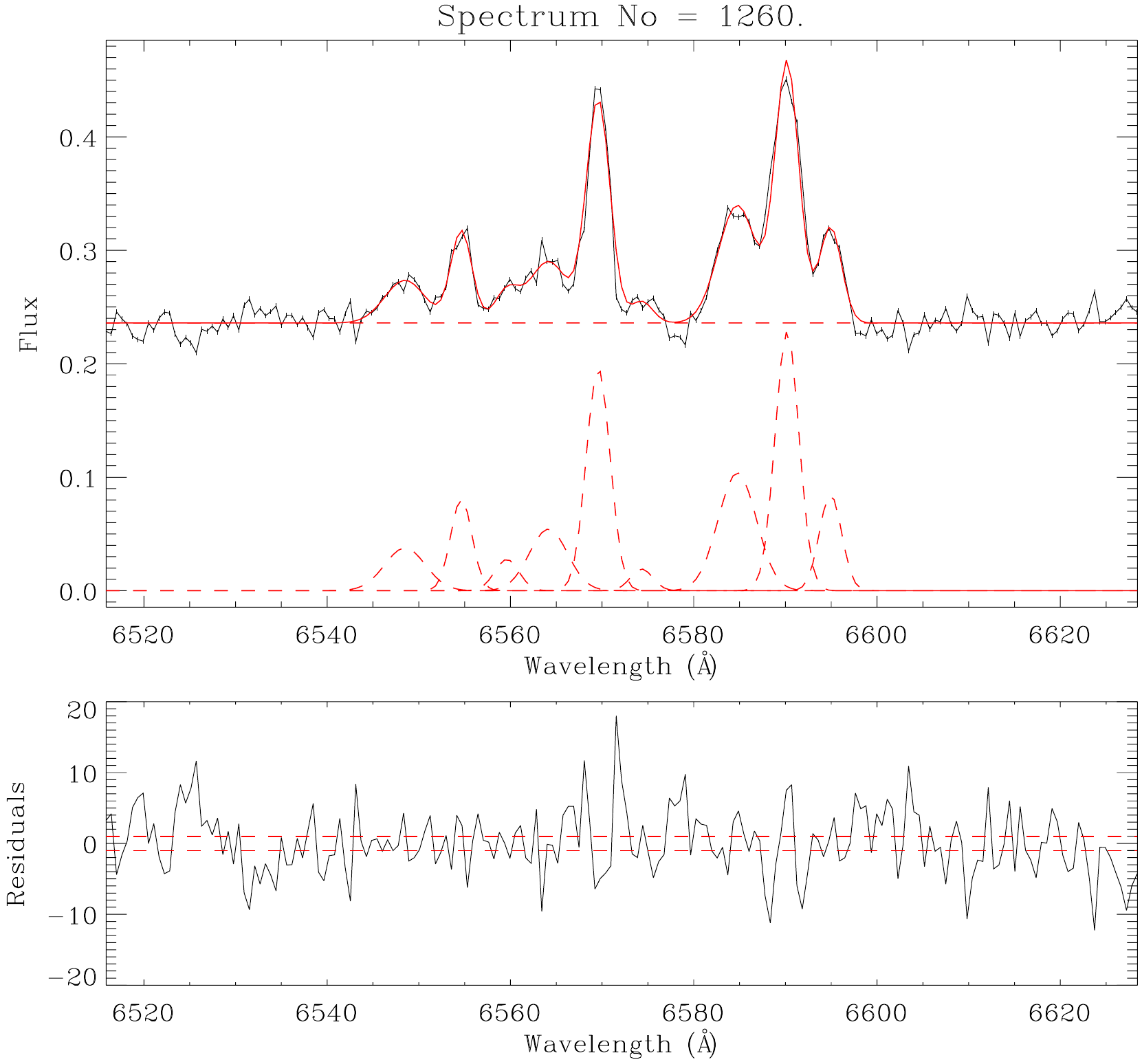}
\put(13,82){\large (d)}
\end{overpic}
\end{minipage}
\begin{minipage}{7cm}
\vspace{0.5cm}
\begin{overpic}[width=7cm]
{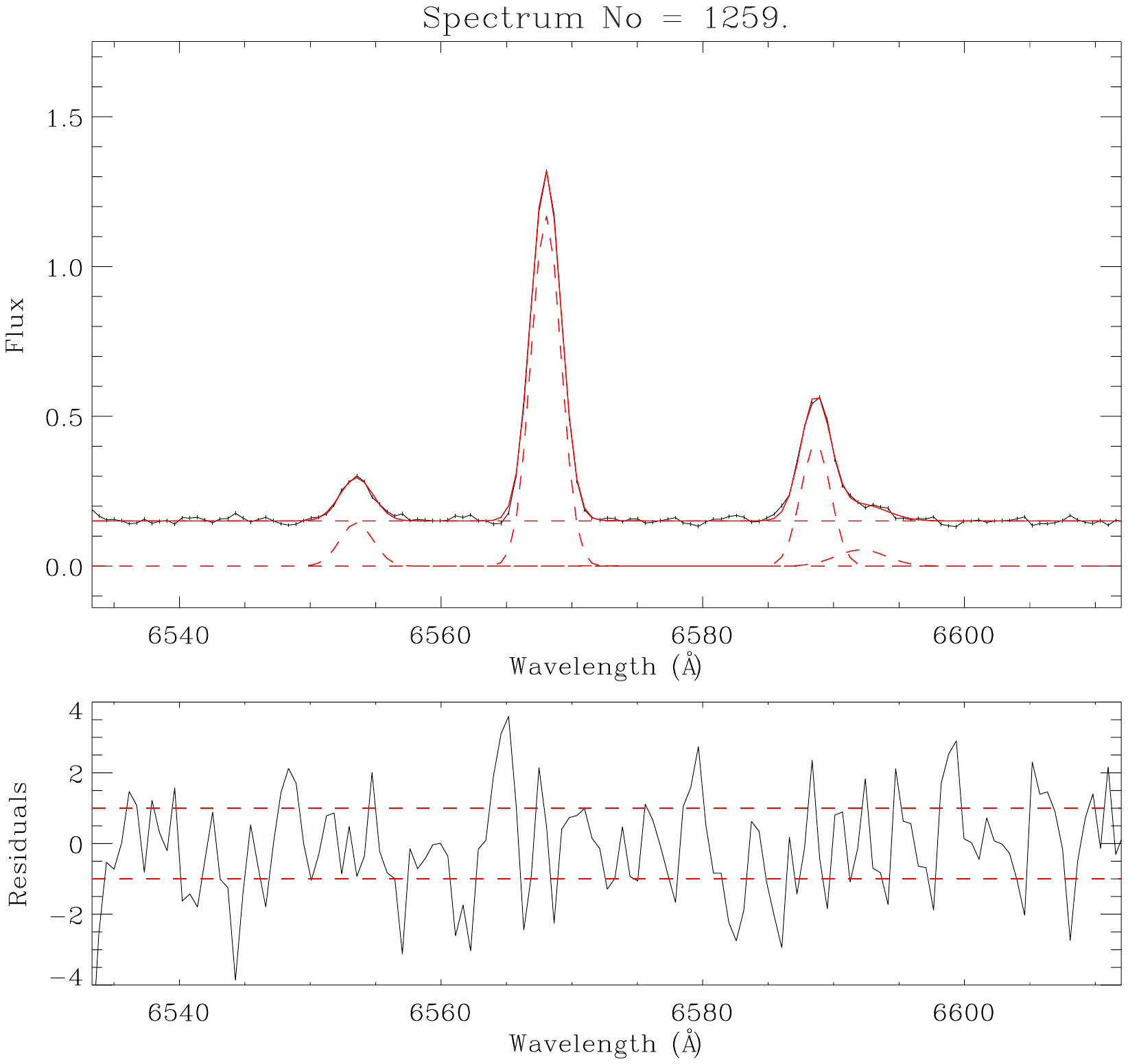}
\put(13,82){\large (e)}
\end{overpic}
\end{minipage}
\hspace{0.8cm}
\begin{minipage}{7cm}
\vspace{0.5cm}
\begin{overpic}[width=7cm]
{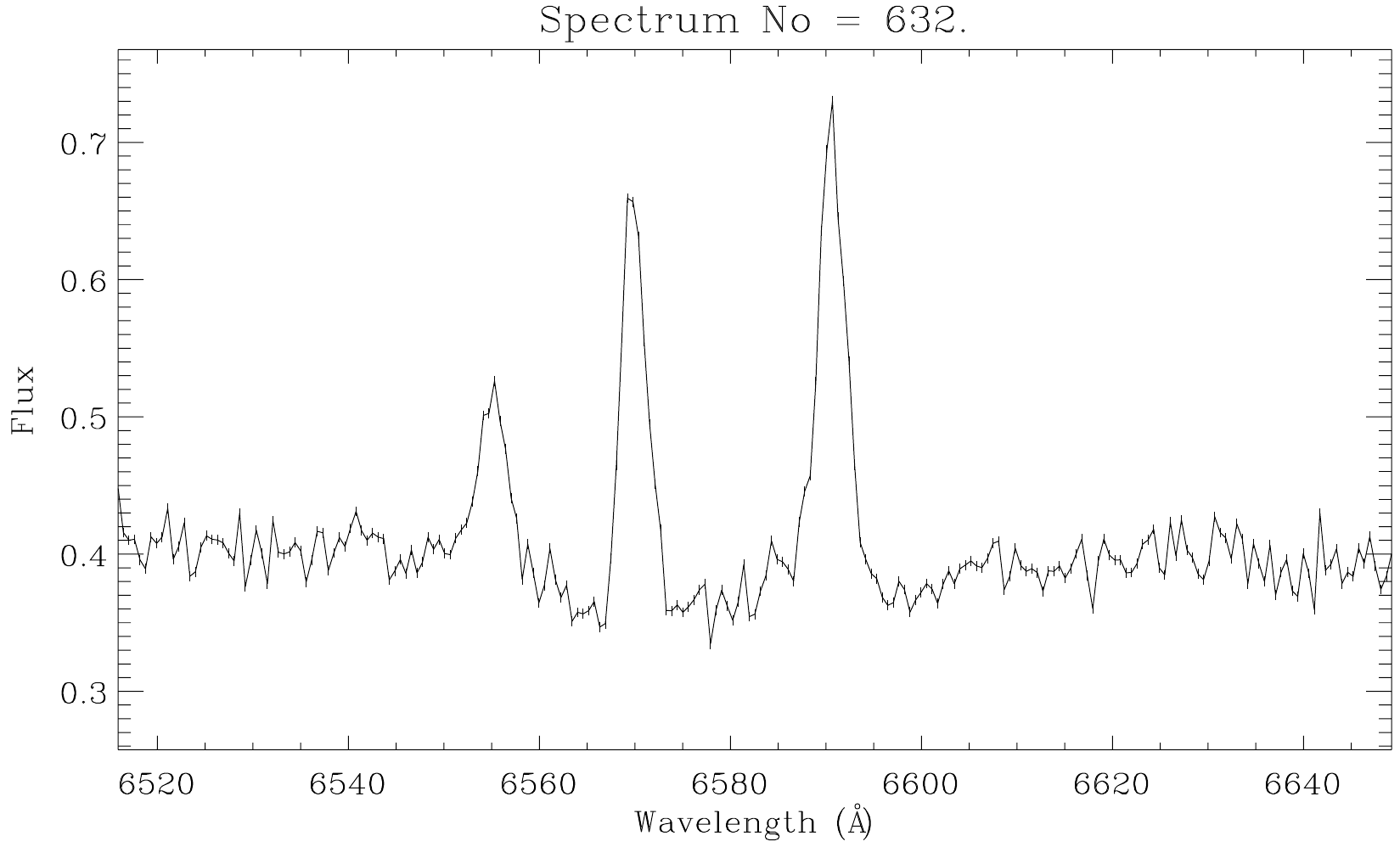}
\put(13,50){\large (f)}
\end{overpic}
\end{minipage}
\caption{Example H$\alpha$+[N\two] line profiles extracted from the locations shown in Fig.~\ref{fig:Ha_flux}, chosen to represent the main types of profile shapes observed over the field. Observed data are shown by a solid black line, individual Gaussian fits by dashed red lines (including the straight-line continuum level fit), and the summed model profile in solid red. Flux units are $10^{-16}$~erg~s$^{-1}$~cm$^{-2}$~\AA$^{-1}$~spaxel$^{-1}$. Below each spectrum is the residual plot (in units of $\sigma$; see text for further information).
(a) An example line fit from position 1 showing a redshifted broad (FWHM$\sim$320~\kms) component;
(b) a profile from position 2 showing a blueshifted broad (FWHM$\sim$390~\kms) component;
(c) example of a clear split-line profile with velocity separation of $\sim$260~\kms\ from position 3;
(d) example of a triple component profile from position 4;
(e) a spectrum from position 5 showing a broad component in [N\two] and not H$\alpha$;
(f) example of a spectrum from position 4 with strong H$\alpha$ absorption.}
\label{fig:eg_fits}
\end{figure*}

%%%%%%%%%%%%%%%%
\subsection{WIYN/SparsePak observations} \label{sect:SP_obs}
Complementary observations of NGC 253 were obtained with the SparsePak instrument \citep{bershady04} on the WIYN 3.5-m telescope. SparsePak is a ``formatted field unit'' similar in design to a traditional IFU, except that its 82 fibres are arranged in a sparsely packed grid, with a small, nearly-integral core \citep{bershady04}. Each fibre has a diameter of 500\,$\upmu$m, corresponding to $4\farcs69$ on the sky; the formatted field has approximate dimensions of $72\times 71.3$~arcsecs, including seven fibres located on the north and west side of the main array separated by $\sim$$25''$ which are designed to sample the sky background. SparsePak is connected to the Hydra bench-mounted echelle spectrograph which at that time used a T2KA $2048\times 2048$ CCD detector.

We observed NGC 253 at two locations, one centred on the nuclear region at $\alpha$=$0^{\rm h} 47^{\rm m} 32\fsec7$, $\delta$=$-25^{\circ} 17' 19\farcs0$ (J2000) (hereafter SP1) and one on the northern disk at $\alpha$=$0^{\rm h} 47^{\rm m} 29\fsec5$, $\delta$=$-25^{\circ} 16' 19\farcs9$ (hereafter SP2) on the night of 21st November 2005 with exposure times of 3$\times$500 and 3$\times$1000~secs, respectively. The position of the SparsePak footprints (together with the VIMOS footprints) are shown in Fig.~\ref{fig:finder_large}, overlaid on the WFI H$\alpha$ image. Using the 860@30.9 grating with an angle of 50.9$^{\circ}$ (giving a spectral coverage of 6150--7050\,\AA\ and dispersion of 0.44\,\Apix), we were able to cover the nebular emission lines of H$\alpha$, [N\two]$\lambda\lambda 6548,6583$, and [S\two]$\lambda\lambda 6716,6731$. A number of bias frames, flat-fields and arc calibration exposures were also taken together with the science frames. Since the seven ``sky fibres'' of the array fell on the galaxy itself, we took offset sky observations with exposure times equal to those of the corresponding science frames to facilitate sky subtraction. We therefore treated the ``sky fibres'' as normal science fibres and processed them accordingly.

Basic reduction was achieved using the {\sc ccdproc} task within the {\sc noao} {\sc iraf} package. Instrument-specific reduction was then achieved using the {\sc hydra} tasks also within the {\sc noao} package. Cosmic-rays were cleaned by use of a simple minmax rejection as the individual frames were combined using {\sc imcombine}. The separate offset sky observations were then scaled and subtracted from the corresponding summed science frames. The final datafile contained 82 reduced, wavelength calibrated and sky-subtracted spectra.

%An example reduced and labelled spectrum is shown in the top panel of Fig.~\ref {fig:spec_labelled}.

In order to determine an accurate measurement of the instrumental contribution to the spectrum broadening, we selected two high S/N and isolated sky lines in one of the wavelength calibrated sky frames and fitted them with single Gaussians in all 82 apertures. We find the average instrumental width at H$\alpha$ is 1.77$\pm$0.17~\AA\ or 81$\pm$5~\kms\ corresponding to a spectral resolution of R$\approx$3700.

Again, the emission line profiles of H$\alpha$, [N\two] and [S\two] were fitted with multiple Gaussians as described in Appendix~\ref{sect:line_profiles}. Overall, we find that each line could be decomposed into $\le$3 components. The resulting emission line maps showing the flux, FWHM and radial velocity distributions are described in Section~\ref{sect:sp_maps}.

\begin{figure*}
\centering
\includegraphics[width=\textwidth]{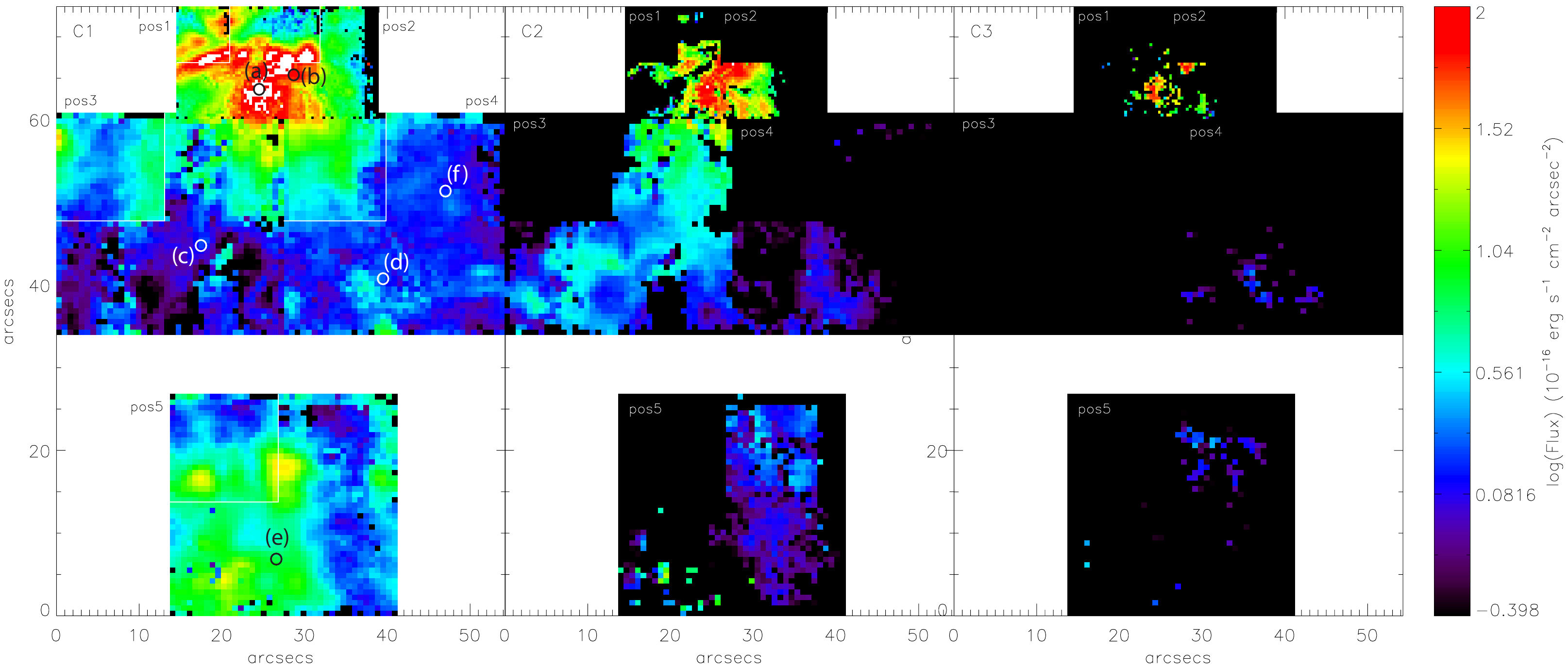}
\caption{H$\alpha$ flux maps in the three line components. The C1 map includes the integrated line fluxes measured in Q3, where the focussing problems meant that the line profiles could not be decomposed accurately. A scale bar is given in units of $10^{-16}$~erg~s$^{-1}$~cm$^{-2}$~arcsec$^{-2}$. The spaxels from which the example line profiles shown in Fig.~\ref{fig:eg_fits} were extracted are labelled with the corresponding letters.}
\label{fig:Ha_flux}
\end{figure*}
\begin{figure*}
\centering
\includegraphics[width=\textwidth]{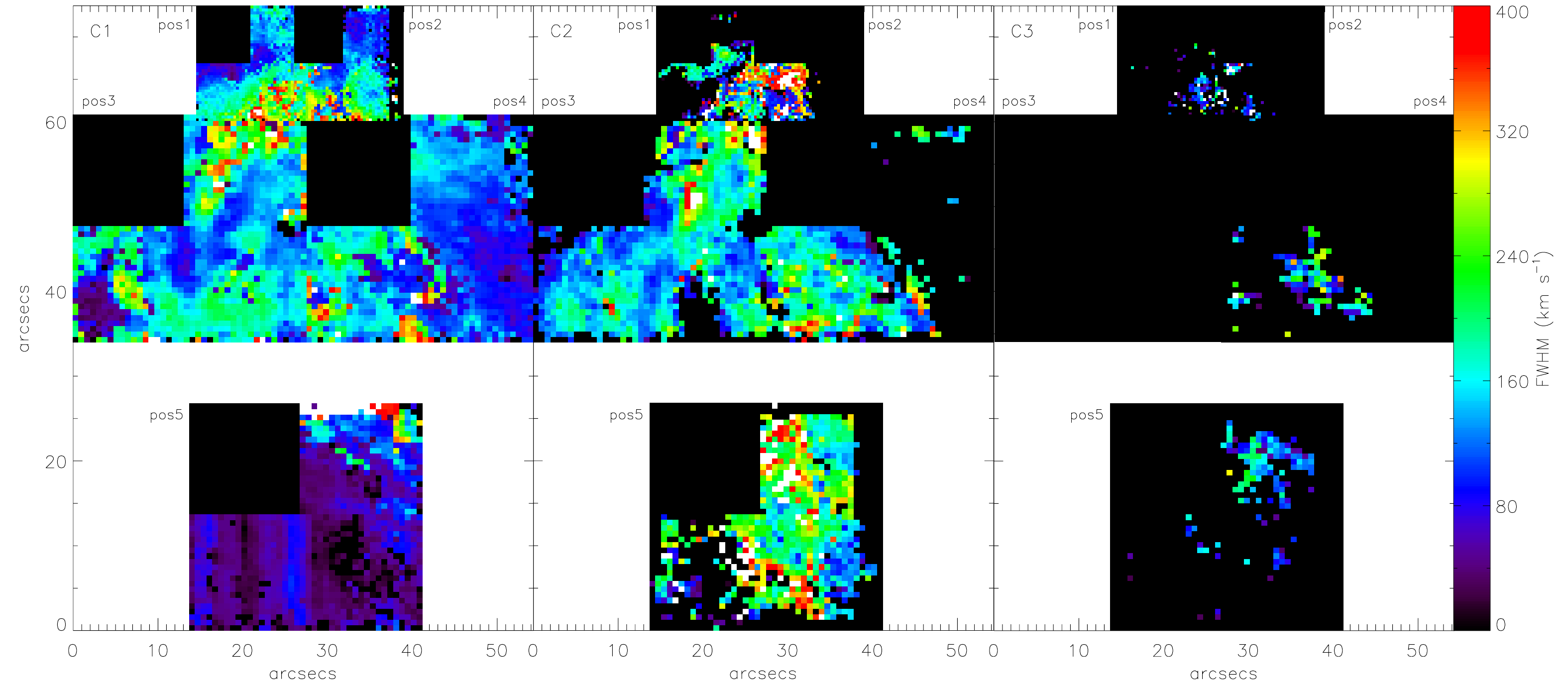}
\caption{H$\alpha$ FWHM maps in the three H$\alpha$ line components. \emph{Left:} C1, \emph{centre:} C2 and \emph{right:} C3. A scale bar is given in units of \kms, corrected for instrumental broadening. The vertical bands towards the bottom-left of position 5 are almost certainly instrumental in origin, although their precise cause has eluded us.}
\label{fig:Ha_fwhm}
\end{figure*}
%\begin{figure*}
%\centering
%\includegraphics[width=\textwidth]{Ha_fwhm_SINFONI-Brg.pdf}
%\caption{H$\alpha$ FWHM maps in the three H$\alpha$ line components. \emph{Left:} C1, \emph{centre:} C2 and \emph{right:} C3. A scale bar is given in units of \kms, corrected for instrumental broadening. The vertical bands towards the bottom-left of position 5 are instrumental in origin, although their precise cause has eluded us. The Br$\gamma$ dispersion map from M\"{u}ller-S\'{a}nchez et al.\ (in prep.) is shown for comparison, but offset to the top to avoid obscuring the VIMOS data. The grey outline shows the actual footprint of the SINFONI observations.}
%\label{fig:Ha_fwhm}
%\end{figure*}
\begin{figure*}
\centering
\includegraphics[width=\textwidth]{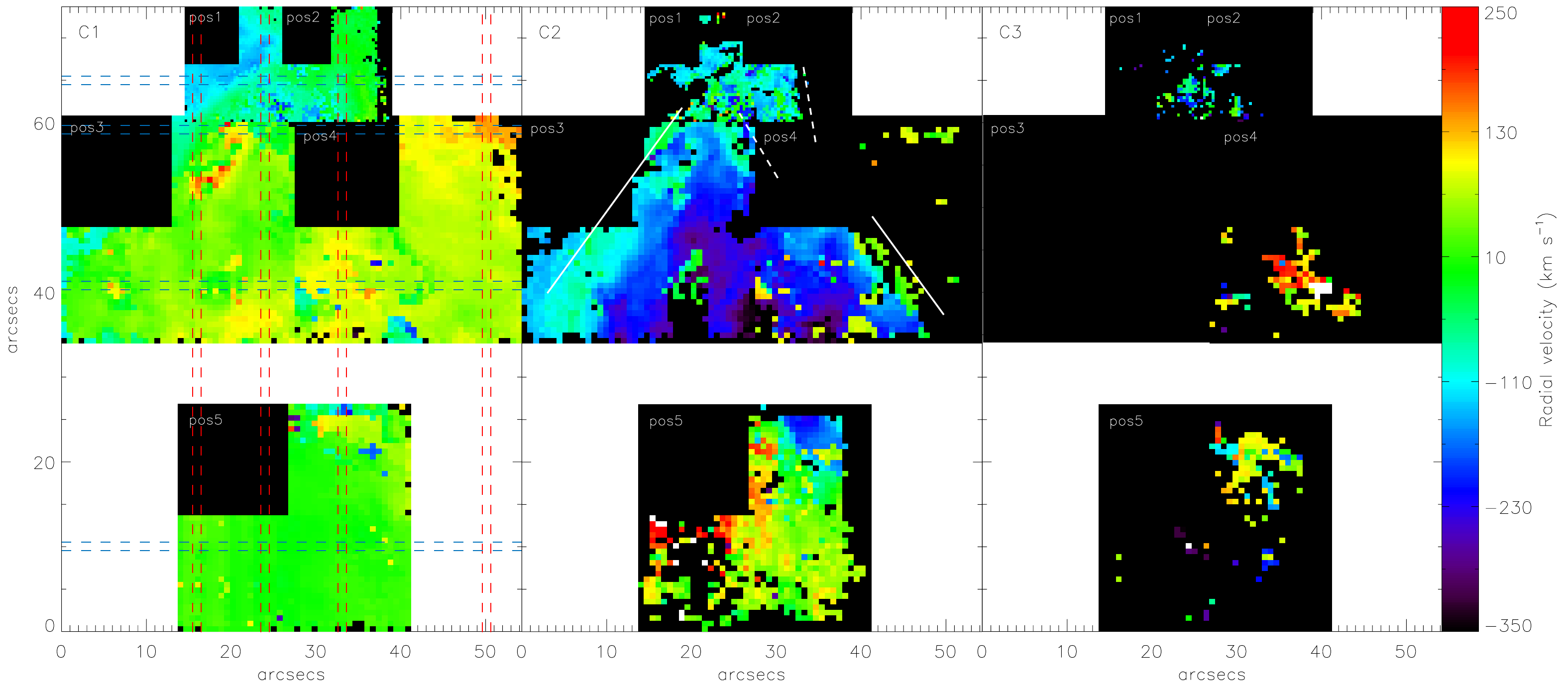}
\caption{H$\alpha$ radial velocity maps in the three line components. \emph{Left:} C1; \emph{centre:} C2 and \emph{right:} C3. A scale bar is given in units of \kms, relative to $v_{\rm sys}$. The dashed lines on the left panel indicate the positions and widths of a number of pseudo-slits from which the position-velocity data shown in of Figs.~\ref{fig:pv_vertical} and \ref{fig:pv_horizontal} were extracted. The solid and dashed white lines on the central panel highlight the approximate cone morphology determined by eye from the diverging H$\alpha$ velocity components.}
\label{fig:Ha_vel}
\end{figure*}

%%%%%%%%%%%%%%%%%%%%%%%%%%%%%%%%
\section{Results} \label{sect:results}

%%%%%%%%%%%%%%%%%%%%%%%%%%%%%%%%
\subsection{VIMOS emission line maps} \label{sect:VIMOS_maps}
In this section we present and describe the VIMOS emission line maps created from line profile fits of the data described above. As mentioned above, each emission line in each spectrum was fit with a multi-Gaussian model, resulting in measurements of the flux, FWHM and radial velocity for each component (see Appendix~\ref{sect:line_profiles}). It is very important when fitting multi-Gaussian models to line profiles to assign each component to a particular map in such a way as to limit the confusion that might arise during analysis of the results, such as discontinuous spatial regions arising from incorrect component assignments. We assigned the brightest component of H$\alpha$ to C1, and the faintest to C2 (or C3 if three components were identified). However, position 3 made more sense if, where two components were detected, the redder was assigned to C1 and the bluer to C2, regardless of their fluxes. For [S\two], as with H$\alpha$, we assigned the brightest component in each case to C1, except for position 3 where we assigned the redder to C1. No evidence of a third component was found in [S\two].

Fig.~\ref{fig:eg_fits} shows a selection of observed H$\alpha$+[N\two] line profiles and best-fitting Gaussian models from a number of regions within our VIMOS fields (located with the corresponding letters in Fig.~\ref{fig:Ha_flux}). This sample represents the variety of different profile shapes we find across the IFU fields, and demonstrates the high quality of the spectra and the accuracy of the line-fitting. Below each plot are the corresponding residuals, $r_{\rm i}$, calculated using the following formula:
\begin{equation}
r_{\rm i} = \frac{y_{\rm i}^{\rm fit} - y_{\rm i}^{\rm data}}{\sigma_{\rm i}}
\end{equation}
where $\sigma_{\rm i}$ are the uncertainties on $y_{\rm i}^{\rm data}$.

As mentioned in Section~\ref{sect:VIMOS_obs}, the spectra in Q3 of each VIMOS field were blurred due to a focussing problem, meaning we could not perform a Gaussian decomposition of the profile shape. We therefore simply integrated the fluxes over each emission line of interest (H$\alpha$, [N\two]$\lambda$6583, [S\two]$\lambda$6717 and $\lambda$6731). The continuum level was then subtracted from each to recover the total line fluxes.

Fig.~\ref{fig:Ha_flux} shows the line flux maps for the three identified H$\alpha$ line components over the five IFU fields. The C1 map (left panel) also includes the integrated line flux measured in Q3, where the focussing problems (see Section~\ref{sect:VIMOSreduction}) meant that the line profiles could not be decomposed accurately. Since the spaxels in positions 1 and 2 are of a different size to those in positions 3, 4 and 5, we have made sure to present the fluxes in arcsec$^{-2}$ units. Comparing Fig.~\ref{fig:finder}a and \ref{fig:Ha_flux} reveals that the depth of our IFU H$\alpha$ spectroscopy and the continuum subtracted WFI H$\alpha$ imaging is approximately the same. Both are considerably deeper than the available \textit{HST} H$\alpha$ imaging.

\begin{figure*}
\centering
\includegraphics[width=\textwidth]{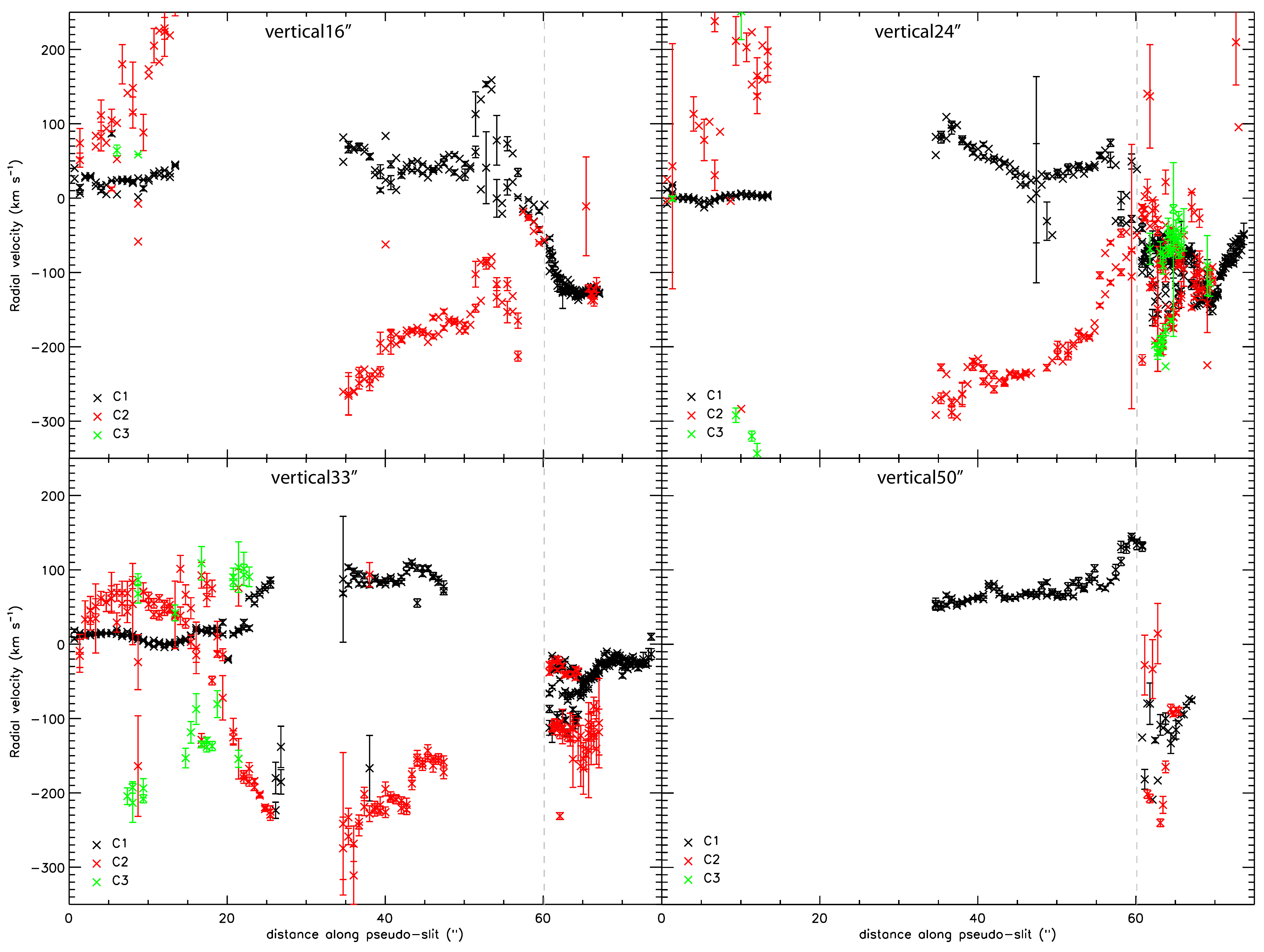}
\caption{Position-velocity plots extracted from $1''$ wide pseudo-slits aligned with the galaxy minor axis at various positions along the dataset. The plots are labelled ``vertical$\#\#$'', where $\#\#$ represents the $x$-axis intercept of the pseudo-slit on the map of Fig.~\ref{fig:Ha_vel}. On these graphs, distance along pseudo-slit=0 (i.e.\ $x$=0) corresponds to $y$-axis=0 in Fig.~\ref{fig:Ha_vel}. The velocities are given relative to $v_{\rm sys}$ (250~\kms). The vertical grey dashed line indicates the transition between positions 1/2 and 3/4.}
\label{fig:pv_vertical}
\end{figure*}

\begin{figure*}
\centering
\includegraphics[width=\textwidth]{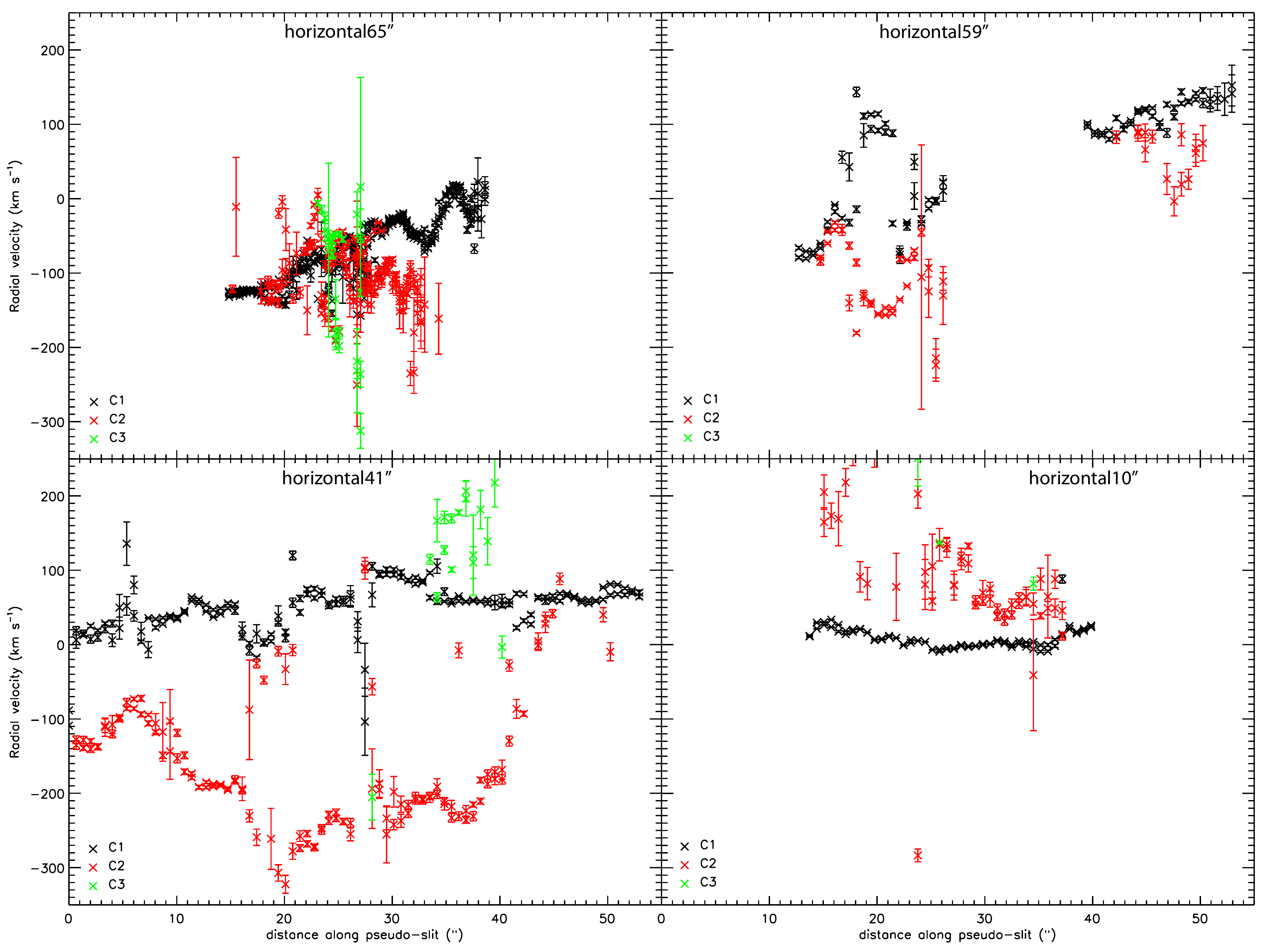}
\caption{Position-velocity plots extracted from $1''$-wide pseudo-slits aligned with the galaxy major axis at various positions along the dataset. The plots are labelled ``horizontal$\#\#$'', where $\#\#$ represents the $y$-axis intercept of the pseudo-slit on the map of Fig.~\ref{fig:Ha_vel}. On these graphs, distance along pseudo-slit=0 (i.e.\ $x$=0) corresponds to $x$-axis=0 in Fig.~\ref{fig:Ha_vel}. Again, the velocities are given relative to $v_{\rm sys}$.}
\label{fig:pv_horizontal}
\end{figure*}

FWHM maps for the three identified H$\alpha$ line components are shown in Fig.~\ref{fig:Ha_fwhm}, and the radial velocity maps are shown in Fig.~\ref{fig:Ha_vel}. Where a comparison is possible, the velocities we measure are consistent with those measured by \citet{schulz92} (given their significantly lower spatial resolution and the fact that they only fitted single Gaussians). Figs.~\ref{fig:pv_vertical} and \ref{fig:pv_horizontal} show position-velocity plots extracted from a number of vertical (minor axis; PA=140$^{\circ}$) and horizontal (major axis; PA=50$^{\circ}$) 1$''$-wide pseudo-slits whose positions and widths are shown with dashed lines in Fig.~\ref{fig:Ha_vel}. Hereafter, we refer to these as horizontal$\#\#$ or vertical$\#\#$, where $\#\#$ represents the $y$ or $x$ axis offsets of the pseudo-slit in arcsecs, respectively. The horizontal65 and vertical24 plots are in good agreement with those presented in \citet{prada98}.

Maps of the electron density distribution (as derived from the flux ratio of the [S\two]$\lambda\lambda$6717,6731 doublet, assuming $T_{\rm e}=10^4$~K) are shown in Fig.~\ref{fig:elecdens}. Before fitting the [S\two] lines, the data cube was smoothed in the spatial directions by a 3$\times$3 rolling summation to increase S/N. A third line component could not be detected in [S\two] in any of the five fields, so only the C1 and C2 maps are shown.

The forbidden/recombination line flux ratios of [S\two]($\lambda$6717+$\lambda$6731)/H$\alpha$ and [N\two]$\lambda$6583/H$\alpha$ can be used as indicators of the radiation field strength and therefore the dominant ionization mechanism \citep{veilleux87,dopita95,dopita00,dopita06b}. [S\two]/H$\alpha$ is particularly sensitive to shock ionization because relatively cool, high-density regions form behind shock fronts which emit strongly in [S\two] thus producing an enhancement in [S\two]/H$\alpha$ \citep{dopita97, oey00}. Figs.~\ref{fig:SII_Ha} and \ref{fig:NII_Ha} show maps of the [S\two]/H$\alpha$ and [N\two]/H$\alpha$ flux ratios, respectively, for the identified line components. Markers on the scale bar indicate values of log[S\two]/H$\alpha$ = $-0.4$ and log[N\two]/H$\alpha$ = $-0.2$, representing fiducial thresholds for the two ratios above which non-photoionized emission is thought to play a dominant role \citep{dopita95, kewley01, calzetti04}. In environments such as this, the most likely excitation mechanism after photoionization is that of shocks. Our line ratio maps are in agreement with those presented by \citet{matsubayashi09} and \citet{sharp10}, given that we decompose our line profile shapes, and have a higher spatial resolution.
% These ratios are plotted against each other in graph form in Fig.~\ref{fig:diagnostic}.

In the following sections we describe what these maps and plots show, and comment on a number of the salient features in the area covered by our observations. Discussions of the meaning of some of these findings is presented in Section~\ref{sect:disc}.


\subsection{Line shape distribution and general observations}
We detect at least two H$\alpha$ line components over the majority of the fields observed. In some regions in positions 1 and 2, 4 and 5, we detect a third component, which, in most cases is narrower and/or fainter than the first two. Through examination of both the FWHM and radial velocity maps together (Figs.~\ref{fig:Ha_fwhm} and \ref{fig:Ha_vel}), the H$\alpha$ line profile can be classified into two main shapes/configurations: (a) a narrow component superimposed on a broader, fainter component at approximately the same velocity (e.g.\ Fig.~\ref{fig:eg_fits}a or b); (b) a split narrow line where the two components are of approximately the same width (e.g.\ Fig.~\ref{fig:eg_fits}c). Type (a) is mainly found in position 2 just south of the nucleus (where we see very broad C2 widths; Fig.~\ref{fig:Ha_fwhm}, middle panel), and position 1 just below and to the right of the missing Q3 section. Here we see C2 line widths of $>$400~\kms. This type is also found towards the lower-right (south) of position 1, but here the broad component is also the brightest and has therefore been assigned to C1. Type (b) is found over most of the rest of the fields, indicating the widespread presence of dynamically expanding material. The red and blue components vary in strength relative to one another (see below), and vary in width. In places, the two dynamical components both become very broad, reaching $>$300~\kms. Compared to positions 3 and 4, the C2 component in position 5 remains surprisingly broad over a large fraction of the field, where C1 remains reasonably narrow (i.e.\ a type (a) narrow+broad configuration). However, unlike the type (a) shape found in position 2, here the broad C2 is clearly kinematically associated with the outflow cone structure (see below).

In position 1, the wedge-shaped blueshifted region with its apex pointing to the nucleus (see Fig.~\ref{fig:Ha_vel}, left panel) is indicative and consistent with the known disk rotation kinematics -- although no obvious evidence for the redshifted counterpart is seen in position 2, even after re-scaling the colour bar.

\begin{figure*}
\centering
\includegraphics[width=\textwidth]{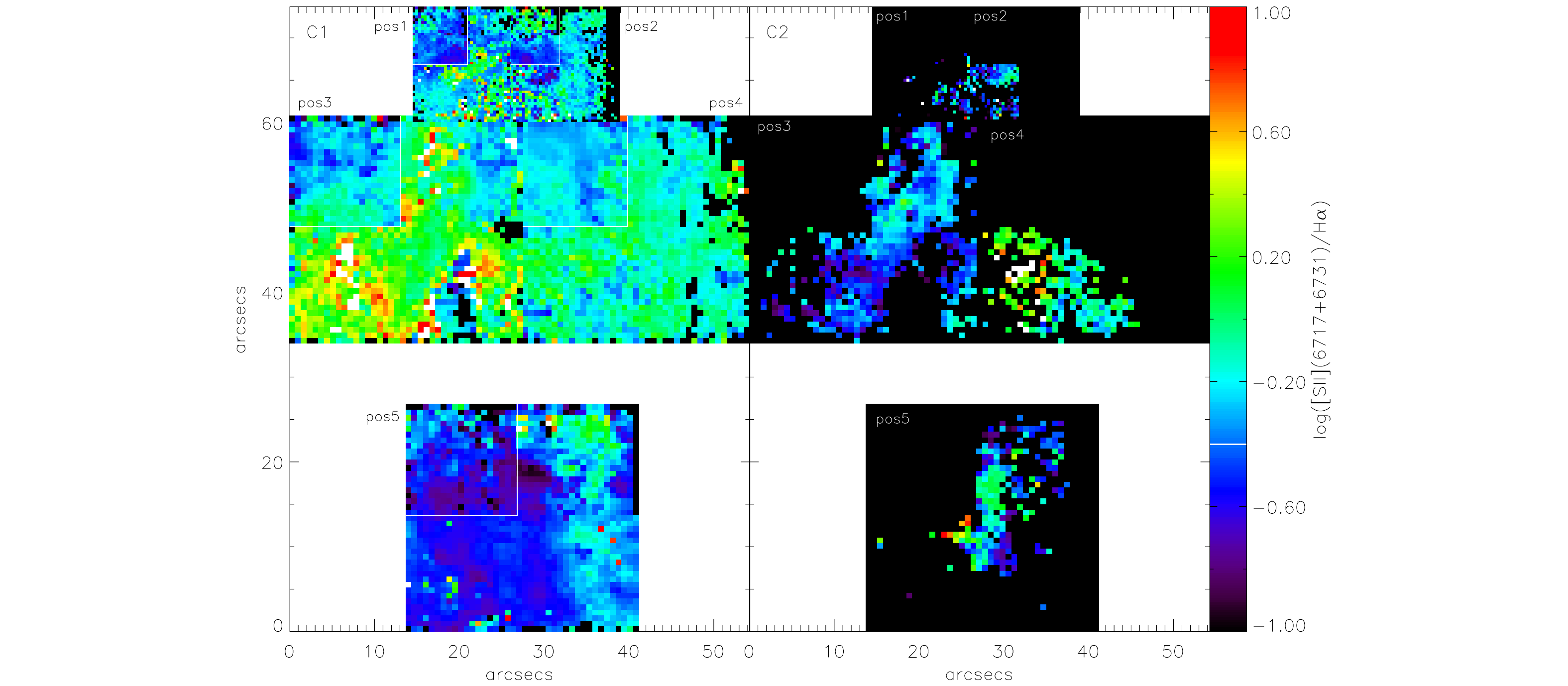}
\caption{[S\two]($\lambda$6717+$\lambda$6731)/H$\alpha$ flux ratio maps for C1 (left) and C2 (right). C3 was not identified in [S\two]. The marker in the colour-bar at $-0.4$ represents the fiducial ratio above which the excitation is likely to be dominated by non-photoionization processes (see text).}
\label{fig:SII_Ha}
\end{figure*}

\begin{figure*}
\centering
\includegraphics[width=\textwidth]{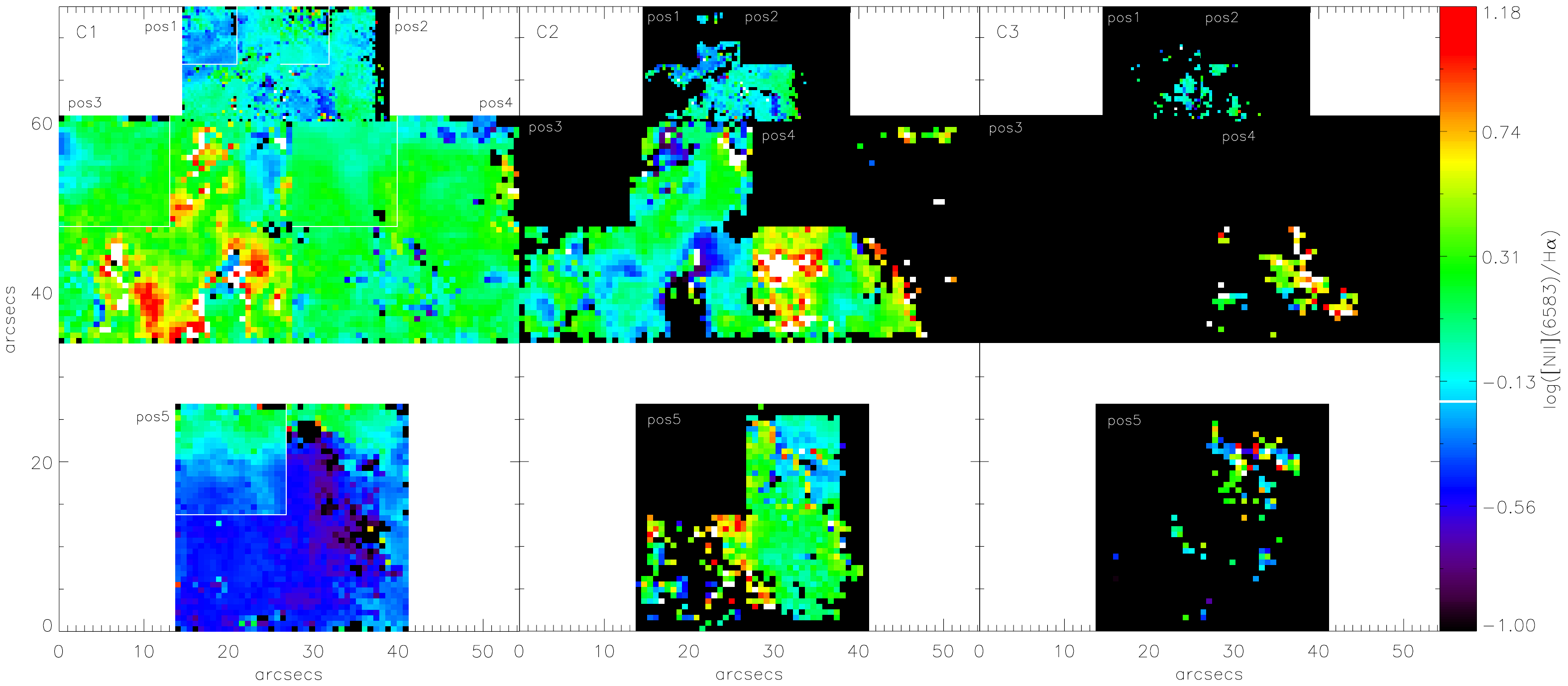}
\caption{[N\two]$\lambda$6583/H$\alpha$ flux ratio maps in the three line components. \emph{Left:} C1, \emph{centre:} C2; \emph{right:} C3. The marker in the colour-bar at $-0.2$ represents the fiducial ratio above which the excitation is likely to be dominated by non-photoionization processes (see text).}
\label{fig:NII_Ha}
\end{figure*}

Much of both the C1- and C2-emitting gas has [N\two]/H$\alpha$ and [S\two]/H$\alpha$ line ratios (Figs.~\ref{fig:SII_Ha} and \ref{fig:NII_Ha}) that fall above the fiducial non-photoionization threshold, suggesting a significant contribution to the ionization from shocks. Overall, C1 exhibits higher ratios, particularly in [S\two]. However, we also see localised regions of much higher line ratios, most likely representing pockets/filaments of strongly shocked gas (log([S\two]/H$\alpha$)$>$0.6 and log([N\two]/H$\alpha$)$>$0.7). These are correlated with fainter line fluxes, a relationship that we have seen in similar environments in other galaxies \citep{westm07a, westm09a}.

\begin{figure*}
\centering
\begin{overpic}[width=\textwidth]
{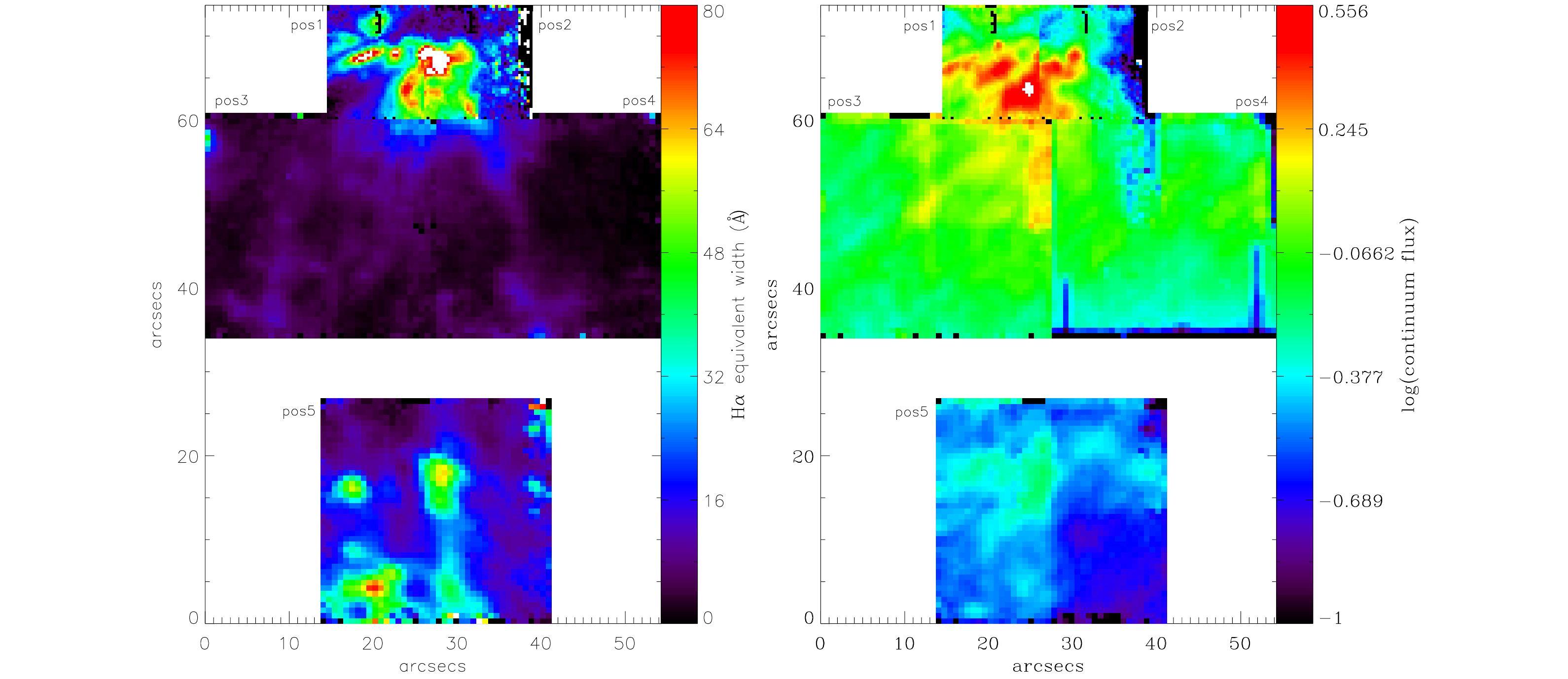}
\put(14,40){(a)}
\put(54,40){(b)}
\end{overpic}
\caption{(a) Equivalent width of the integrated H$\alpha$ line flux; (b) continuum flux level measured near H$\alpha$ in units of $10^{-16}$ erg~s$^{-1}$~cm$^{-2}$~arcsec$^{-2}$. Some residual instrumental artefacts are evident around the field edges.}
\label{fig:ew}
\end{figure*}

Since we are observing the H$\alpha$-emitting gas superimposed onto the background disk of the galaxy, stellar H$\alpha$ absorption is expected, and indeed identified in some spaxels (see Fig.~\ref{fig:eg_fits}f). It is therefore possible that the regions of enhanced [N\two]/H$\alpha$ and [S\two]/H$\alpha$ line ratios seen in positions 3 and 4 are simply the result of H$\alpha$ absorption affecting the line ratios. Given the spectral resolution and S/N of our data, it is very difficult to quantify the level of H$\alpha$ absorption in each spectrum. In order to investigate whether the regions of enhanced line ratios are due to H$\alpha$ absorption, we have measured and mapped out the equivalent width of the integrated H$\alpha$ line flux, and the continuum flux level near H$\alpha$. These maps are shown in Fig.~\ref{fig:ew}. No correlation is seen between the regions of enhanced line ratios and either of these maps, leading us to conclude that the effect of H$\alpha$ absorption is only slight, and our measured line ratios are real.

\begin{figure*}
\centering
\includegraphics[width=\textwidth]{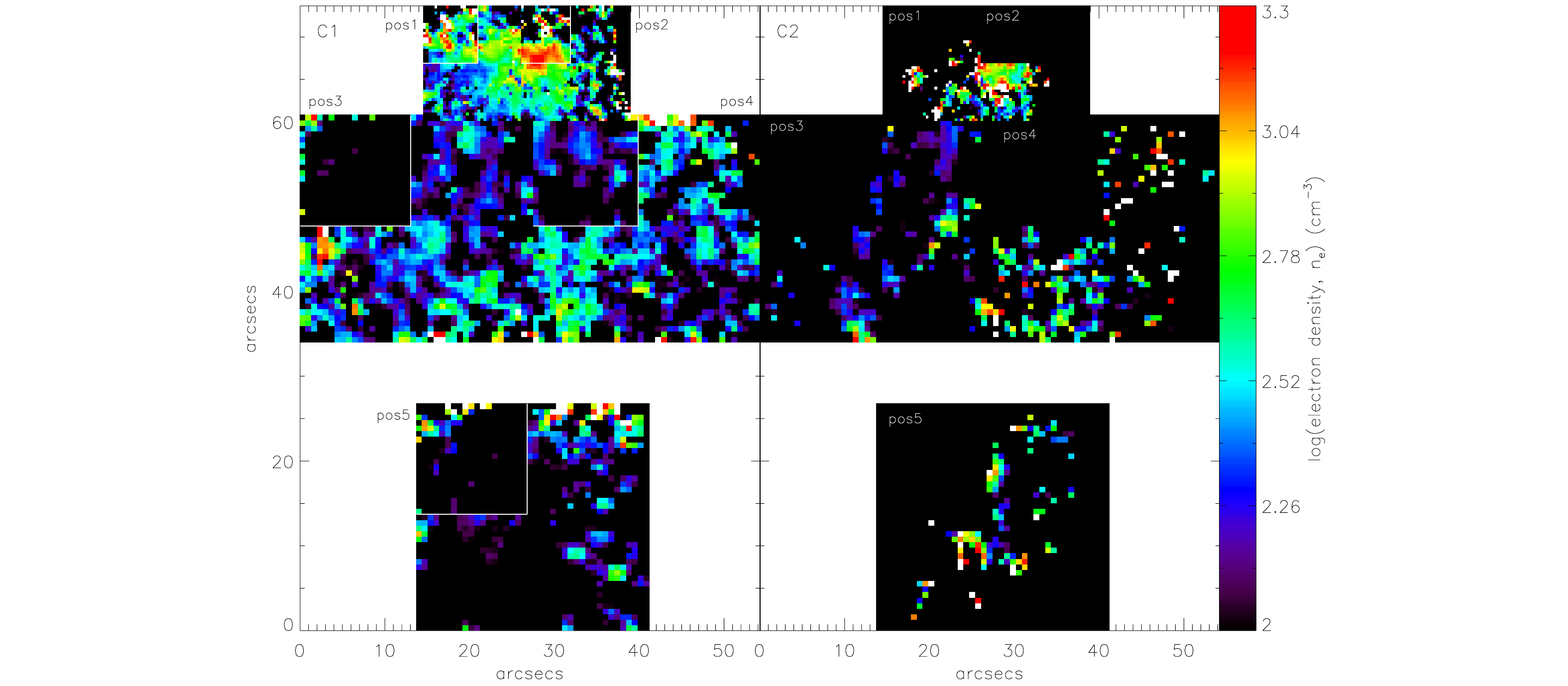}
\caption{Electron density maps in the two identified [S\two] line components. \emph{Left:} C1, \emph{right:} C2. The datacube was smoothed in the spatial directions by a 3x3 rolling summation to increase S/N before the [S\two] lines were fit.}
\label{fig:elecdens}
\end{figure*}

The [S\two]-derived electron densities (Fig.~\ref{fig:elecdens}) are high in the nuclear region, peaking at $\sim$2000~\cmt, and remain $>$700~\cmt\ in the whole central disk. By a radius of $\sim$10$''$ the electron densities have dropped dramatically to close to the low density limit (few 100~\cmt). Localised regions of increased density are seen, with values up to 500--700~\cmt. The density morphology does not follow any of the other patterns seen in the H$\alpha$ kinematics.

\subsection{Outflow cone}\label{sect:obs_cone}
As described in the introduction, deep X-ray and H$\alpha$ imaging have revealed that the inner $\sim$1~kpc of the starburst outflow within NGC~253 forms a limb-brightened cone \citep{strickland00}. This structure can also be clearly identified both in our IFU maps and WFI H$\alpha$ imaging. In the following, we describe its properties as derived from these observations. A simple kinematic model of the outflow cone is presented and discussed in Section~\ref{sect:cone}.

The outflow cone is clearly seen in the morphology of the diverging H$\alpha$ velocity components in the VIMOS radial velocity maps of Fig.~\ref{fig:Ha_vel}, where the two components trace out a triangular shape extending south-east from the nuclear region along the minor axis. We have attempted to highlight this formation, which has an opening angle of $\sim$60--65$^{\circ}$, with white lines on the C2 map. That the H$\alpha$ cone is contained within the minor-axis spurs of H$_2$ emission seen in the NICMOS image of \citet{sugai03}, implies that molecular H$_2$ is entrained by the outflow.

The P-V diagrams derived from the vertically-oriented pseudo-slits (Fig.~\ref{fig:pv_vertical}) quite clearly show diverging H$\alpha$ velocity components emerging from a chaotic central region at disk heights of $\gtrsim$15$''$. The coherence of the H$\alpha$ velocities in the outflow region demonstrates that the $10^4$\,K, H$\alpha$-emitting gas layers must be quite thin and dynamically stable. The horizontal59 and horizontal41 (Fig.~\ref{fig:pv_horizontal}) show how the cone expands from $\sim$10$''$ in width at a minor axis distance of 10$''$ from the nucleus, to $>$40$''$ at a projected distance of 30$''$. The morphology of the FWHM maps also broadly follows the cone shape, although the line widths vary dramatically, with many small-scale regions of much broader emission. Generally, the line widths remain above $\sim$200~\kms\ in both C1 and C2. The very broad components suggest there may be material filling the cone in some regions.


\subsection{Kinematically distinct features in the flow}
The flux, FWHM, and radial velocity morphologies are suggestive of small-scale, kinematically distinct, shell-like or filamentary features embedded within, and aligned with the outflow. For example, position 3 contains what appears to be a filament of material that is redshifted in both C1 and C2 compared to the surrounding material (this shows up in the vertical16 P-V plot, Fig.~\ref{fig:pv_vertical}, at $x$=50--55$''$). From the radial velocity map, this feature appears to extend along the wind outflow direction, radially away from the nucleus. Another region of redshifted material appears in C3 in position 4. This region is again clearly kinematically distinct from the surrounding gas (see the horizontal41 P-V plot, Fig.~\ref{fig:pv_horizontal}, at $x$=35--40$''$).

These distinct filaments/blobs of gas may be material in the foreground that is moving separately to the gas in the outflow. Alternatively, the redward velocity shifts may suggest that we are seeing through gaps in the foreground material, allowing the line-of-sight to penetrate either to a point deeper within the outflow, or gas associated with the background disk of the galaxy.

These features are reminiscent of other systems with ``bubbling winds'' such as the LIRG NGC~3256 \citep[][see also \citealt{lipari04b}]{lipari04a}. These type of flows seem to produce fragmented shells where the cool gas is already accelerated, and thus differ somewhat from the radial streamers predicted to come from stationary clumps in a steady outflow \citep[e.g.][]{pittard05, cooper09}.

%As mentioned above, the FWHM map also shows that C2 in position 5 behaves differently to C1. Here C1 is narrow ($<$100~\kms) across the majority of the field, whereas C2 remains quite broad (200--300~\kms).

%\subsection{Gas outside the cone}
%Outside of outflow cone, where only one H$\alpha$ component is detected (positions 1, 2 and 4), the line widths are much narrower ($\sim$60--120~\kms). Here the velocity field is also quiescent.
%H$\alpha$ filaments to the NE of position 1 exhibit high densities and line ratios here clearly not enhanced and consistent with just photoionization.

\begin{figure*}
\centering
\includegraphics[width=\textwidth]{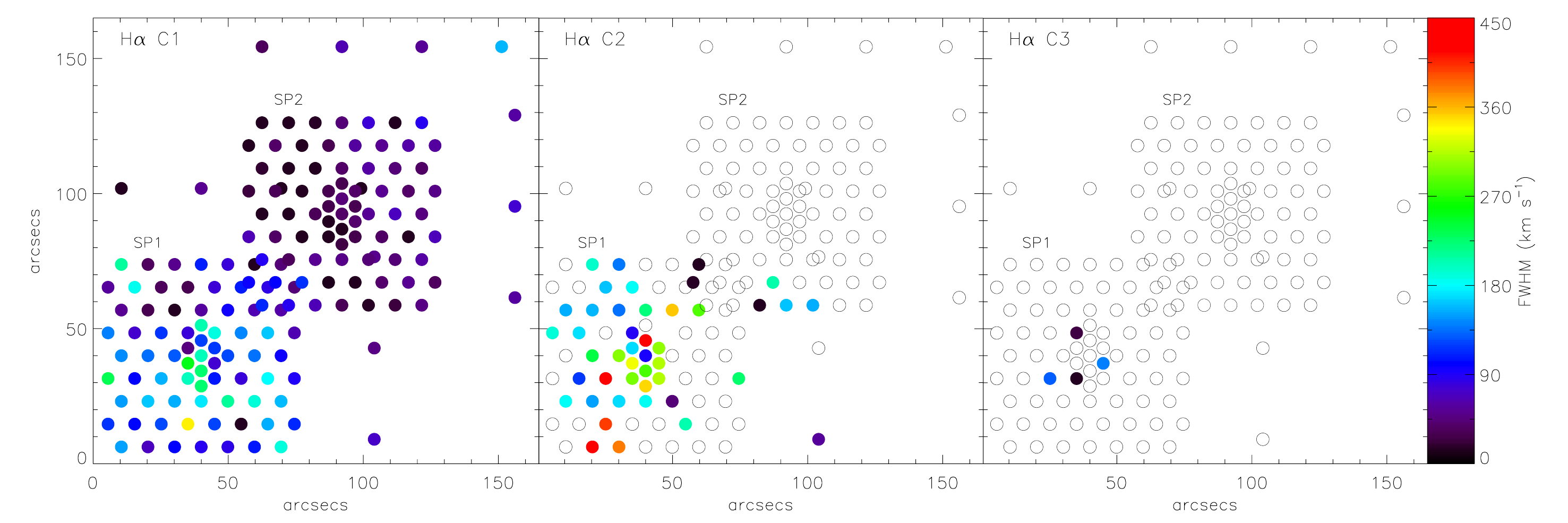}
\caption{SparsePak H$\alpha$ FWHM maps in the three H$\alpha$ line components. \emph{Left:} C1, \emph{centre:} C2 and \emph{right:} C3. A scale bar is given in units of \kms, corrected for instrumental broadening.}
\label{fig:sp_Ha_fwhm}
\end{figure*}
\begin{figure*}
\centering
\includegraphics[width=\textwidth]{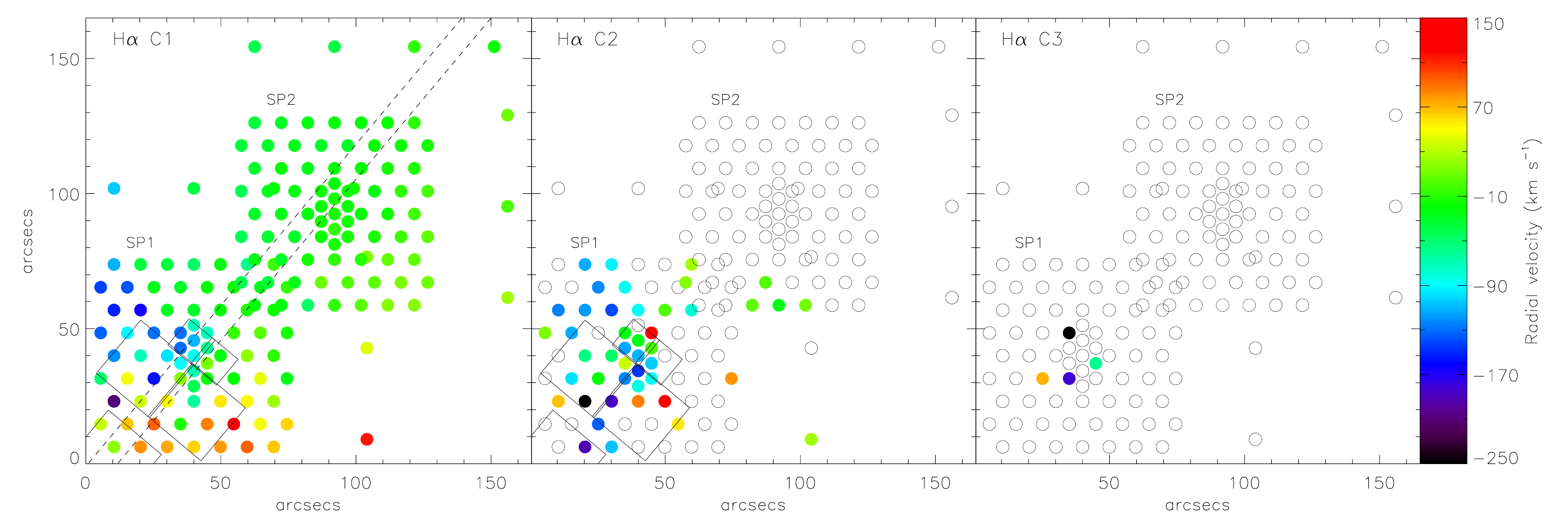}
\caption{SparsePak H$\alpha$ radial velocity maps in the three line components. \emph{Left:} C1; \emph{centre:} C2 and \emph{right:} C3. A scale bar is given in units of \kms, relative to $v_{\rm sys}$. The VIMOS IFU footprints are overlaid. The diagonal dashed lines indicate the position and width of a pseudo-slit from which the position-velocity data shown in of Fig.~\ref{fig:pv_vertical_sp} were extracted.}
\label{fig:sp_Ha_vel}
\end{figure*}

\begin{figure*}
\centering
\begin{minipage}{7cm}
\includegraphics[width=7cm]{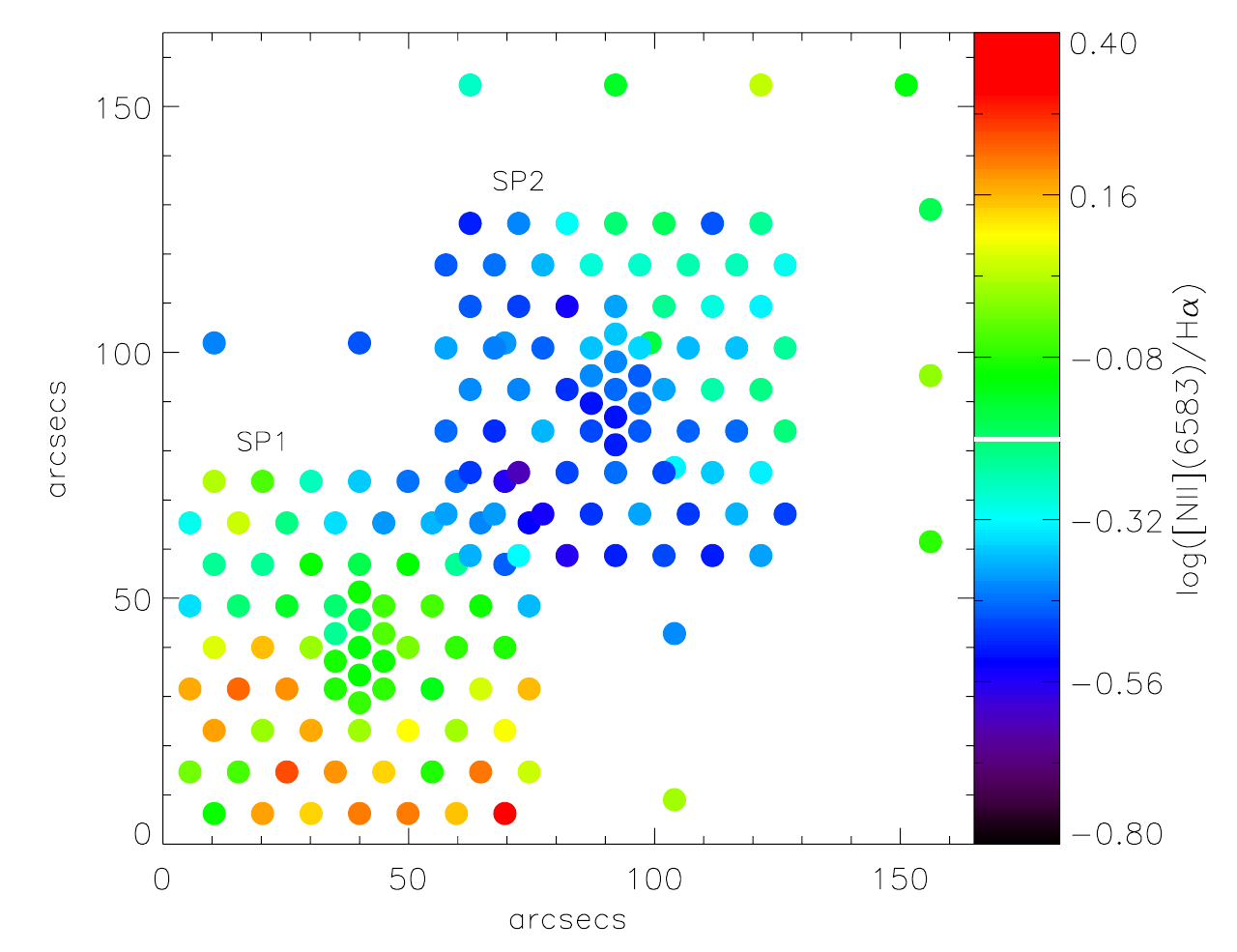}
\end{minipage}
\hspace{0.1cm}
\begin{minipage}{7cm}
\includegraphics[width=7cm]{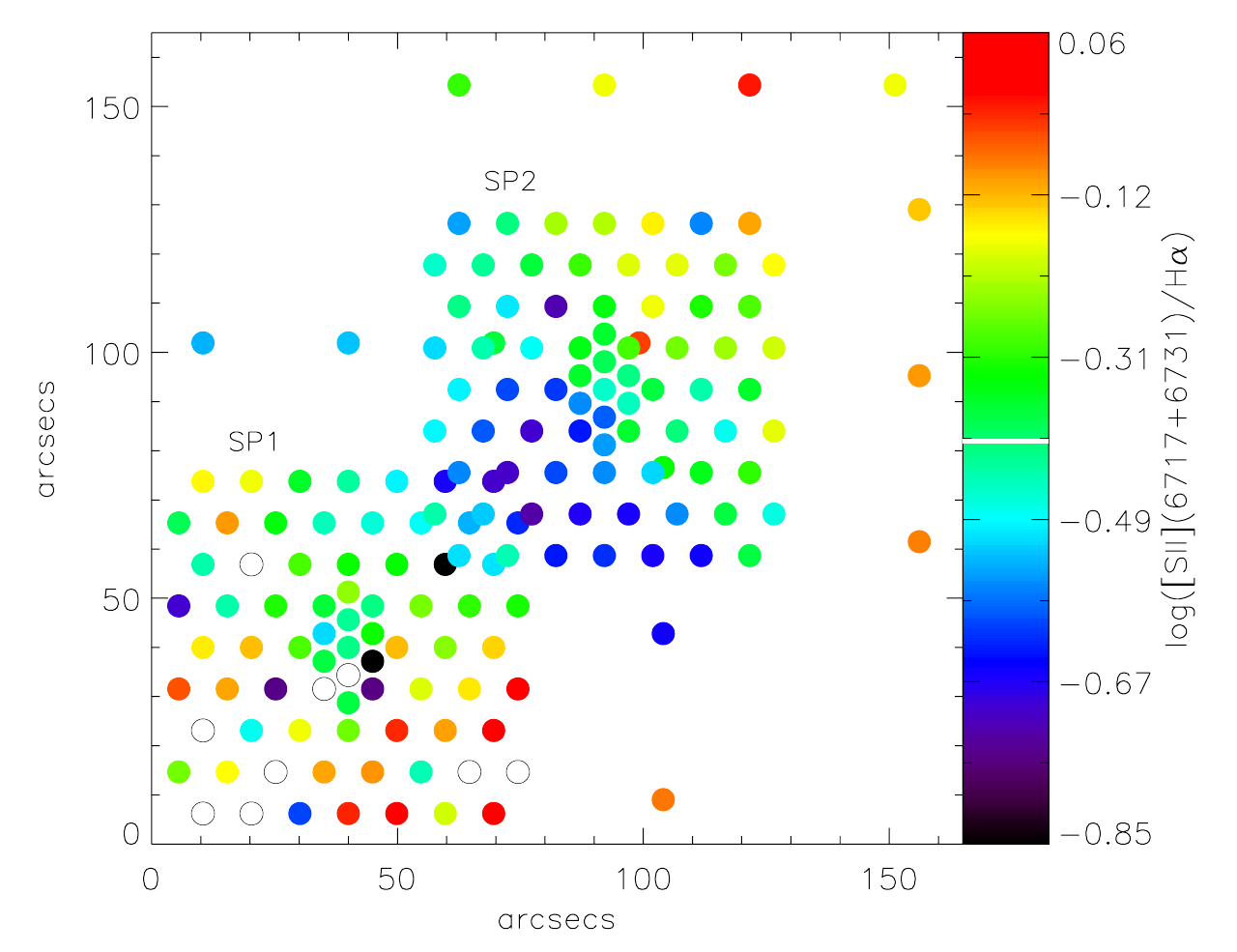}
\end{minipage}
\caption{\textit{Left:} SparsePak [N\two]$\lambda$6583/H$\alpha$ flux ratio maps from summed line fluxes over all identified components; \textit{right:} SparsePak [S\two]($\lambda$6717+$\lambda$6731)/H$\alpha$ flux ratio maps from summed line fluxes over all identified components. The marker in the colour-bars represents the respective fiducial ratios above which the excitation is likely to be dominated by non-photoionization processes.}
\label{fig:sp_NII_Ha}
\end{figure*}

\begin{figure}
\centering
\includegraphics[width=0.49\textwidth]{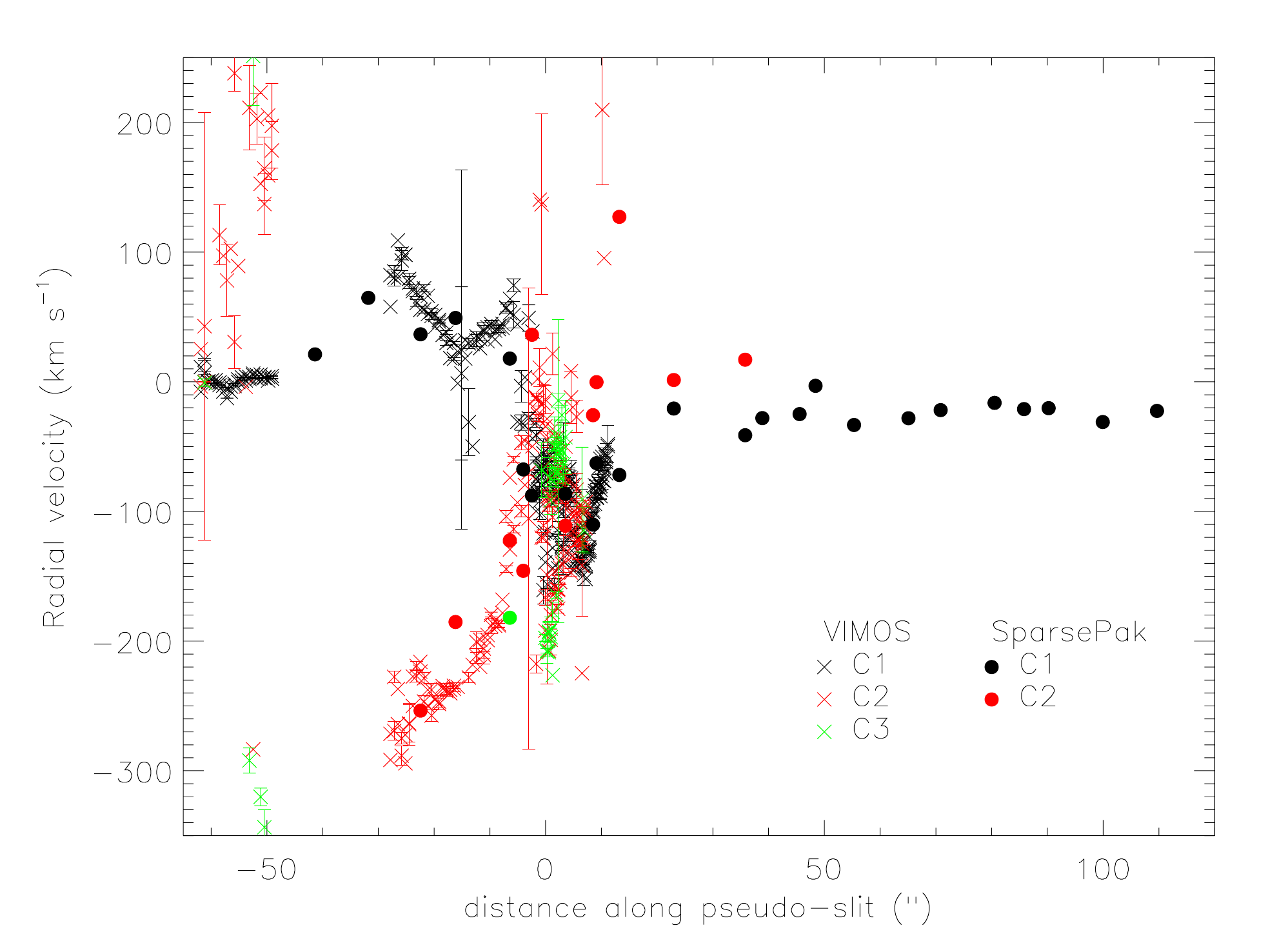}
\caption{Minor axis position-velocity plot including data from both the VIMOS and SparsePak datasets. Crosses represent radial velocities extracted from the vertical24 VIMOS pseudo-slit shown in Fig.~\ref{fig:pv_vertical}, whereas filled circles represent SparsePak velocities extracted from a $10''$-wide pseudo-slit aligned with minor axis as shown on Fig.~\ref{fig:sp_Ha_vel}. The offsets are calibrated such that $x=0$ corresponds to the nucleus. Velocities are corrected to the heliocentric reference frame, and given relative to $v_{\rm sys}$ (=250~\kms).}
\label{fig:pv_vertical_sp}
\end{figure}

%%%%%%%%%
\subsection{SparsePak emission line maps}\label{sect:sp_maps}
%\textit{Unfortunately the SparsePak data doesn't really add a huge amount!}

Figs.~\ref{fig:sp_Ha_fwhm} and \ref{fig:sp_Ha_vel} show the SparsePak FWHM and radial velocity maps, respectively. The [N\two]$\lambda$6583/H$\alpha$ and [S\two]($\lambda$6717+$\lambda$6731)/H$\alpha$ line ratio maps are shown in Fig.~\ref{fig:sp_NII_Ha}. As mentioned above, the footprints of the SparsePak observations are shown in Fig.~\ref{fig:finder_large}.
%The electron density (derived from the [S\two]$\lambda\lambda$6717,6731 doublet)

Two H$\alpha$ line components are detected in the central disk and southern outflow regions. In the spaxels covering the nuclear regions and a few associated with the southern outflow, the line shape is of the type (a) configuration (narrow+broad), with broad component widths of 200--450~\kms\ (Fig.~\ref{fig:sp_Ha_fwhm}). The remaining spaxels show type (b) line shapes (two split line components of $\sim$same width). After running our tests and filters to remove incorrect or unphysical line fits, we assigned the brightest component to C1 in all cases. A third line component can be identified in a few spaxels in the nuclear region. Here we assigned the additional narrower component to C3.

As modelled by \citet{sorai00} and \citet{das01}, the inner $x_2$ orbits of the stellar bar extend diagonally over the full extent of position SP1 (as shown in Fig.~\ref{fig:finder_large}), meaning that the major axis dynamics are dominated by the inner bar region.

Densities (maps not shown) of $>$300~\cmt\ are found in places along the major axis (disk and starburst), but otherwise they remain close to or below the low density limit ($\lesssim$100~\cmt; as expected in the halo). Where the SparsePak and VIMOS observations overlap, the electron densities measured with both datasets are consistent (once the coarser spatial resolution of SparsePak is taken into account).

[N\two]/H$\alpha$ and [S\two]/H$\alpha$ line ratios (Fig.~\ref{fig:sp_NII_Ha}) are again consistent with those measured from our VIMOS data, where the coverage of the two datasets overlap. The highest line ratios are seen across the south-east of position SP1, coincident with the southern outflow cone. Line ratios fall dramatically just to the north-west of the nucleus, along a band parallel to the galaxy major axis. They then gradually rise again towards the north-west of position SP2, out to $>$100$''$ ($>$2~kpc) from the nucleus.

To check consistency between our VIMOS and SparsePak datasets, we extracted the velocities along a $10''$-wide pseudo-slit aligned with minor axis as shown on Fig.~\ref{fig:sp_Ha_vel}, and plotted them together with the VIMOS radial velocities from the vertical24 pseudo-slit in Fig.~\ref{fig:pv_vertical_sp}. This plot shows how the measurements from the two datasets correspond very well. These velocities are also consistent with those measured by \citet{schulz92}.

%%%%%%%%%%%%%%%%%%%%%%%%%%%%%%%%
\section{Discussion} \label{sect:disc}
%\textit{From proposal:} Questions we hope to address here include: does the outflow become supersonic (where is the sonic point)? Does the flow behave differently within and without of this point (as we find in M82)? Are the coherent structures themselves fast moving, or are they simply ablation trails \citep{melioli05}? How/where is the outflow collimated, and does this relate to the geometry of the starburst?

In this section we discuss the implications of our results on our understanding of the NGC~253 superwind. We then discuss the relationships between the known properties of the central starburst, the disk, the molecular gas distribution, and the larger-scale properties of the aforementioned diffuse halo, to the observed characteristics of the inner wind. A number of the features that we refer to in the following sections are illustrated schematically on Fig.~\ref{fig:schematic}, introduced at the end of Section~\ref{sect:inner_wind}, to aid the reader in following our discussion.

%Although the line splitting is clear, SBH98 found that the optical emission is by no means distributed evenly over the surfaces of the cones, nor emanates smoothly from the entire starburst region. -- N253 similar to M82.

%%
\subsection{Kinematic modelling of inner wind cone}\label{sect:cone}
We modelled the inner wind cone as described in Section~\ref{sect:obs_cone} using the 3D morpho-kinematical modelling tool \textsc{shape} \citep{steffen10}. We restricted our fit to the region between $R$=15--35$''$ from the nucleus (or $y$=35--55$''$ on Fig.~\ref{fig:Ha_vel}) where a coherent signature of an expanding conical outflow is observed. This shape therefore forms a conic frustum. Beyond this outer radius, the gap in our IFU coverage restricts our analysis, and it is not clear what happens to the cone structure here.

We chose to model the particle velocities as flowing tangentially to the cone surface since this provided a good match to the observations, and was most easily understood physically in terms of material entrained by the high-velocity wind. Furthermore, a number of constraints to the model can be made from the observations. The fact that the two velocity components continue to diverge with radius implies that the outflow is accelerating. The increasing velocity of the redshifted component indicates that the far-side of the cone is angled away from the line-of-sight (assuming the gas flows tangentially to the cone surface). A smaller opening angle not only does not fit the morphology of the velocity maps of Fig.~\ref{fig:pv_vertical}, but requires excessively large accelerations to fit the data. It was not required to change the inclination of the cone from that of normal to the galaxy disk (i.e.\ $i = 12^{\circ}$).

\begin{figure}
\centering
\begin{minipage}{7cm}
\begin{overpic}[width=7cm]
{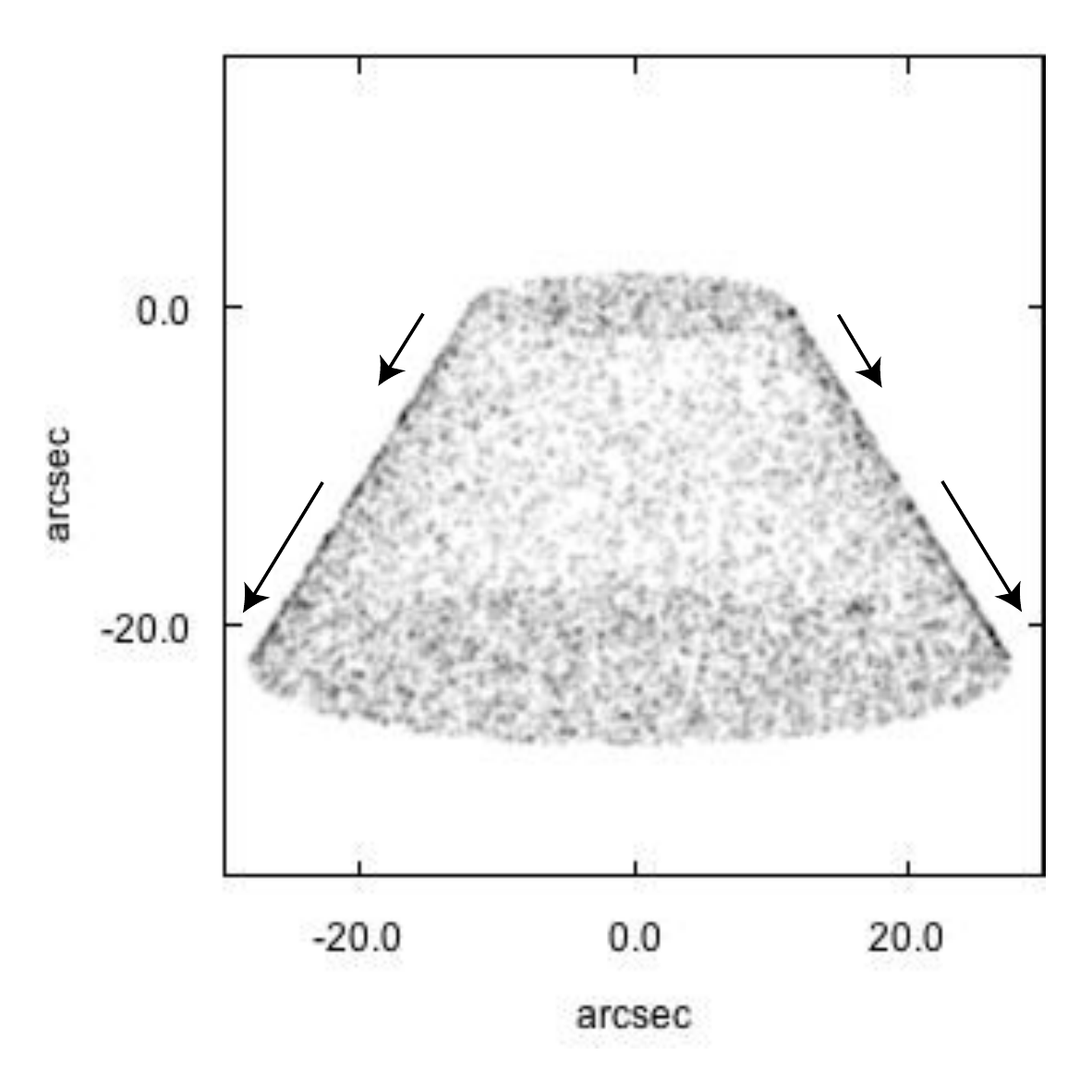}
\put(25,87){\large (a)}
\end{overpic}
\end{minipage}
\begin{minipage}{7cm}
\begin{overpic}[width=7cm]
{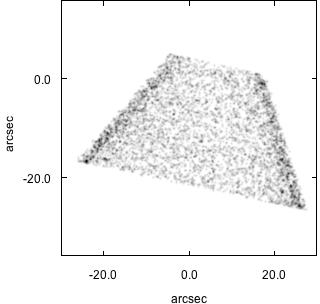}
\put(25,85){\large (b)}
\end{overpic}
\end{minipage}
\caption{Best fitting kinematic model of outflow cone from Shape as seen (a) projected onto the plane of the sky, and (b) side on. The arrows in the top panel illustrate the direction and magnitude of the velocity vectors: along the cone walls, increasing with radius.}
\label{fig:kinematic_model}
\end{figure}

\begin{figure}
\centering
\begin{minipage}{7cm}
\includegraphics[width=7cm]{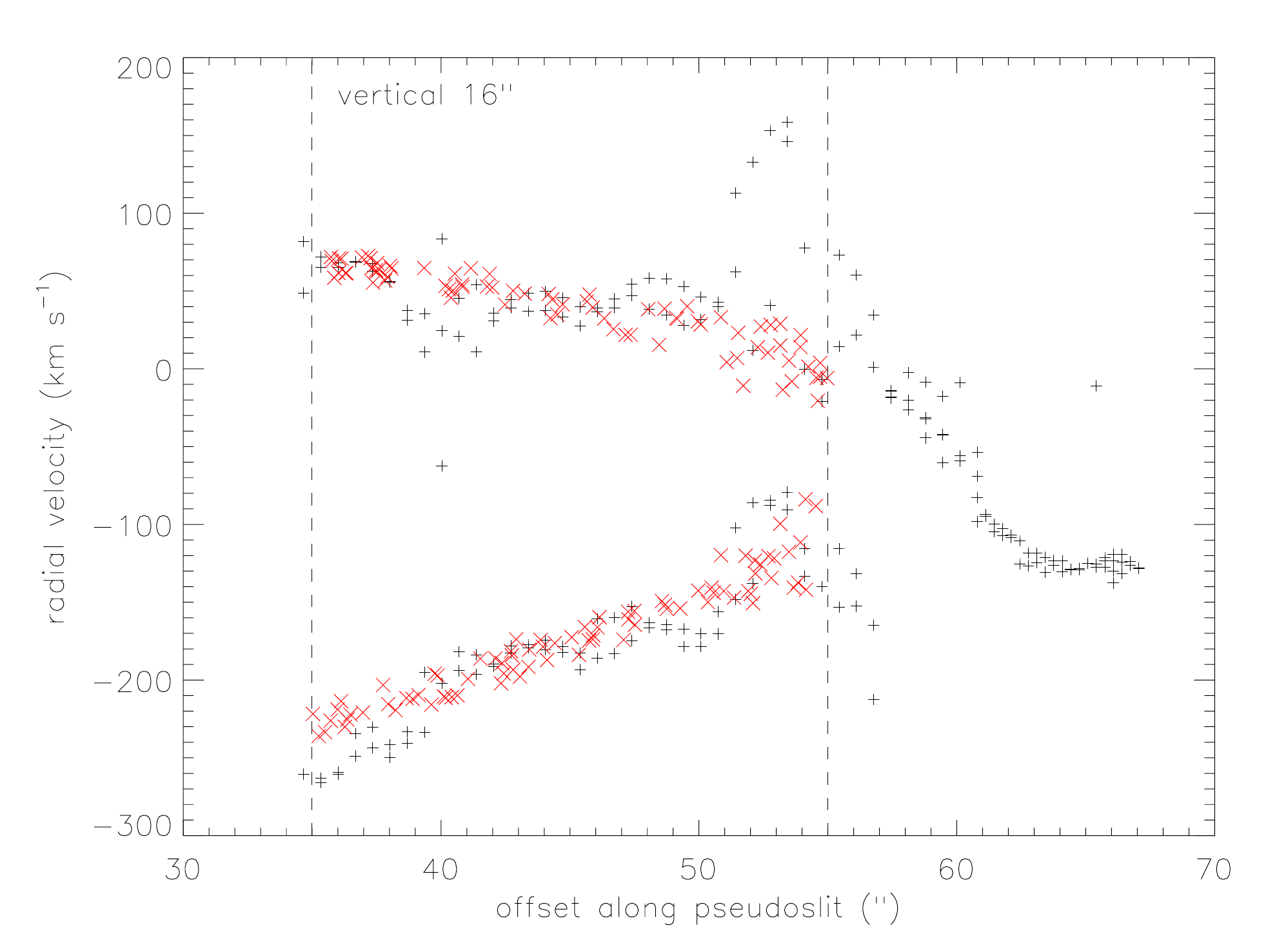}
\end{minipage}
%\vspace{0.5cm}
\begin{minipage}{7cm}
\includegraphics[width=7cm]{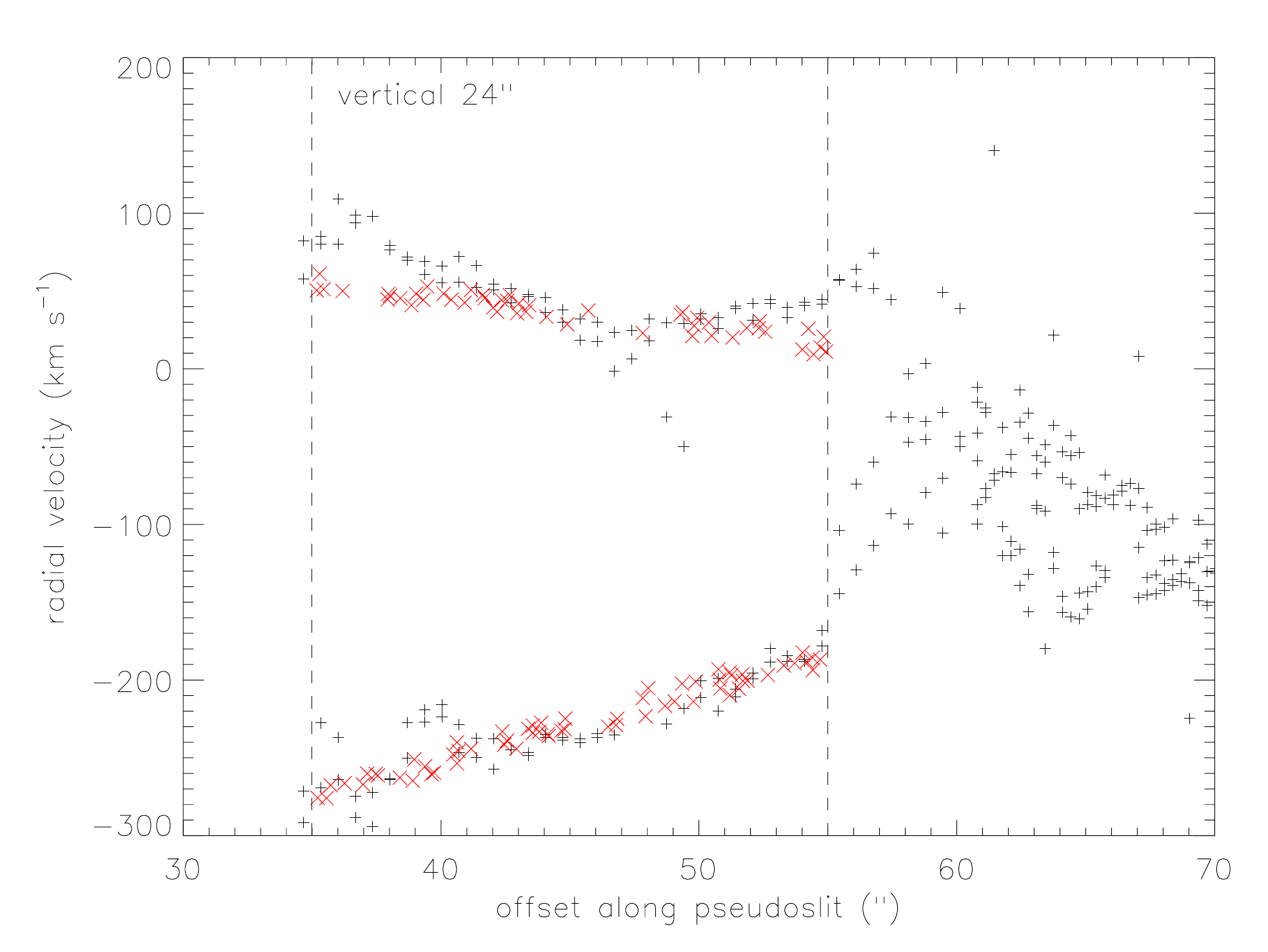}
\end{minipage}
%\vspace{0.5cm}
\begin{minipage}{7cm}
\includegraphics[width=7cm]{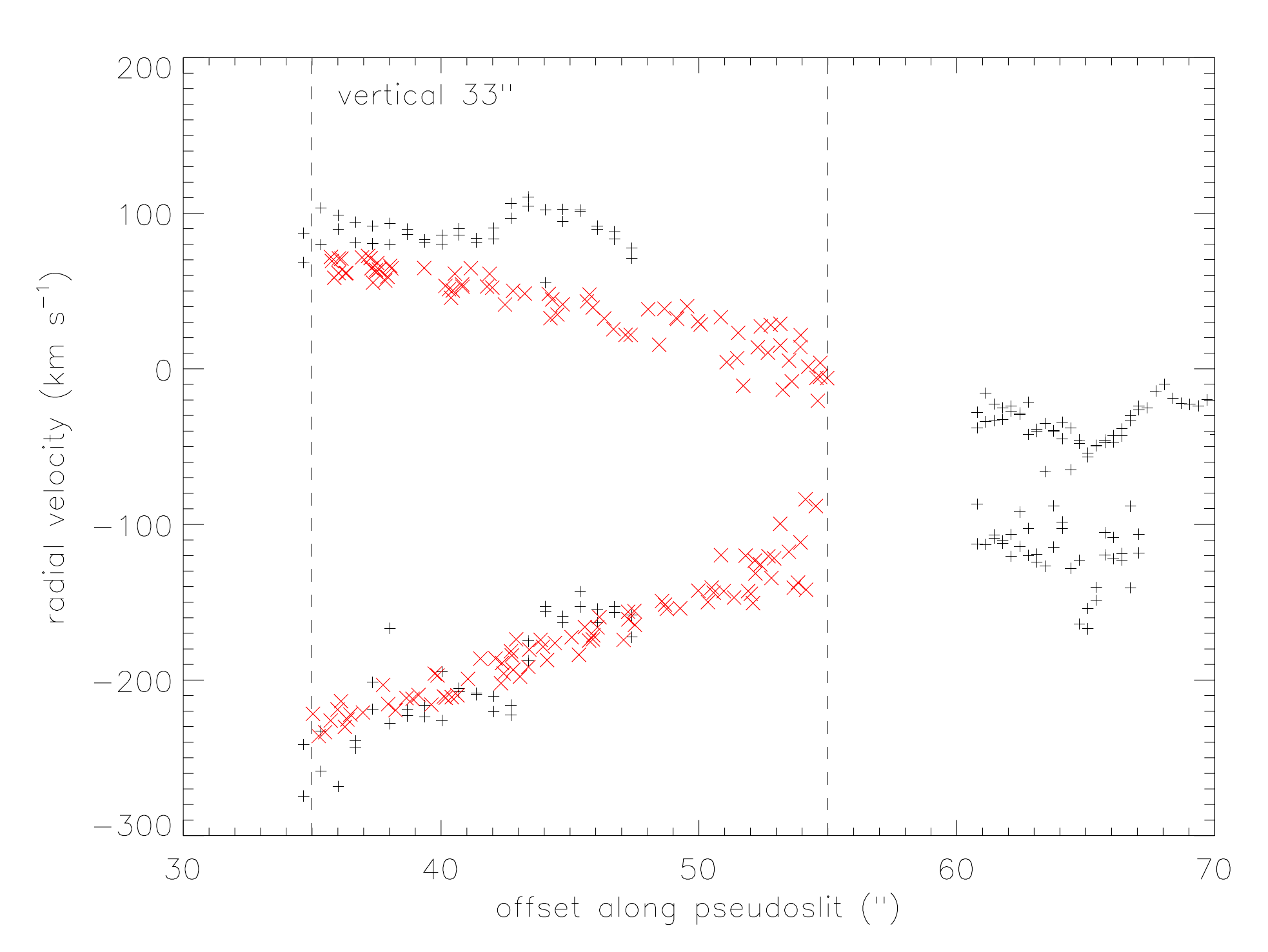}
\end{minipage}
\caption{Position-velocity plots from Fig.~\ref{fig:pv_vertical} (black +) with the best-fit model points overlaid (red $\times$). The $x$-axis scale is the same as in Fig.~\ref{fig:pv_vertical}. The vertical dashed line indicate the region within which we modelled the outflow cone. An offset of $-30$~\kms\ has been applied to the model data in the vertical24 plot in order to better match the data.}
\label{fig:pv_sims}
\end{figure}

After performing simulations over a range of frustum sizes, inclination angles, opening angles, and velocity laws, trying to match both the morphology of the H$\alpha$ radial velocity maps, and the position-velocity diagrams of Fig.~\ref{fig:pv_vertical}, we derived a best-fitting model whose parameters are given in Table~\ref{tbl:cone_params}. Fig.~\ref{fig:kinematic_model} shows the model, both as seen projected onto the plane of the sky and side-on, where the arrows show the direction of the velocity vectors and illustrate how the flow speed increases with radius. Fig.~\ref{fig:pv_sims} shows three of the observed P-V diagrams from Fig.~\ref{fig:pv_vertical} together with the simulated data from the best-fitting model, illustrating how well this simple model fits the data. There are indications, however, that there are distortions from a simple cone shape and/or simple velocity law. For example, an offset of $-30$~\kms\ needed to be applied to the model data in the vertical24 plot in order to match the observations. Additionally, the velocity ellipse in horizontal41 (Fig.~\ref{fig:pv_horizontal}) does not appear to close on the left side, again indicating deviations from a simple geometry.

The inferred diameter of the base of the outflow model (the top end of the cone) is $\sim$20$''$ ($\sim$380~pc). This, however, does not reflect the diameter of the energy injection zone (the starburst region), but simply the width of the optically visible cone as it emerges from the veil of obscuration surrounding the nuclear regions. The best-fitting velocity law tells us that the H$\alpha$-emitting gas, as it enters this top end, is travelling at a deprojected outflow speed of 110~\kms, which then accelerates at a rate of $5z$ (where $z$ is the non-projected physical distance along the cone in arcsecs). If the gas accelerates at a constant rate from its launch site, then the flow would become supersonic with respect to gas at $10^4$\,K (sonic speed $\sim$10~\kms) almost immediately.

By the time the gas reaches the end of the cone model (35$''$), it is travelling at 210~\kms. If this velocity law continues to hold to larger radii, it would be travelling at 360~\kms\ at 50$''$ ($\sim$1~kpc). This is much slower than the deprojected outflow speed of 390~\kms\ estimated for material at $R<16''$ by \citet{schulz92}, and the $\sim$600~\kms\ outflow speed of the H$\alpha$ gas in M82 \citep{shopbell98}. Since the escape velocity of NGC~253 is also estimated to be a few hundred \kms\ \citep{heesen09a}, it is not clear whether the H$\alpha$ gas would be able to escape the galaxy's gravitational field. The hot X-ray emitting gas, however, must have an outflow speed which is faster than this, implying that the hottest material has the potential of escaping the galaxy.

The best-fitting model has an opening angle of $\alpha=60^{\circ}$ between radii of 15--35$''$, constrained quite tightly by the morphology of the H$\alpha$ C2 velocity map (Fig.~\ref{fig:Ha_vel}). This is much larger than the 20--25$^{\circ}$ or 32$^{\circ}$ opening angles measured by \citet{strickland00} and \citet{pietsch00}, respectively, although these were measured at larger radii (30--55$''$). This is also much larger than the $\sim$$25^{\circ}$ opening angle of the M82 outflow \citep{shopbell98}.

%We can now make an assessment of whether the kinetic energy in the H$\alpha$-emitting phase of the outflow gas is enough to
To measure the kinetic energy in the inner wind from our model, we assume the volume of the model conical frustum as described above is $2.1\times 10^7$~pc$^3$ (assuming a cone wall thickness of 2$''$ $\sim$ 40~pc), and the electron density in the cone walls is $\lesssim$100~\cmt. This gives a mass of H$\alpha$-emitting gas of $\sim$$5\times 10^7$\,$f$ \Msol\ (where $f$ is the filling factor). To check this we can also apply the method of \citet{pottasch80}: an observed H$\alpha$ surface brightness of $\sim$$1\times 10^{-16}$~erg~s$^{-1}$~cm$^{-2}$~arcsec$^{-2}$ and electron density of 100~\cmt, results in a mass of $6\times 10^{6}$~\Msol, which is in reasonable agreement if the filling factor $f$$\sim$0.1. Thus, assuming a mass of $10^7$~\Msol\ and an outflow velocity of $\sim$200~\kms, the H$\alpha$-emitting phase has a kinetic energy of $\sim$$4\times 10^{54}$~ergs.

The SNe rate in NGC~253 has been variously estimated to lie between 0.03--0.3~yr$^{-1}$ \citep{mattila01, ulvestad97}. Following the arguments of \citet{strickland02}, assuming the lowest SNe rate estimate gives a mechanical energy injection rate of $\sim$$5\times 10^{41}$~$\eta_{\rm therm}$~erg~s$^{-1}$ (where $\eta_{\rm therm}$ is the SN energy thermalisation efficiency). Thus a starburst of even 5~Myr duration is able to produce enough energy ($\approx$ $8\times 10^{54}$ ergs, with a conservative choice of $\eta_{\rm therm}$=10\%) to power the observed H$\alpha$ outflow in the inner wind. This result is in agreement with the findings of \citet{strickland00} and \citet{matsubayashi09}.

\begin{table}
\centering
\caption{Parameters of the conic frustum model that best reproduces the observed kinematics of the southern outflow in NGC~253.}
\label{tbl:cone_params}
\begin{tabular}{l l}
\hline
Parameter & Best-fit value \\
\hline
Projected distance of frustum start, $R_1$$^{\rm a}$ & 15$''$ (280~pc) \\
Projected distance of frustum end, $R_2$ & $>$35$''$ ($>$660~pc)$^{\rm b}$ \\
Diameter of the base of the outflow, $D_1$ & 20$''$ (380~pc) \\
%Minimum radius, $R_1$ & 10$''$ (190~pc) \\
%Maximum radius, $R_2$ & 24$''$ (450~pc) \\
Opening angle, $\alpha$ & $60^{\circ}$ \\
Inclination angle, $i$ & $12^{\circ}$ \\
Position angle (P.A.) & $140^{\circ}$ \\
Velocity law, $v(z)^{\rm a}$ (\kms) ($''$) & $110 + 5z$ \\
\hline
\end{tabular}
\begin{tabular}{p{8cm}}
$^{\rm a}$ Here we define $R$ as the projected distance below the plane of the disk, and $z$ as the physical distance below the plane in the $z$-direction. \\
$^{\rm b}$ This distance is constrained by the limit of our IFU coverage. The cone is known to extend further out (see text for discussion).  \\
\end{tabular}
\end{table}

%\textit{Some evidence that flow becomes more collimated with distance, evolving from $\alpha=60^{\circ}$ to $\sim$30$^{\circ}$ from 15--55$''$? Does this make physical sense?} 

%%
\subsection{The origins and evolution of the inner wind} \label{sect:inner_wind}
The wind cone modelled above does not seem to be centred on the known nuclear starburst (located in the missing Q3 section of position 2), but instead on a region to the north-east of the nucleus. This displacement is also evident in both the deep H$\alpha$ image of \citet{strickland00} and the NICMOS H$_2$ image of \citet{sugai03}. What, then, is driving the outflow? If it is the nuclear starburst, how does the gas escape the nuclear region? What is collimating the flow? What are the conditions in the gas at the base of the flow?

We can see from the vertically-aligned P-V diagrams of Fig.~\ref{fig:pv_vertical} that, in H$\alpha$, the outflow emerges at $\sim$15$''$ from a dynamically chaotic nuclear region. That the extinction in the nuclear region is know to be high \citep[$A_{V}\gtrsim 10$;][]{kornei09, fernandez-ontiveros09} indicates that within $\sim$15$''$ we see, in the optical, only the surface layers of the gas. This is consistent with the fact that we measure peak [S\two]-derived electron densities of only $\sim$2000~\cmt, when observations of the high-density molecular gas tracer, HCN, imply densities of $\gtrsim$$10^5$~\cmt\ \citep{knudsen07}. Drawing any conclusions about the gas conditions within the nucleus from optical observations is therefore not possible.

IR and mm-wave imaging, however, shows that the molecular gas in the starburst region is distributed in a toroidal shape \citep{engelbracht98, garcia-burillo00, paglione95, paglione04}, and kinematic analysis suggests that molecular gas is channelled inward through the action of the bar \citep{sorai00}. This is presumably how the starburst is being fuelled, but how does it affect the outflowing gas? The first piece of evidence comes from the SiO ($v$=0, $J$=2$\rightarrow$1) observations made with the IRAM interferometer \citep{garcia-burillo00}, and shown as contours overlaid on the {HST} F814W ($I$-band) image in Fig.~\ref{fig:finder}b. The SiO emission reveals two nested molecular tori in the circumnuclear disk that are aligned with the PA of the stellar bar. The location and size-scales of these tori suggest that they may play a significant role in confining or collimating the outflow on the smallest scales. In support of this idea, \citet{t-t98} discuss how the inflow of matter to fuel the starburst can lead to a stationary hydrodynamic flow. In their model, the infalling material produces a ram pressure that easily exceeds that of the disk thermal pressure, and, when confronted with the mechanical luminosity from the starburst (i.e.\ the wind ram pressure), acts to channel the expanding starburst wind and collimate it into a kpc-scale, filamentary, biconical outflow.

The second piece of evidence comes from the gas motions seen in Br$\gamma$ in near-IR VLT/SINFONI IFU observations (M\"{u}ller-S\'{a}nchez et al.\ in prep.; see also \citealt{muller-sanchez10}). These observations reveal that there are multiple distinct outflow channels associated with the different components of the nuclear starburst, and that the outflows first exit the nuclear region in an easterly direction, then quickly divert towards the direction of the minor axis (M\"{u}ller-S\'{a}nchez et al.\ in prep.). This diversion or curving of the flow, also visible in our deep WFI H$\alpha$ imaging (Fig.~\ref{fig:finder}a), is supported by the changing orientation of the nuclear starburst clumps and extraplanar dust lanes seen on the \textit{HST} F814W image with distance. Within $\sim$550~pc of the galaxy mid-plane, the dust lanes are oriented in an easterly direction (particularly on the north-eastern side of the nucleus). Beyond this radius, the orientation of the dust lanes change to being consistent with the minor axis. Since the toroidal circumnuclear molecular disk(s) and the larger-scale stellar bar is oriented at PA $\approx$ 70$^{\circ}$, we hypothesise that these components have a strong influence on the collimation and direction of the flow in these inner regions, and that the hot wind flows not only down the pressure gradient but down any local gas density gradients as well. Given that the bar is rotating, the flow from within the nucleus must therefore change direction every few Myr (analogous to a garden sprinkler), and it is not clear how this would allow a kpc-scale outflow structure to survive.

\begin{figure*}
\centering
\includegraphics[width=1\textwidth]{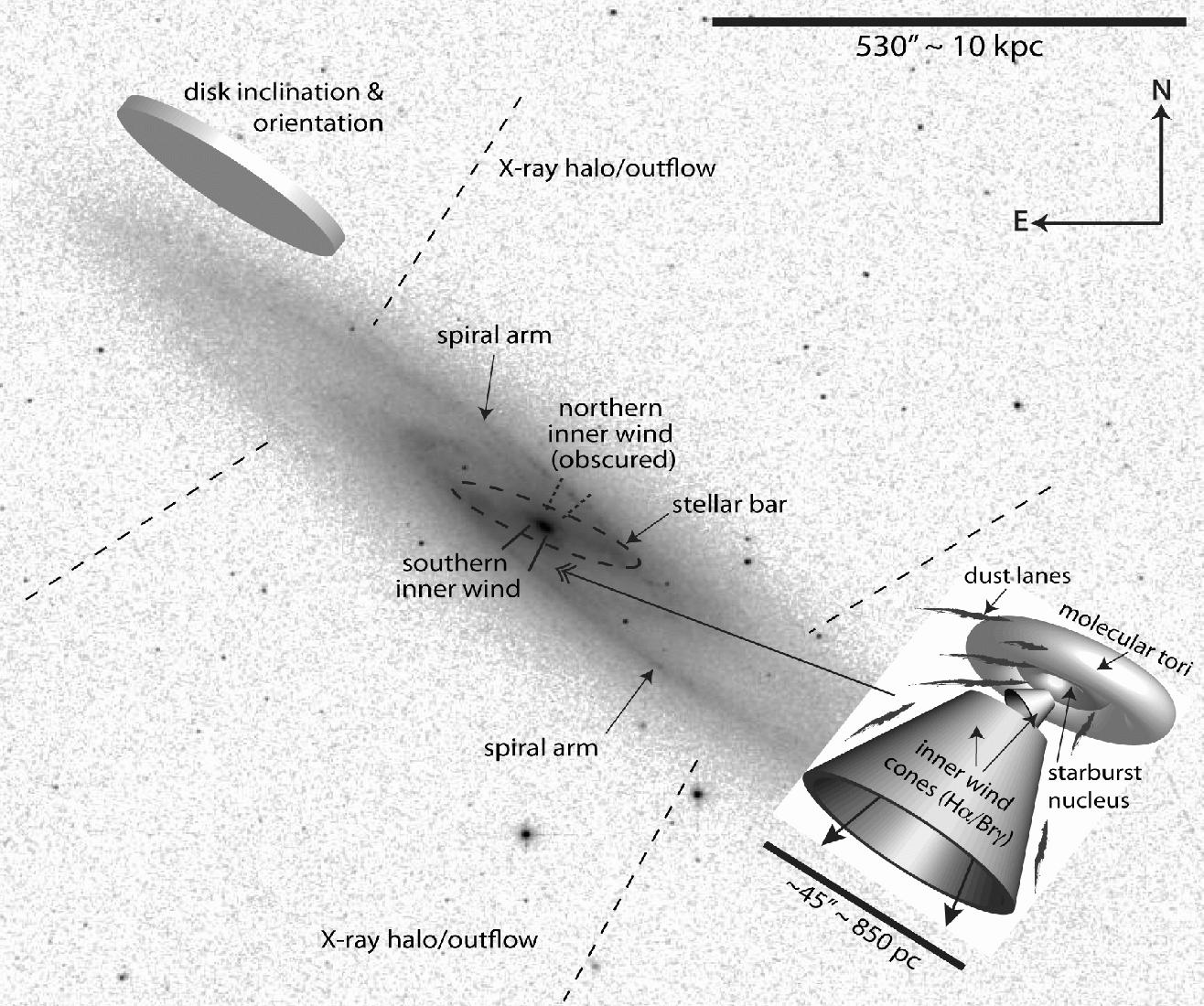}
\caption{Labelled 2MASS $K_{S}$-band image of NGC~253 \citep{jarrett03} showing the main features of the galaxy relevant to our discussion and conclusions. The inset shows a zoomed-in cartoon depicting the configuration of the nuclear regions and inner wind.}
\label{fig:schematic}
\end{figure*}

A schematic overview of the salient points of NGC~253 as discussed above is shown in Fig.~\ref{fig:schematic}. Here we use the 2MASS $K_{S}$-band image \citep{jarrett03} as the basis to illustrate the main large-scale features of the galaxy, as this clearly shows the stellar disk, spiral arms and bar. Further to this, we include an inset, zoomed-in cartoon depicting the configuration of the nuclear regions and inner wind. This shows the circumnuclear molecular tori, and the on-sky orientation and configuration of the inner wind outflow as implied by the VIMOS H$\alpha$, SINFONI Br$\gamma$ spectroscopy and the dust lanes seen on the \textit{HST} F475W and F814W imaging.

Finally, we know that the the superwind must entrain substantial amounts of cooler material (a certain proportion of which also mass-loads the hot flow) through the detection of PAH emission \citep{tacconi05} and far-IR filamentary dust emission extending out to 6--9~kpc \citep{kaneda09}. Furthermore, observations of near-IR H$_2$ emission with \textit{HST} reveal sharply delineated filamentary structures extending out to $\sim$130~pc from the disk that follow the edges of soft X-ray emission \citep{sugai03}. This spatial coincidence suggests that the molecular emission originates in an interaction zone between superwind fluid and surrounding colder molecular gas, again suggesting entrainment of cool material.

%Observations indicate that the circumnuclear molecular disk is highly disturbed on scales of $<$100~pc, and a number of individual 100~pc-scale superbubbles have been detected in the circumnuclear disk in CO \citep{sakamoto06} and NH$_3$ \citep{ott05}. 

%\textbf{To investigate the dynamical evolution of the gas in highly obscured central regions further, we have analysed the Br$\gamma$ emission line properties revealed by the near-IR VLT/SINFONI IFU observations of M\"{u}ller-S\'{a}nchez et al.\ (in prep.; see also \citealt{muller-sanchez10}). While we reserve a detailed description of our analysis to a forthcoming paper, our multi-component line fits reveal that the broadest Br$\gamma$ lines ($\sim$500~\kms\ FWHM) are found near the IR peak and the radio nucleus, where we also see evidence for inflowing ionized gas. They also reveal two localised expanding structures with one appearing to be associated with the radio nucleus region, and one with the IR peak region, suggesting that we are seeing multiple distinct outflow channels developing meaning that the superwind is driven by more than one starburst region.}

%%
\subsection{Evidence for a partial break-out of the cone}
The vertical33 P-V cut (Fig.~\ref{fig:pv_vertical}) indicates that the two main H$\alpha$ line components converge again at a distance of $\sim$55$''$ from the nucleus ($x$$\approx$20$''$ on Fig.~\ref{fig:pv_vertical}). However, the radial velocity map (Fig.~\ref{fig:Ha_vel}) in the region where the vertical33 pseudo-slit cuts through position 5 suggests this convergence may only be a localised phenomenon. The point at which we see the convergence is at the limit of the limb-brightened cone region identified from the deep H$\alpha$ images and soft \textit{Chandra} X-ray emission by \citet{strickland00} (Fig.~\ref{fig:finder}). This therefore indicates that we may be seeing a remnant shell of the outflow. 

However, the outflow is known to extend to $>$10$'$ ($>$12~kpc) from the large-scale soft X-ray emission morphology seen by \textit{ROSAT} \citep{pietsch00, strickland02}, implying that the wind must have flowed out past this inner region in the past. This suggests that the wind is only quasi-steady, and that successive waves of outflowing gas driven by distinct episodes of nuclear star formation (perhaps also linked to nuclear gas morphology and/or infalling gas kinematics) result in nested superbubbles that take time to rupture. At 55$''$ our model predicts a deprojected outflow speed of 385~\kms, giving a dynamical age of $\sim$2.5~Myr, meaning that this bubble must have been driven by the most recent burst. Given the small number of massive star clusters that can be identified ($\sim$10--12), and their compact arrangement within the $\sim$200~pc nuclear region, it would be quite surprising if the starburst did not have quiescent and active periods.

Further evidence for the remnants of a ruptured superbubble may lie in the series of H$\alpha$ clumps identified by \citet{strickland00} in their deep H$\alpha$ image (that they refer to as ``additional clouds''), \citet[][that they collectively refer to as H\two\ region `B']{matsubayashi09} and our reduction of the ESO WFI H$\alpha$ image (Fig.~\ref{fig:finder}a). These also clearly show up in the position 5 H$\alpha$ C1 flux map (although some of the knots fall in the missing Q3 section). The corresponding C1 velocity map shows no evidence for any velocity variations in these regions, meaning that the clumps are kinematically indistinct from their surroundings. (We do see spatially coincident redshifted gas in the C2 map, but this component is very faint in this region.) \citet{matsubayashi09} argue that these clumps (their H\two\ region B) are not part of the wind, but H\two\ regions within the disk, because their [N\two]/H$\alpha$ and [S\two]/H$\alpha$ line ratios are consistent with similar H\two\ regions in other disk galaxies. Further investigation at higher spatial resolution and S/N is required to distinguish whether these clumps are really associated with the wind or simply part of the galaxy disk.

\subsection{The northern disk-halo interface}
The SparsePak observations (Section~\ref{sect:sp_maps}) show that the H$\alpha$ emission from the region to the north-west of the nucleus (north-western part of position SP1 and all of SP2) remains quiescent and well-behaved. Interestingly, no evidence for a northern counterpart to the southern outflow is found in H$\alpha$: only one component can be detected (Figs.~\ref{fig:sp_Ha_vel} and \ref{fig:pv_vertical_sp}), and its line widths remain $<$80~\kms, with most below $<$50~\kms\ (Fig.~\ref{fig:sp_Ha_fwhm}). Radial velocities are instead consistent with simple disk rotation, with a smooth gradient being seen across the field between $-40$ and +25~\kms\ from the north-east to south-west (although this gradient does not show up on the map due to the colour bar scaling). Since there is clear evidence for the northern outflow in soft X-rays \citep{pietsch00}, why might there be no signatures of it in H$\alpha$? In the X-ray map, absorption by the inclined disk (the inclination of NGC~253 is such that the northern outflow is tipped away from the observer and the disk is in the foreground) dramatically reduces the X-ray emission flux up to $R$$\sim$3~kpc \citep{pietsch00} from the nucleus. Since our SparsePak observations are well within this 3~kpc, we attribute the lack of any outflow signatures in H$\alpha$ also to obscuration by the foreground disk.

%Line ratios fall dramatically just to the north-west of the nucleus, along a band parallel to the galaxy major axis (Fig.~\ref{fig:sp_NII_Ha}). 

The gradual rise in line ratios towards the north-west of position SP2, out to $>$100$''$ ($>$2~kpc) from the nucleus (Fig.~\ref{fig:sp_NII_Ha}) is interesting. The rise is particularly prominent in [S\two]/H$\alpha$, although both ratios are high enough to be consistent with a significant contribution from shock excitation. However, since there is no kinematic evidence for the northern outflow, shock excitation would seem unlikely. It is known that diffuse ionized gas (DIG) in normal spirals can have high [N\two]/H$\alpha$ and [S\two]/H$\alpha$ ratios \citep{collins01, otte02}. That this is not due to shock heating comes from high resolution studies of the Galactic DIG, which has the the same emission line ratio characteristics, but where very precise line widths can be measured \citep[e.g.][]{reynolds77, reynolds85}. Thus, we conclude that the high line ratios result from excited DIG in the halo, with some contribution from underlying disk emission.
%More probable is that underlying stellar H$\alpha$ absorption is removing flux from the nebular emission component, resulting in higher line ratios. Inspection of a number of H$\alpha$ profiles from the north-west of position 2 does not reveal any obvious evidence of an absorption component, although the S/N here is fairly low, and there is some residual contamination from Galactic H$\alpha$ emission that was not fully removed by the sky subtraction. We conclude that H$\alpha$ absorption must be the culprit for the elevated line ratios.

%%
\subsection{The connection between the inner wind and the kpc-scale X-ray superbubbles}\label{sect:connection}
As mentioned previously, apart from the nuclear outflow (inner wind) focussed on in this paper, ``horn-shaped'' lobes of diffuse UV, X-ray and far-IR emission are observed to reach out to a much larger scale ($>$8~kpc) perpendicular to the disk \citep{pietsch00, strickland02, bauer08, kaneda09}. In this section we briefly discuss how the two components of the outflow may be connected. 
%Are the large-scale lobes powered by the conical inner wind flow forming a type of super mushroom cloud? Or, more specifically, does the nuclear starburst power the large-scale bubbles?
Does the nuclear starburst-driven inner wind flow examined in this paper power the large-scale lobes, or are they formed through another mechanism?

%Given the wide spacing of the X-ray ``horns'' compared to the size of the conical inner wind flow,  what can we say about increasing wind radial expansion with increasing vertical height?  E.g. to what degree is it trying to form a  with a spherical shell expanding into the N253 halo?

Assessing all the evidence, it is not clear to us that there is a link between the obviously nuclear starburst-driven inner wind and the large-scale X-ray lobes. Firstly there appears to be a large disparity in scales between the collimated, conical inner wind flow and the widely spaced X-ray ``horns'', suggesting an unphysically rapid expansion in the radial direction with increasing vertical height. Without further confinement, it is not clear how an X-shaped morphology could form. Recent observations of the synchrotron halo of NGC~253 suggest that a cosmic-ray (CR) driven ``disk-wind'' may play a significant part in driving material out into the halo and create the large-scale lobes \citep{heesen09a, heesen09b}. The fact that these authors find a remarkably constant vertical CR bulk speed over the entire extent of the disk shows that the CRs do not originate from a localised (centrally located) source, but from the whole disk. Furthermore, the vertical bulk speed of the CR-driven gas is large ($\sim$280~\kms), meaning the kinetic energy of CR-driven gas is certainly high enough to expel material out of the galactic plane.

On large-scales, the magnetic field structure of NGC~253 resembles the X-shape of the H$\alpha$ and soft X-ray lobes \citep{strickland02, heesen09a}. \citet{heesen09b} found that both a disk wind model or a superwind model in conjunction with a large-scale dynamo action could explain the X-shaped halo magnetic field. However, they argue, such X-shaped magnetic field structures have been observed in several edge-on galaxies, not all with a central starburst and superwind. This suggests that the X-shaped halo field might be due solely to the disk wind. The superwind may then be (partly) collimated by the halo field on large-scales, or perhaps the superwind may act further to amplify and align the halo field by compression of the lobes of the expanding superbubbles.

As mentioned in the introduction, \citet{boomsma05} interpret the presence of a substantial mass of extraplanar H\one\ as a result of stellar feedback. \citet{heesen09b} argue that the asymmetric distribution of the gas makes a minor merger origin very unlikely, and go on to surmise that a disk wind is a good candidate for lifting this H\one\ gas out into the halo. However, to lift that amount of material to such large distances implies that a significant injection of energy would be needed ($\sim$$5\times 10^{55}$~erg) over a reasonably long time ($\sim$50~Myr). The disk does indeed contain a great many compact, bright H\two\ regions as shown in our WFI H$\alpha$ image, and the existence of strong radio emission outside the 200~pc starburst region \citep{carilli92, beck94, carilli96, heesen09a} suggests a significant amount of star formation. \citet{ulvestad00} estimates that the region outside the central starburst may account for at least 20\% of the recent star formation in NGC~253. Energetically, therefore, it seems feasible that the plumes are powered by a disk wind and not the starburst-driven outflow. Recent, deep imaging by \citet{davidge10} show that the north east portion of the NGC~253 disk seems to have experienced excess star formation for the last few 10s--100s Myr. Although the `disk-wind' mechanism was not mentioned specifically, the suggestion of a causal relationship between the elevated SFRs and the gas emission through feedback from a `boiling disk' \citep{sofue94}� was made.

%If the CR-driven wind follows similar dynamics to the CRs themselves (i.e.\ an outflow speed that is constant over the whole disk) then angular momentum (a.m.) is transported to large galactocentric radii, and that this a.m.\ loss is much more effective than in the case of just a superwind \citep{zirakashvili96, heesen09b}.

%The extended, minor axis morphology of both the lower ionization state X-ray gas (below 0.5 keV) and the O\,{\sc viii} emission \citep{bauer07} is not limb-brightened, suggesting that this emission may be tracing the hot wind outflow itself through clumps embedded in the flow, and therefore that the wind is mass-loaded.

%%%%%%%%%%%%%%%%%%%%%%%%%%%%%%%%
\section{Summary} \label{sect:summary}
In this paper we have presented high resolution optical integral field unit (IFU) observations and associated deep H$\alpha$ imaging of ionized gas in one of the nearest nuclear starburst galaxies with a large-scale superwind, NGC 253 (3.9~Mpc), in an attempt to understand the outflow in more detail. Our main results and conclusions can be summarised as follows:

\begin{itemize}
  \item We confirm the presence of an expanding outflow cone between radii of 280--660~pc from the nucleus, oriented along the minor axis. This structure is well-defined in the H$\alpha$ morphology and kinematics. Dynamical modelling indicates a wide opening angle ($\sim$60$^{\circ}$), an inclination consistent with being normal to the disk (12$^{\circ}$), and outflow speeds of a few 100~\kms.
  \item The [N\two]/H$\alpha$ and [S\two]/H$\alpha$ line ratio maps suggest a great deal of the gas in the wind is shocked, with localised regions of high line ratios most likely representing pockets/filaments of strongly shocked gas. H$\alpha$ line widths of $>$400~\kms\ (FWHM) are measured in regions associated with the wind cone, suggesting that there is material filling the cone in some regions. 
  \item By comparison to VLT/SINFONI IFU near-IR emission line data (M\"{u}ller-S\'{a}nchez et al.\ in prep.), we can trace the outflow back to the nuclear driving sources, and find evidence for multiple distinct outflow channels implying that the superwind is driven by more than one starburst region. From the kinematics, the cone appears partially closed in one place. These two pieces of evidence together strongly suggest that the wind is only quasi-steady, and that successive waves of outflowing gas have been driven by distinct episodes of nuclear star formation to form the structure we see today.
  \item The location and size-scales of the nested tori seen in SiO emission by \citet{garcia-burillo00} suggest that they may play a significant role in initially confining or collimating the outflow. It is thought that molecular gas is channelled inwards through the action of the bar \citep{sorai00}. Thus, the ram pressure of this infalling material acting against the ram pressure of the wind outflow may also act to collimate the flow somewhat, as discussed by \citet{t-t98}.% hydrodynamical models suggest that . This therefore leads us to believe that this is the mechanism by which the inner wind is initially confined.
  \item The SINFONI near-IR emission line measurements show that the outflow first exits the nuclear region in an easterly direction, then quickly diverts towards the direction of the minor axis. This deflection, also visible in our deep WFI H$\alpha$ imaging, is supported by the changing orientation of the extraplanar dust lanes seen on the \textit{HST} F814W image with distance. Since the toroidal circumnuclear molecular disk(s) and the larger-scale stellar bar is oriented at PA $\approx$ 70$^{\circ}$, we hypothesise that these components have a strong influence on the direction of the flow in these inner regions, and that the hot wind flows not only down the pressure gradient but down any local gas density gradients as well.
  \item No evidence is found in our optical H$\alpha$ SparsePak observations for an outflow on the north-western side out to $>$2~kpc. We surmise this is due to obscuration by the foreground disk \citep[c.f.][]{pietsch00}.
  \item The connection between the inner ($R$$<$1~kpc) H$\alpha$-bright conical outflow focussed on in this paper and the large-scale ($R$$\lesssim$10~kpc) X-ray ``horns'' is also discussed in the context of whether or not the inner starburst-driven wind powers the large-scale lobes. \citet{heesen09a, heesen09b} find evidence for a cosmic-ray wind driven by the disk itself at speeds of $\sim$280~\kms. We conclude that the X-ray lobes are powered by a cosmic-ray driven disk wind \citep{heesen09a, heesen09b}. Since the disk contains a great many compact, bright H\two\ regions as shown in our WFI H$\alpha$ image, and strong radio emission from the disk suggests a significant amount of star formation is occurring outside the central starburst \citep{carilli92, beck94, carilli96, ulvestad00, heesen09a}, we favour the explanation that the large-scale plumes are powered by a disk wind, with, if anything, only a small contribution from the starburst-driven inner wind.
\end{itemize}

%%%%%%%%%%%%%%%%%%%%%%%%%%%%%%%
\section*{Acknowledgments}
We thank the anonymous referee for comments that led to an improvement of the structure and clarity of this paper. 
MSW would like to thank Rob Sharp for discussions regarding specific IFU reduction techniques, Mischa Schirmer for help with the WFI data reduction, Francisco M\"{u}ller-S\'{a}nchez for discussions about the nuclear region and outflow origins, P.\ Knezek and K.\ Dellenbusch for carrying out the SparsePak observations, and the staff at the WIYN Observatory for their support during their observing run.
JSG's research on starbursts was partially funded by the National Science Foundation through grant AST-0708967 to the University of Wisconsin-Madison.

%%%%%%%%%%%%%%%%%%%%%%%%%%%%%%%%%%
% Bibliography
%%%%%%%%%%%%%%%%%%%%%%%%%%%%%%%%%%
\bibliographystyle{mn2e}
\bibliography{/home/mwestmoq/Dropbox/references}

\begin{thebibliography}{}

\bibitem[\protect\citeauthoryear{{Alonso-Herrero}, {Rieke}, {Rieke} \&
  {Kelly}}{{Alonso-Herrero} et~al.}{2003}]{alonso03}
{Alonso-Herrero} A.,  {Rieke} G.~H.,  {Rieke} M.~J.,    {Kelly} D.~M.,  2003,
  \aj, 125, 1210

\bibitem[\protect\citeauthoryear{{Baade}, {Meisenheimer}, {Iwert}, {Alonso},
  {Augusteijn}, {Beletic}, {Bellemann}, {Benesch}, {B{\"o}hm}, {B{\"o}hnhardt}
  \& {Brewer}}{{Baade} et~al.}{1999}]{baade99}
{Baade} D.,  {Meisenheimer} K.,  {Iwert} O.,  {Alonso} J.,  {Augusteijn} T.,
  {Beletic} J.,  {Bellemann} H.,  {Benesch} W.,  {B{\"o}hm} A.,
  {B{\"o}hnhardt} H.,    {Brewer} J.,  1999, The Messenger, 95, 15

\bibitem[\protect\citeauthoryear{{Bauer}, {Pietsch}, {Trinchieri},
  {Breitschwerdt}, {Ehle}, {Freyberg} \& {Read}}{{Bauer}
  et~al.}{2008}]{bauer08}
{Bauer} M.,  {Pietsch} W.,  {Trinchieri} G.,  {Breitschwerdt} D.,  {Ehle} M.,
  {Freyberg} M.~J.,    {Read} A.~M.,  2008, \aap, 489, 1029

\bibitem[\protect\citeauthoryear{{Beck}, {Carilli}, {Holdaway} \&
  {Klein}}{{Beck} et~al.}{1994}]{beck94}
{Beck} R.,  {Carilli} C.~L.,  {Holdaway} M.~A.,    {Klein} U.,  1994, \aap,
  292, 409

\bibitem[\protect\citeauthoryear{{Bershady}, {Andersen}, {Harker}, {Ramsey} \&
  {Verheijen}}{{Bershady} et~al.}{2004}]{bershady04}
{Bershady} M.~A.,  {Andersen} D.~R.,  {Harker} J.,  {Ramsey} L.~W.,
  {Verheijen} M.~A.~W.,  2004, \pasp, 116, 565

\bibitem[\protect\citeauthoryear{{Boomsma}, {Oosterloo}, {Fraternali}, {van der
  Hulst} \& {Sancisi}}{{Boomsma} et~al.}{2005}]{boomsma05}
{Boomsma} R.,  {Oosterloo} T.~A.,  {Fraternali} F.,  {van der Hulst} J.~M.,
  {Sancisi} R.,  2005, \aap, 431, 65

\bibitem[\protect\citeauthoryear{{Brunthaler}, {Castangia}, {Tarchi}, {Henkel},
  {Reid}, {Falcke} \& {Menten}}{{Brunthaler} et~al.}{2009}]{brunthaler09}
{Brunthaler} A.,  {Castangia} P.,  {Tarchi} A.,  {Henkel} C.,  {Reid} M.~J.,
  {Falcke} H.,    {Menten} K.~M.,  2009, \aap, 497, 103

\bibitem[\protect\citeauthoryear{{Calzetti}, {Harris}, {Gallagher}, {Smith},
  {Conselice}, {Homeier} \& {Kewley}}{{Calzetti} et~al.}{2004}]{calzetti04}
{Calzetti} D.,  {Harris} J.,  {Gallagher} J.~S.,  {Smith} D.~A.,  {Conselice}
  C.~J.,  {Homeier} N.,    {Kewley} L.,  2004, \aj, 127, 1405

\bibitem[\protect\citeauthoryear{{Carilli}}{{Carilli}}{1996}]{carilli96}
{Carilli} C.~L.,  1996, \aap, 305, 402

\bibitem[\protect\citeauthoryear{{Carilli}, {Holdaway}, {Ho} \& {de
  Pree}}{{Carilli} et~al.}{1992}]{carilli92}
{Carilli} C.~L.,  {Holdaway} M.~A.,  {Ho} P.~T.~P.,    {de Pree} C.~G.,  1992,
  \apjl, 399, L59

\bibitem[\protect\citeauthoryear{{Chevalier} \& {Clegg}}{{Chevalier} \&
  {Clegg}}{1985}]{cc85}
{Chevalier} R.~A.,  {Clegg} A.~W.,  1985, \nat, 317, 44

\bibitem[\protect\citeauthoryear{{Collins} \& {Rand}}{{Collins} \&
  {Rand}}{2001}]{collins01}
{Collins} J.~A.,  {Rand} R.~J.,  2001, \apj, 551, 57

\bibitem[\protect\citeauthoryear{{Conselice}}{{Conselice}}{2003}]{conselice03}
{Conselice} C.~J.,  2003, \apjs, 147, 1

\bibitem[\protect\citeauthoryear{{Cooper}, {Bicknell}, {Sutherland} \&
  {Bland-Hawthorn}}{{Cooper} et~al.}{2009}]{cooper09}
{Cooper} J.~L.,  {Bicknell} G.~V.,  {Sutherland} R.~S.,    {Bland-Hawthorn} J.,
   2009, \apj, 703, 330

\bibitem[\protect\citeauthoryear{{Das}, {Anantharamaiah} \& {Yun}}{{Das}
  et~al.}{2001}]{das01}
{Das} M.,  {Anantharamaiah} K.~R.,    {Yun} M.~S.,  2001, \apj, 549, 896

\bibitem[\protect\citeauthoryear{{Davidge}}{{Davidge}}{2010}]{davidge10}
{Davidge} T.~J.,  2010, \apj, 725, 1342

\bibitem[\protect\citeauthoryear{{de Vaucouleurs}}{{de
  Vaucouleurs}}{1978}]{devaucouleurs78}
{de Vaucouleurs} G.,  1978, \apj, 224, 710

\bibitem[\protect\citeauthoryear{{de Vaucouleurs}, {de Vaucouleurs}, {Corwin}
  Jr., {Buta}, {Paturel} \& {Fouque}}{{de Vaucouleurs}
  et~al.}{1991}]{devaucouleurs91}
{de Vaucouleurs} G.,  {de Vaucouleurs} A.,  {Corwin} Jr. H.~G.,  {Buta} R.~J.,
  {Paturel} G.,    {Fouque} P.,  1991, {Third Reference Catalogue of Bright
  Galaxies}

\bibitem[\protect\citeauthoryear{{Demoulin} \& {Burbidge}}{{Demoulin} \&
  {Burbidge}}{1970}]{demoulin70}
{Demoulin} M.~H.,  {Burbidge} E.~M.,  1970, \apj, 159, 799

\bibitem[\protect\citeauthoryear{{Dimeo}}{{Dimeo}}{2005}]{dimeo}
{Dimeo} R.,  2005, PAN User Guide, ﻿ftp://ftp.ncnr.nist.gov/pub/staff/dimeo/pandoc.pdf

\bibitem[\protect\citeauthoryear{{Dopita}}{{Dopita}}{1997}]{dopita97}
{Dopita} M.~A.,  1997, \apjl, 485, L41

\bibitem[\protect\citeauthoryear{{Dopita}, {Fischera}, {Sutherland}, {Kewley},
  {Leitherer}, {Tuffs}, {Popescu}, {van Breugel} \& {Groves}}{{Dopita}
  et~al.}{2006}]{dopita06b}
{Dopita} M.~A.,  {Fischera} J.,  {Sutherland} R.~S.,  {Kewley} L.~J.,
  {Leitherer} C.,  {Tuffs} R.~J.,  {Popescu} C.~C.,  {van Breugel} W.,
  {Groves} B.~A.,  2006, \apjs, 167, 177

\bibitem[\protect\citeauthoryear{{Dopita}, {Kewley}, {Heisler} \&
  {Sutherland}}{{Dopita} et~al.}{2000}]{dopita00}
{Dopita} M.~A.,  {Kewley} L.~J.,  {Heisler} C.~A.,    {Sutherland} R.~S.,
  2000, \apj, 542, 224

\bibitem[\protect\citeauthoryear{{Dopita} \& {Sutherland}}{{Dopita} \&
  {Sutherland}}{1995}]{dopita95}
{Dopita} M.~A.,  {Sutherland} R.~S.,  1995, \apj, 455, 468

\bibitem[\protect\citeauthoryear{{Engelbracht}, {Rieke}, {Rieke}, {Kelly} \&
  {Achtermann}}{{Engelbracht} et~al.}{1998}]{engelbracht98}
{Engelbracht} C.~W.,  {Rieke} M.~J.,  {Rieke} G.~H.,  {Kelly} D.~M.,
  {Achtermann} J.~M.,  1998, \apj, 505, 639

\bibitem[\protect\citeauthoryear{{Fabbiano} \& {Trinchieri}}{{Fabbiano} \&
  {Trinchieri}}{1984}]{fabbiano84}
{Fabbiano} G.,  {Trinchieri} G.,  1984, \apj, 286, 491

\bibitem[\protect\citeauthoryear{{Fern{\'a}ndez-Ontiveros}, {Prieto} \&
  {Acosta-Pulido}}{{Fern{\'a}ndez-Ontiveros}
  et~al.}{2009}]{fernandez-ontiveros09}
{Fern{\'a}ndez-Ontiveros} J.~A.,  {Prieto} M.~A.,    {Acosta-Pulido} J.~A.,
  2009, \mnras, 392, L16

\bibitem[\protect\citeauthoryear{{Forbes}, {Polehampton}, {Stevens}, {Brodie}
  \& {Ward}}{{Forbes} et~al.}{2000}]{forbes00}
{Forbes} D.~A.,  {Polehampton} E.,  {Stevens} I.~R.,  {Brodie} J.~P.,    {Ward}
  M.~J.,  2000, \mnras, 312, 689

\bibitem[\protect\citeauthoryear{{Garc{\'{\i}}a-Burillo},
  {Mart{\'{\i}}n-Pintado}, {Fuente} \& {Neri}}{{Garc{\'{\i}}a-Burillo}
  et~al.}{2000}]{garcia-burillo00}
{Garc{\'{\i}}a-Burillo} S.,  {Mart{\'{\i}}n-Pintado} J.,  {Fuente} A.,
  {Neri} R.,  2000, \aap, 355, 499

\bibitem[\protect\citeauthoryear{{Heesen}, {Beck}, {Krause} \&
  {Dettmar}}{{Heesen} et~al.}{2009a}]{heesen09a}
{Heesen} V.,  {Beck} R.,  {Krause} M.,    {Dettmar} R.,  2009a, \aap, 494, 563

\bibitem[\protect\citeauthoryear{{Heesen}, {Krause}, {Beck} \&
  {Dettmar}}{{Heesen} et~al.}{2009b}]{heesen09b}
{Heesen} V.,  {Krause} M.,  {Beck} R.,    {Dettmar} R.,  2009b, \aap, 506, 1123

\bibitem[\protect\citeauthoryear{{Jarrett}, {Chester}, {Cutri}, {Schneider} \&
  {Huchra}}{{Jarrett} et~al.}{2003}]{jarrett03}
{Jarrett} T.~H.,  {Chester} T.,  {Cutri} R.,  {Schneider} S.~E.,    {Huchra}
  J.~P.,  2003, \aj, 125, 525

\bibitem[\protect\citeauthoryear{{Kaneda}, {Yamagishi}, {Suzuki} \&
  {Onaka}}{{Kaneda} et~al.}{2009}]{kaneda09}
{Kaneda} H.,  {Yamagishi} M.,  {Suzuki} T.,    {Onaka} T.,  2009, \apjl, 698,
  L125

\bibitem[\protect\citeauthoryear{{Karachentsev}, {Grebel}, {Sharina},
  {Dolphin}, {Geisler}, {Guhathakurta}, {Hodge}, {Karachentseva}, {Sarajedini}
  \& {Seitzer}}{{Karachentsev} et~al.}{2003}]{karachentsev03}
{Karachentsev} I.~D.,  {Grebel} E.~K.,  {Sharina} M.~E.,  {Dolphin} A.~E.,
  {Geisler} D.,  {Guhathakurta} P.,  {Hodge} P.~W.,  {Karachentseva} V.~E.,
  {Sarajedini} A.,    {Seitzer} P.,  2003, \aap, 404, 93

\bibitem[\protect\citeauthoryear{{Kewley}, {Dopita}, {Sutherland}, {Heisler} \&
  {Trevena}}{{Kewley} et~al.}{2001}]{kewley01}
{Kewley} L.~J.,  {Dopita} M.~A.,  {Sutherland} R.~S.,  {Heisler} C.~A.,
  {Trevena} J.,  2001, \apj, 556, 121

\bibitem[\protect\citeauthoryear{{Knudsen}, {Walter}, {Weiss}, {Bolatto},
  {Riechers} \& {Menten}}{{Knudsen} et~al.}{2007}]{knudsen07}
{Knudsen} K.~K.,  {Walter} F.,  {Weiss} A.,  {Bolatto} A.,  {Riechers} D.~A.,
   {Menten} K.,  2007, \apj, 666, 156

\bibitem[\protect\citeauthoryear{{Kornei} \& {McCrady}}{{Kornei} \&
  {McCrady}}{2009}]{kornei09}
{Kornei} K.~A.,  {McCrady} N.,  2009, \apj, 697, 1180

\bibitem[\protect\citeauthoryear{{Le F{\`e}vre}, {Saisse}, {Mancini},
  {Brau-Nogue}, {Caputi}, {Castinel}, {D'Odorico}, {Garilli}, {Kissler-Patig},
  {Lucuix}, {Mancini}, {Pauget}, {Sciarretta}, {Scodeggio}, {Tresse} \&
  {Vettolani}}{{Le F{\`e}vre} et~al.}{2003}]{le-fevre03}
{Le F{\`e}vre} O.,  {Saisse} M.,  {Mancini} D.,  {Brau-Nogue} S.,  {Caputi} O.,
   {Castinel} L.,  {D'Odorico} S.,  {Garilli} B.,  {Kissler-Patig} M.,
  {Lucuix} C.,  {Mancini} G.,  {Pauget} A.,  {Sciarretta} G.,  {Scodeggio} M.,
  {Tresse} L.,    {Vettolani} G.,  2003, in {M.~Iye \& A.~F.~M.~Moorwood} ed.,
  Society of Photo-Optical Instrumentation Engineers (SPIE) Conference Series
  Vol.~4841 of Presented at the Society of Photo-Optical Instrumentation
  Engineers (SPIE) Conference, {Commissioning and performances of the VLT-VIMOS
  instrument}.
pp 1670--1681

\bibitem[\protect\citeauthoryear{{L{\'{\i}}pari}, {Mediavilla},
  {Garcia-Lorenzo}, {D{\'{\i}}az}, {Acosta-Pulido}, {Ag{\"u}ero}, {Taniguchi},
  {Dottori} \& {Terlevich}}{{L{\'{\i}}pari} et~al.}{2004b}]{lipari04b}
{L{\'{\i}}pari} S.,  {Mediavilla} E.,  {Garcia-Lorenzo} B.,  {D{\'{\i}}az}
  R.~J.,  {Acosta-Pulido} J.,  {Ag{\"u}ero} M.~P.,  {Taniguchi} Y.,  {Dottori}
  H.,    {Terlevich} R.,  2004b, \mnras, 355, 641

\bibitem[\protect\citeauthoryear{{L{\'{\i}}pari}, {D{\'{\i}}az}, {Forte},
  {Terlevich}, {Taniguchi}, {Ag{\"u}ero}, {Alonso-Herrero}, {Mediavilla} \&
  {Zepf}}{{L{\'{\i}}pari} et~al.}{2004a}]{lipari04a}
{L{\'{\i}}pari} S.~L.,  {D{\'{\i}}az} R.~J.,  {Forte} J.~C.,  {Terlevich} R.,
  {Taniguchi} Y.,  {Ag{\"u}ero} M.~P.,  {Alonso-Herrero} A.,  {Mediavilla} E.,
    {Zepf} S.,  2004a, \mnras, 354, L1

\bibitem[\protect\citeauthoryear{{Malin} \& {Hadley}}{{Malin} \&
  {Hadley}}{1997}]{malin97}
{Malin} D.,  {Hadley} B.,  1997, \pasa, 14, 52

\bibitem[\protect\citeauthoryear{{Martin} \& {Bouch{\'e}}}{{Martin} \&
  {Bouch{\'e}}}{2009}]{martin09}
{Martin} C.~L.,  {Bouch{\'e}} N.,  2009, \apj, 703, 1394

\bibitem[\protect\citeauthoryear{{Matsubayashi}, {Sugai}, {Hattori}, {Kawai},
  {Ozaki}, {Kosugi}, {Ishigaki} \& {Shimono}}{{Matsubayashi}
  et~al.}{2009}]{matsubayashi09}
{Matsubayashi} K.,  {Sugai} H.,  {Hattori} T.,  {Kawai} A.,  {Ozaki} S.,
  {Kosugi} G.,  {Ishigaki} T.,    {Shimono} A.,  2009, \apj, 701, 1636

\bibitem[\protect\citeauthoryear{{Mattila} \& {Meikle}}{{Mattila} \&
  {Meikle}}{2001}]{mattila01}
{Mattila} S.,  {Meikle} W.~P.~S.,  2001, \mnras, 324, 325

\bibitem[\protect\citeauthoryear{{Mauersberger}, {Henkel}, {Wielebinski},
  {Wiklind} \& {Reuter}}{{Mauersberger} et~al.}{1996}]{mauersberger96}
{Mauersberger} R.,  {Henkel} C.,  {Wielebinski} R.,  {Wiklind} T.,    {Reuter}
  H.,  1996, \aap, 305, 421

\bibitem[\protect\citeauthoryear{{McCarthy}, {van Breugel} \&
  {Heckman}}{{McCarthy} et~al.}{1987}]{mccarthy87}
{McCarthy} P.~J.,  {van Breugel} W.,    {Heckman} T.,  1987, \aj, 93, 264

\bibitem[\protect\citeauthoryear{{Mohan}, {Anantharamaiah} \& {Goss}}{{Mohan}
  et~al.}{2002}]{mohan02}
{Mohan} N.~R.,  {Anantharamaiah} K.~R.,    {Goss} W.~M.,  2002, \apj, 574, 701

\bibitem[\protect\citeauthoryear{{M{\"u}ller-S{\'a}nchez},
  {Gonz{\'a}lez-Mart{\'{\i}}n}, {Fern{\'a}ndez-Ontiveros}, {Acosta-Pulido} \&
  {Prieto}}{{M{\"u}ller-S{\'a}nchez} et~al.}{2010}]{muller-sanchez10}
{M{\"u}ller-S{\'a}nchez} F.,  {Gonz{\'a}lez-Mart{\'{\i}}n} O.,
  {Fern{\'a}ndez-Ontiveros} J.~A.,  {Acosta-Pulido} J.~A.,    {Prieto} M.~A.,
  2010, \apj, 716, 1166

\bibitem[\protect\citeauthoryear{{Munoz-Tunon}, {Vilchez} \&
  {Castaneda}}{{Munoz-Tunon} et~al.}{1993}]{m-t93}
{Munoz-Tunon} C.,  {Vilchez} J.~M.,    {Castaneda} H.~O.,  1993, \aap, 278, 364

\bibitem[\protect\citeauthoryear{{Nakai}, {Inoue}, {Miyazawa}, {Miyoshi} \&
  {Hall}}{{Nakai} et~al.}{1995}]{nakai95}
{Nakai} N.,  {Inoue} M.,  {Miyazawa} K.,  {Miyoshi} M.,    {Hall} P.,  1995,
  \pasj, 47, 771

\bibitem[\protect\citeauthoryear{{Oey}, {Dopita}, {Shields} \& {Smith}}{{Oey}
  et~al.}{2000}]{oey00}
{Oey} M.~S.,  {Dopita} M.~A.,  {Shields} J.~C.,    {Smith} R.~C.,  2000, \apjs,
  128, 511

\bibitem[\protect\citeauthoryear{{Ott}, {Weiss}, {Henkel} \& {Walter}}{{Ott}
  et~al.}{2005}]{ott05}
{Ott} J.,  {Weiss} A.,  {Henkel} C.,    {Walter} F.,  2005, \apj, 629, 767

\bibitem[\protect\citeauthoryear{{Otte}, {Gallagher} III \& {Reynolds}}{{Otte}
  et~al.}{2002}]{otte02}
{Otte} B.,  {Gallagher} III J.~S.,    {Reynolds} R.~J.,  2002, \apj, 572, 823

\bibitem[\protect\citeauthoryear{{Paglione}, {Tosaki} \& {Jackson}}{{Paglione}
  et~al.}{1995}]{paglione95}
{Paglione} T.~A.~D.,  {Tosaki} T.,    {Jackson} J.~M.,  1995, \apjl, 454, L117+

\bibitem[\protect\citeauthoryear{{Paglione}, {Yam}, {Tosaki} \&
  {Jackson}}{{Paglione} et~al.}{2004}]{paglione04}
{Paglione} T.~A.~D.,  {Yam} O.,  {Tosaki} T.,    {Jackson} J.~M.,  2004, \apj,
  611, 835

\bibitem[\protect\citeauthoryear{{Pettini}, {Shapley}, {Steidel}, {Cuby},
  {Dickinson}, {Moorwood}, {Adelberger} \& {Giavalisco}}{{Pettini}
  et~al.}{2001}]{pettini01}
{Pettini} M.,  {Shapley} A.~E.,  {Steidel} C.~C.,  {Cuby} J.-G.,  {Dickinson}
  M.,  {Moorwood} A.~F.~M.,  {Adelberger} K.~L.,    {Giavalisco} M.,  2001,
  \apj, 554, 981

\bibitem[\protect\citeauthoryear{{Pietsch}, {Vogler}, {Klein} \&
  {Zinnecker}}{{Pietsch} et~al.}{2000}]{pietsch00}
{Pietsch} W.,  {Vogler} A.,  {Klein} U.,    {Zinnecker} H.,  2000, \aap, 360,
  24

\bibitem[\protect\citeauthoryear{{Pittard}, {Dyson}, {Falle} \&
  {Hartquist}}{{Pittard} et~al.}{2005}]{pittard05}
{Pittard} J.~M.,  {Dyson} J.~E.,  {Falle} S.~A.~E.~G.,    {Hartquist} T.~W.,
  2005, \mnras, 361, 1077

\bibitem[\protect\citeauthoryear{{Pottasch}}{{Pottasch}}{1980}]{pottasch80}
{Pottasch} S.~R.,  1980, \aap, 89, 336

\bibitem[\protect\citeauthoryear{{Prada}, {Gutierrez} \& {McKeith}}{{Prada}
  et~al.}{1998}]{prada98}
{Prada} F.,  {Gutierrez} C.~M.,    {McKeith} C.~D.,  1998, \apj, 495, 765

\bibitem[\protect\citeauthoryear{{Protassov}, {van Dyk}, {Connors}, {Kashyap}
  \& {Siemiginowska}}{{Protassov} et~al.}{2002}]{protassov02}
{Protassov} R.,  {van Dyk} D.~A.,  {Connors} A.,  {Kashyap} V.~L.,
  {Siemiginowska} A.,  2002, \apj, 571, 545

\bibitem[\protect\citeauthoryear{{Rekola}, {Richer}, {McCall}, {Valtonen},
  {Kotilainen} \& {Flynn}}{{Rekola} et~al.}{2005}]{rekola05}
{Rekola} R.,  {Richer} M.~G.,  {McCall} M.~L.,  {Valtonen} M.~J.,  {Kotilainen}
  J.~K.,    {Flynn} C.,  2005, \mnras, 361, 330

\bibitem[\protect\citeauthoryear{{Reynolds}}{{Reynolds}}{1985}]{reynolds85}
{Reynolds} R.~J.,  1985, \apjl, 298, L27

\bibitem[\protect\citeauthoryear{{Reynolds}, {Roesler} \& {Scherb}}{{Reynolds}
  et~al.}{1977}]{reynolds77}
{Reynolds} R.~J.,  {Roesler} F.~L.,    {Scherb} F.,  1977, \apj, 211, 115

\bibitem[\protect\citeauthoryear{{Rieke}, {Lebofsky}, {Thompson}, {Low} \&
  {Tokunaga}}{{Rieke} et~al.}{1980}]{rieke80}
{Rieke} G.~H.,  {Lebofsky} M.~J.,  {Thompson} R.~I.,  {Low} F.~J.,
  {Tokunaga} A.~T.,  1980, \apj, 238, 24

\bibitem[\protect\citeauthoryear{{Sakamoto}, {Ho}, {Iono}, {Keto}, {Mao},
  {Matsushita}, {Peck}, {Wiedner}, {Wilner} \& {Zhao}}{{Sakamoto}
  et~al.}{2006}]{sakamoto06}
{Sakamoto} K.,  {Ho} P.~T.~P.,  {Iono} D.,  {Keto} E.~R.,  {Mao} R.,
  {Matsushita} S.,  {Peck} A.~B.,  {Wiedner} M.~C.,  {Wilner} D.~J.,    {Zhao}
  J.,  2006, \apj, 636, 685

\bibitem[\protect\citeauthoryear{{Schulz} \& {Wegner}}{{Schulz} \&
  {Wegner}}{1992}]{schulz92}
{Schulz} H.,  {Wegner} G.,  1992, \aap, 266, 167

\bibitem[\protect\citeauthoryear{{Scoville}, {Soifer}, {Neugebauer},
  {Matthews}, {Young} \& {Yerka}}{{Scoville} et~al.}{1985}]{scoville85}
{Scoville} N.~Z.,  {Soifer} B.~T.,  {Neugebauer} G.,  {Matthews} K.,  {Young}
  J.~S.,    {Yerka} J.,  1985, \apj, 289, 129

\bibitem[\protect\citeauthoryear{{Shapley}, {Steidel}, {Pettini} \&
  {Adelberger}}{{Shapley} et~al.}{2003}]{shapley03}
{Shapley} A.~E.,  {Steidel} C.~C.,  {Pettini} M.,    {Adelberger} K.~L.,  2003,
  \apj, 588, 65

\bibitem[\protect\citeauthoryear{{Sharp} \& {Bland-Hawthorn}}{{Sharp} \&
  {Bland-Hawthorn}}{2010}]{sharp10}
{Sharp} R.~G.,  {Bland-Hawthorn} J.,  2010, \apj, 711, 818

\bibitem[\protect\citeauthoryear{{Shopbell} \& {Bland-Hawthorn}}{{Shopbell} \&
  {Bland-Hawthorn}}{1998}]{shopbell98}
{Shopbell} P.~L.,  {Bland-Hawthorn} J.,  1998, \apj, 493, 129

\bibitem[\protect\citeauthoryear{{Sofue}, {Wakamatsu} \& {Malin}}{{Sofue}
  et~al.}{1994}]{sofue94}
{Sofue} Y.,  {Wakamatsu} K.,    {Malin} D.~F.,  1994, \aj, 108, 2102

\bibitem[\protect\citeauthoryear{{Sorai}, {Nakai}, {Kuno}, {Nishiyama} \&
  {Hasegawa}}{{Sorai} et~al.}{2000}]{sorai00}
{Sorai} K.,  {Nakai} N.,  {Kuno} N.,  {Nishiyama} K.,    {Hasegawa} T.,  2000,
  \pasj, 52, 785

\bibitem[\protect\citeauthoryear{{Steffen}, {Koning}, {Wenger}, {Morisset} \&
  {Magnor}}{{Steffen} et~al.}{2010}]{steffen10}
{Steffen} W.,  {Koning} N.,  {Wenger} S.,  {Morisset} C.,    {Magnor} M.,
  2010, ArXiv e-prints

\bibitem[\protect\citeauthoryear{{Strickland}, {Heckman}, {Weaver}, {Hoopes} \&
  {Dahlem}}{{Strickland} et~al.}{2002}]{strickland02}
{Strickland} D.~K.,  {Heckman} T.~M.,  {Weaver} K.~A.,  {Hoopes} C.~G.,
  {Dahlem} M.,  2002, \apj, 568, 689

\bibitem[\protect\citeauthoryear{{Strickland} \& {Stevens}}{{Strickland} \&
  {Stevens}}{2000}]{strickland00}
{Strickland} D.~K.,  {Stevens} I.~R.,  2000, \mnras, 314, 511

\bibitem[\protect\citeauthoryear{{Sugai}, {Davies} \& {Ward}}{{Sugai}
  et~al.}{2003}]{sugai03}
{Sugai} H.,  {Davies} R.~I.,    {Ward} M.~J.,  2003, \apjl, 584, L9

\bibitem[\protect\citeauthoryear{{Tacconi-Garman}, {Sturm}, {Lehnert}, {Lutz},
  {Davies} \& {Moorwood}}{{Tacconi-Garman} et~al.}{2005}]{tacconi05}
{Tacconi-Garman} L.~E.,  {Sturm} E.,  {Lehnert} M.,  {Lutz} D.,  {Davies}
  R.~I.,    {Moorwood} A.~F.~M.,  2005, \aap, 432, 91

\bibitem[\protect\citeauthoryear{{Tenorio-Tagle} \&
  {Mu{\~n}oz-Tu{\~n}{\'o}n}}{{Tenorio-Tagle} \&
  {Mu{\~n}oz-Tu{\~n}{\'o}n}}{1998}]{t-t98}
{Tenorio-Tagle} G.,  {Mu{\~n}oz-Tu{\~n}{\'o}n} C.,  1998, \mnras, 293, 299

\bibitem[\protect\citeauthoryear{{Turner} \& {Ho}}{{Turner} \&
  {Ho}}{1985}]{turner85}
{Turner} J.~L.,  {Ho} P.~T.~P.,  1985, \apjl, 299, L77

\bibitem[\protect\citeauthoryear{{Ulrich}}{{Ulrich}}{1978}]{ulrich78}
{Ulrich} M.,  1978, \apj, 219, 424

\bibitem[\protect\citeauthoryear{{Ulvestad}}{{Ulvestad}}{2000}]{ulvestad00}
{Ulvestad} J.~S.,  2000, \aj, 120, 278

\bibitem[\protect\citeauthoryear{{Ulvestad} \& {Antonucci}}{{Ulvestad} \&
  {Antonucci}}{1997}]{ulvestad97}
{Ulvestad} J.~S.,  {Antonucci} R.~R.~J.,  1997, \apj, 488, 621

\bibitem[\protect\citeauthoryear{{van Dokkum}}{{van
  Dokkum}}{2001}]{vandokkum01}
{van Dokkum} P.~G.,  2001, \pasp, 113, 1420

\bibitem[\protect\citeauthoryear{{Veilleux} \& {Osterbrock}}{{Veilleux} \&
  {Osterbrock}}{1987}]{veilleux87}
{Veilleux} S.,  {Osterbrock} D.~E.,  1987, \apjs, 63, 295

\bibitem[\protect\citeauthoryear{{Walsh} \& {Roy}}{{Walsh} \&
  {Roy}}{1990}]{walsh90}
{Walsh} J.~R.,  {Roy} J.~R.,  1990, in {D.~Baade \& P.~J.~Grosbol} ed.,
  European Southern Observatory Conference and Workshop Proceedings Vol.~34 of
  European Southern Observatory Conference and Workshop Proceedings, {Area
  Spectroscopy and Correction for Differential Atmospheric Refraction}.
pp 95--+

\bibitem[\protect\citeauthoryear{{Watson}, {Gallagher} III, {Holtzman},
  {Hester}, {Mould}, {Ballester}, {Burrows}, {Casertano}, {Clarke}, {Crisp},
  {Evans}, {Griffiths}, {Hoessel}, {Scowen}, {Stapelfeldt}, {Trauger} \&
  {Westphal}}{{Watson} et~al.}{1996}]{watson96}
{Watson} A.~M.,  {Gallagher} III J.~S.,  {Holtzman} J.~A.,  {Hester} J.~J.,
  {Mould} J.~R.,  {Ballester} G.~E.,  {Burrows} C.~J.,  {Casertano} S.,
  {Clarke} J.~T.,  {Crisp} D.,  {Evans} R.,  {Griffiths} R.~E.,  {Hoessel}
  J.~G.,  {Scowen} P.~A.,  {Stapelfeldt} K.~R.,  {Trauger} J.~T.,    {Westphal}
  J.~A.,  1996, \aj, 112, 534

\bibitem[\protect\citeauthoryear{{Weaver}, {Heckman}, {Strickland} \&
  {Dahlem}}{{Weaver} et~al.}{2002}]{weaver02}
{Weaver} K.~A.,  {Heckman} T.~M.,  {Strickland} D.~K.,    {Dahlem} M.,  2002,
  \apjl, 576, L19

\bibitem[\protect\citeauthoryear{{Westmoquette}, {Exter}, {Smith} \&
  {Gallagher}}{{Westmoquette} et~al.}{2007a}]{westm07a}
{Westmoquette} M.~S.,  {Exter} K.~M.,  {Smith} L.~J.,    {Gallagher} J.~S.,
  2007a, \mnras, 381, 894

\bibitem[\protect\citeauthoryear{{Westmoquette}, {Gallagher}, {Smith},
  {Trancho}, {Bastian} \& {Konstantopoulos}}{{Westmoquette}
  et~al.}{2009b}]{westm09b}
{Westmoquette} M.~S.,  {Gallagher} J.~S.,  {Smith} L.~J.,  {Trancho} G.,
  {Bastian} N.,    {Konstantopoulos} I.~S.,  2009b, \apj, 706, 1571

\bibitem[\protect\citeauthoryear{{Westmoquette}, {Smith}, {Gallagher} \&
  {Exter}}{{Westmoquette} et~al.}{2007b}]{westm07b}
{Westmoquette} M.~S.,  {Smith} L.~J.,  {Gallagher} J.~S.,    {Exter} K.~M.,
  2007b, \mnras, 381, 913

\bibitem[\protect\citeauthoryear{{Westmoquette}, {Smith}, {Gallagher},
  {Trancho}, {Bastian} \& {Konstantopoulos}}{{Westmoquette}
  et~al.}{2009a}]{westm09a}
{Westmoquette} M.~S.,  {Smith} L.~J.,  {Gallagher} J.~S.,  {Trancho} G.,
  {Bastian} N.,    {Konstantopoulos} I.~S.,  2009a, \apj, 696, 192

\end{thebibliography}
\bsp

%%%%%%%%%%%%%%%%%%%%%%%%%%%%%%%%%%
% Figures
%%%%%%%%%%%%%%%%%%%%%%%%%%%%%%%%%%

%\begin{figure*}
%\centering
%\includegraphics[width=0.8\textwidth]{../../SparsePak_data/new_daisy/pos2_skysub_absn_test.pdf}
%\caption{Spectra of spaxels highlighted in Fig.~\ref{fig:sp_NII_Ha} showing that there is no evidence for H$\alpha$ absorption affecting the [S\two]/H$\alpha$ line ratios. The dashed line indicates the rest-wavelength of H$\alpha$ where we see residual Galactic emission.}
%\label{fig:absn_test}
%\end{figure*}

\clearpage

\clearpage

%%%%%%%%%%%%%%%%%%%%%%%%%%%%%%%%
%%%%%%%%%%%%%%%%%%%%%%%%%%%%%%%%
\appendix
\section{Decomposing the emission line profile shapes} \label{sect:line_profiles}
%%%%%%%%%%%%%%%%%%%%%%%%%%%%%%%%

Following the methodology first presented by \citet{westm07a}, we fitted multiple Gaussian profile models to each emission line seen in the VIMOS (Section~\ref{sect:VIMOS_obs}) and SparsePak (Section~\ref{sect:SP_obs}) datasets using an \textsc{idl} $\chi^{2}$ fitting package called \textsc{pan} \citep[Peak ANalysis;][]{dimeo} to quantify the gas properties observed in each IFU field. The high S/N and spectral resolution of these data have allowed us to quantify the line profile shapes of these lines to a high degree of accuracy.

%In general, we find the emission lines to be composed of a bright, narrow component (hereafter C1; this was sometimes split into two narrow components, hereafter C1 and C3) overlaid on a fainter, broad component (hereafter C2). 

Each line in each of the 1200 spaxels (VIMOS IFU less Q3) and 82 spaxels (SparsePak IFU) was fitted using a single, double, and triple Gaussian component initial guess (no lines were found to need more than three components). Line fluxes were constrained to be positive and widths to be greater than the instrumental contribution to guard against spurious results. 

%In some regions of our DensePak data, we can also identify an H$\alpha$ absorption component of stellar origin (hereafter C4). Here we add an additional Gaussian absorption component, with a flux constrained to be negative. 

Although R$\approx$15000 long slit echelle spectra taken by one of us (JSG) suggest some differences may exist between the H$\alpha$ and [N\two] emission line profiles, these are not detected in our lower resolution data. Therefore, to fit the H$\alpha$ and [N\two]$\lambda 6583$ lines, for each component we constrained the wavelength difference between the two Gaussian models to be equal to the laboratory difference, and the FWHMs equal to one another. This constrained approach, also taken for fitting the [S\two] doublet, dramatically improves the quality of the fits, particularly for spectra with reduced S/N. Multi-component fits were run several times with different initial guess configurations (widths and wavelengths) in order to account for the varied profile shapes, and the one with the lowest $\chi^2$ fit statistic was kept. However, we note that the $\chi^2$ minimisation routine employed by PAN is very robust with respect to the initial guess parameters. 

To determine how many Gaussian components best fit an observed profile (one, two or three), we used a likelihood ratio test to determine whether a fit of $n$+1 components was more appropriate to an $n$-component fit. This test says that if the ratio of the $\chi^{2}$ statistics for the two fits falls above a certain threshold, then the fits are considered statistically distinguishable, and the one with the lower $\chi^{2}$ can be selected. Here we determine the threshold ratio by visual inspection of a range of spectra and fits. This method is a more generalised form of the formal F-test that we have used in previous work, e.g.\ \citet{westm07a, westm09a}. We have chosen to modify our methodology in light of work we have recently become aware of that cautions against the use of the F-test to test for the presence of a line \citep[or additional line component;][]{protassov02}. In fact these authors show that using any kind of likelihood ratio test in this kind of situation is statistically incorrect, but given the absence of a statistically correct alternative that could be applied sensibly to the volume of data presented here, we have chosen to opt for this generalised $\chi^{2}$ ratio test approach.

This test, however, only tells us which of the fits (single, double or triple component) is most appropriate for the corresponding line profile. Experience has taught us that we need to apply a number of additional, physically motivated tests to filter out well-fit but physically improbable results.

Another addition to our methodology was to bootstrap the uncertainties on the fit parameters using Monte Carlo techniques, assuming that the noise is uncorrelated. Using the fit results determined from the above step as the initial guess, we re-fitted each profile 100 times, each time seeding the spectrum with randomly generated noise based on the standard deviation of the continuum. The standard deviation on each parameter (flux, FWHM and radial velocity) over the 100 results was then adopted as the associated uncertainty for that parameter, while the mean of those results was adopted as the actual value. Uncertainties range from 1--30\% depending on the S/N of the spectrum and how many components are fit.

The final step of the fitting procedure was to assign each fit component to a particular map in such a way as to limit the confusion that might arise during analysis of the results, such as discontinuous spatial regions arising from incorrect component assignments. How we assigned the components in each case is described in the relevant results sections (\ref{sect:VIMOS_maps} and \ref{sect:sp_maps}).

\label{lastpage}
\end{document}